\documentclass[useAMS]{mn2e}
\def\mic {\hbox{\,$\umu$m }}
\def\IRAS{\textit{IRAS}}

\usepackage{graphicx}
\usepackage{subfigure} 
\usepackage{times}

\begin{document}
\title[The SCUBA Local Universe Galaxy Survey -- III.]
{The SCUBA Local Universe Galaxy Survey -- III. Dust along the Hubble sequence}
\author[Catherine Vlahakis, Loretta Dunne, Stephen Eales]
{Catherine Vlahakis$^1$\thanks{E-mail: Catherine.Vlahakis@astro.cf.ac.uk; Steve.Eales@astro.cf.ac.uk}\thanks{Present address: Radioastronomisches Institut, Universit\"at Bonn, Auf dem H\"ugel 71, 53121 Bonn, Germany.}, 
Loretta Dunne$^2$ and
Stephen Eales$^1$\footnotemark[1]\\
$^1$School of Physics and Astronomy, Cardiff University,
\\ 5, The Parade, Cardiff, CF24 3YB\\
$^2$School of Physics and Astronomy, University of Nottingham, 
\\University Park, Nottingham, NG7 2RD\\}
\maketitle

\begin{abstract}
We present new results from the SCUBA Local Universe Galaxy Survey (SLUGS), the first large systematic submillimetre survey of the local Universe. Since our initial survey of a sample of 104 \textit{IRAS}-selected galaxies we have now completed a survey of a sample of 81 optically-selected galaxies, observed with the SCUBA camera on the James Clerk Maxwell Telescope. Since SCUBA is sensitive to the 90\% of dust too cold to radiate significantly in the \IRAS\/ bands our new sample represents the first unbiased survey of dust in galaxies along the whole length of the Hubble sequence.

We find little change in the properties of dust in galaxies along the Hubble sequence, except a marginally significant trend for early-type galaxies to be less luminous submillimetre sources than late-types. We nevertheless detected 6 out of 11 elliptical galaxies, although some of the emission may possibly be synchrotron rather than dust emission. As in our earlier work on \IRAS\/ galaxies we find that the \IRAS\/ and submillimetre fluxes are well-fitted by a two-component dust model with dust emissivity index $\beta$=2. The major difference from our earlier work is that we find the ratio of the mass of cold dust to the mass of warm dust is much higher for our optically-selected galaxies and can reach values of $\sim$1000. Comparison of the results for the \IRAS- and optically-selected samples shows that there is a population of galaxies containing a large proportion of cold dust that is unrepresented in the \IRAS\/ sample.

We derive local submillimetre luminosity and dust mass functions, both directly from our optically-selected SLUGS sample, and by extrapolation from the \textit{IRAS} PSCz survey using the method of Serjeant \& Harrison (by extrapolating the spectral energy distributions of the \textit{IRAS} PSCz survey galaxies out to 850\mic we probe a wider range of luminosities than probed directly by the SLUGS samples), and find excellent agreement between the two. We find them to be well-fitted by Schechter functions except at the highest luminosities. We find that as a consequence of the omission of cold galaxies from the \IRAS\/ sample the luminosity function presented in our earlier work is too low by a factor of 2, reducing the amount of cosmic evolution required between the low-z and high-z Universe.
\end{abstract}


\section{Introduction}
\label{intro}
Relatively little is known about the submillimetre properties of `normal' galaxies in the local Universe. The advent of \IRAS\/ in the 1980s brought the first investigations of dust in relatively large samples of galaxies (e.g. Devereux \& Young 1990), yet the limitations of investigating dust at far-IR wavelengths are marked; the strong temperature dependence of thermal emission means that even a small amount of warm dust can dominate the emission from a substantially larger proportion of cold dust, and \textit{IRAS} is only sensitive to dust with \mbox{$T>30$\,K}. \IRAS\/ studies of `normal' galaxies (e.g. Devereux \& Young 1990) found a high value of the gas-to-dust ratio ($\sim$1000), an order of magnitude higher than found for the Milky Way ($\sim$160; Dunne et al. 2000), indicating that \textit{IRAS} may have `missed' $\sim$90\% of the dust in late-type galaxies. \IRAS\/ also revealed relatively little about the dust in early-type galaxies, since only $\sim$15\% of ellipticals were detected by \IRAS\/ (Bregman et al. 1998). 

The next major step in the study of dust in galaxies is to make observations in the submillimetre waveband \mbox{($100\,\micron\le \lambda \le 1$\,mm)} since the 90\% of dust that is too cold to radiate in the far-IR will be producing most of its emission in this waveband. The advent of the SCUBA camera on the James Clark Maxwell Telescope (JCMT)\footnote{The JCMT is operated by the Joint Astronomy Center on behalf of the UK Particle Physics and Astronomy Research Council, the Netherlands Organisation for Scientific Research and the Canadian National Research Council.} (Holland et al. 1999) opened up the submillimetre waveband for astronomy and  made it possible, for the first time, to investigate the submillimetre emission of a large sample of galaxies; prior to SCUBA only a handful of submillimetre measurements had been made of nearby galaxies, using single-element bolometers. In particular, in contrast to the extensive survey work going on at other wavelengths, prior to SCUBA it was not possible to carry out a large survey in the submillimetre waveband. SCUBA has 2 bolometer arrays (850\mic and 450\,\micron) which operate simultaneously with a field of view of $\sim$2 arcminutes. At 850\mic SCUBA is sensitive to thermal emission from dust with fairly cool temperatures \mbox{($T\geq10$\,K)} so crucially, whereas \textit{IRAS} was only sensitive to warmer dust \mbox{($T>30$\,K)}, SCUBA should trace most of the dust mass.

\subsection{A local submillimetre galaxy survey}
\label{local-survey}
A survey of the dust in nearby galaxies is also important because of the need to interpret the results from surveys of the distant Universe. Many deep SCUBA surveys have been carried out (Smail, Ivison \& Blain 1997; Hughes et al. 1998; Barger et al. 1998, 1999; Blain et al. 1999a; Eales et al. 1999; Lilly et al. 1999; Mortier et al. 2005), but studies of the high redshift Universe, and in particular studies of cosmological evolution (Eales et al. 1999; Blain et al. 1999b), have until now depended critically on assumptions about, rather than measurements of, the submillimetre properties of the \textit{local} Universe. Prior to the existence of a direct local measurement of the submillimetre luminosity function (LF) most deep submillimetre investigations have started from a local \textit{IRAS} 60\mic LF, extrapolating out to submillimetre wavelengths by making assumptions about the average FIR-submm SED. However, as shown by Dunne et al. (2000), this underestimates the local submillimetre LF, and thus a direct measurement of the local submillimetre LF is vital for overcoming this significant limitation in the interpretation of the results of high-redshift surveys. 

The ideal method of carrying out a submillimetre survey of the local Universe would be to survey a large area of the sky and then measure the redshifts of all the submillimetre sources found by the survey. However, with current submillimetre instruments such a survey is effectively impossible since, for example, the field of view of SCUBA is only $\sim$2 arcminutes. The alternative method, and the only one that is currently practical, is to carry out targeted submillimetre observations of galaxies selected from statistically complete samples selected in other wavebands. With an important proviso, explained below, it is then possible to produce an unbiased estimate of the submillimetre LF using `accessible volume' techniques (Avni \& Bahcall 1980) (see Section~\ref{lumfun}).

To this end several years ago we began the SCUBA Local Universe Galaxy Survey (SLUGS). In Papers I and II Dunne et al. (2000, hereafter D00) and Dunne \& Eales (2001, hereafter DE01) presented the results of SCUBA observations of a sample selected at 60\mic (the \textit{IRAS}-Selected sample, hereafter the IRS sample). This paper presents the results of SCUBA observations of an optically-selected sample (hereafter the OS sample).

The accessible volume method will produce unbiased estimates of the LF provided that no class of galaxy is unrepresented in the samples used to construct the LF. D00 produced the first direct measurements of the submillimetre LF and dust mass function (the space-density of galaxies as a function of dust mass) using the IRS sample, but this LF would be biased if there exists a `missed' population of submillimetre-emitting galaxies, i.e. a population \emph{that is not represented at all in the IRS sample}. In this earlier work we found that the slope of the submillimetre LF at lower luminosities was steeper than $-$2 (a submillimetre `Olbers' Paradox'), which  indicated that the IRS sample may not be fully representative of all submillimetre-emitting sources in the local Universe. This `missed' population could consist of cold-dust-dominated galaxies, i.e. galaxies containing large amounts of `cold' dust (at \mbox{$T<25$\,K}), which would be strong emitters at 850\mic but weak 60\,\micron-emitters. The OS sample is selected on the basis of the optical emission from the galaxies and, unlike the IRS sample which was biased towards warmer dust, the OS sample should be free from dust temperature selection effects. The results from the OS sample will therefore test the idea that our earlier IRS sample LF was an underestimate.

\subsection{Previous investigations of cold dust in galaxies}
\label{cold-dust}
The paradigm for dust in galaxies is that there are two main components: (i) a warm component \mbox{(T\,$>$\,30\,K)} arising from dust grains near to star-forming regions and heated by young (OB) stars, and (ii) a cool `cirrus' component \mbox{(T\,=\,15--25\,K)} arising from diffuse dust associated with the HI and heated by the general interstellar radiation field (ISRF) (Cox, Kr\"ugel \& Mezger 1986; Lonsdale Persson \& Helou 1987; Rowan-Robinson \& Crawford 1989). \textit{IRAS} would only have detected the warm component, hence using \textit{IRAS} fluxes alone to estimate dust temperature would result in an overestimate of the dust temperature and an underestimate of the dust mass. Conversely, using the submillimetre to estimate dust masses has clear advantages. The flux is more sensitive to the mass of the emitting material and less sensitive to temperature in the Rayleigh-Jeans part of the Planck function, which is sampled when looking at longer submillimetre wavelengths. 

Studies at the longer wavelengths (170--850\,\micron; e.g. ISO, SCUBA) have confirmed the existence of cold dust components \mbox{($15<T_{d}<25$\,K)}, in line with the theoretical prediction of grain heating by the general ISRF (Cox et al. 1986), both in nearby spiral galaxies and in more IR-luminous/interacting systems (Gu\'elin et al. 1993, 1995; Sievers et al. 1994; Sodroski et al. 1994; Neininger et al. 1996; Braine et al. 1997; Dumke et al. 1997; Alton et al. 1998a,b, 2001; Haas et al. 1998; Davies et al. 1999; Frayer et al. 1999; Papadopoulos \& Seaquist 1999; Xilouris et al. 1999; Haas et al. 2000; DE01; Popescu et al. 2002; Spinoglio et al. 2002; Hippelein et al. 2003; Stevens, Amure \& Gear 2005). Many of these authors find an order of magnitude more dust than \IRAS\/ observations alone would indicate. Alton et al. (1998a), for example, find, by comparing their 200\mic images of nearby galaxies to B-band images, that the cold dust has a greater radial extent than the stars, and conclude that \textit{IRAS} `missed' the majority of dust grains lying in the spiral disks. Other studies find evidence of cold dust components in a large proportion of galaxies. Contursi et al. (2001) find evidence of a cold dust component \mbox{($T\sim22$\,K)} for most of their sample of late-type galaxies; Stickel et al. (2000) find a large fraction of sources in their 170\mic survey have high $S_{170}/S_{100}$ flux ratios and suggest this indicates a cold dust component \mbox{($T\leq20$\,K)} exists in many galaxies; Popescu et al. (2002) find, for their sample of late-type (later than S0) galaxies in the Virgo Cluster, that 30 out of 38 galaxies detected in all three observed wavebands (60, 100 and 170\,\micron) exhibit a cold dust component.

An additional property of dust that can be investigated with submillimetre measurements is the dust emissivity index $\beta$. Dust radiates as a modified Planck function (a `grey-body'), modified by the emissivity term such that $Q_{em}\propto v^{\beta}$. Until recently the value of $\beta$ was quite uncertain, with suggested values lying between 1 and 2 (Hildebrand 1983). Recent multi-wavelength studies of galaxies including submillimetre observations, however,  have consistently found $1.5\le\beta\le2$ with $\beta$=2 tending to be favoured (Chini et al. 1989; Chini \& Kr\"ugel 1993; Braine et al. 1997; Alton et al. 1998b; Bianchi et al. 1998; Frayer et al. 1999; DE01). This agrees with the values found in \textit{COBE}/FIRAS studies of the diffuse ISM in the Galaxy (Masi et al. 1995; Reach et al. 1995; Sodroski et al. 1997).

\subsection{The scope of this paper}
This paper presents the results from the SCUBA Local Universe Galaxy Survey (SLUGS) optically-selected sample. This OS sample is taken from the Center for Astrophysics (CfA) optical redshift survey (Huchra et al. 1983), and includes galaxies drawn from right along the Hubble sequence.

In Section~\ref{data-red} we discuss our observation and data reduction techniques. Section~\ref{results} presents the sample and the results. Section~\ref{properties} presents an analysis of the submillimetre properties of the sample. In Section~\ref{lumfun} we present the local submillimetre luminosity and dust mass functions. We assume a Hubble constant \mbox{$H_{0}$=75 km\,s$^{-1}$ Mpc$^{-1}$} throughout.

\begin{table*}
\centering
\begin{minipage}{14.5cm}
\caption{\label{fluxtab}\small{850\mic flux densities and isothermal SED parameters. (Notes on individual objects are listed in Section~\ref{maps}).}}
\begin{tabular}{lllrrrrrllr}
\hline
(1) & (2) & (3) & (4) & (5) & (6) & (7) & (8) & (9) & (10) & (11) \\
Name & R.A. & Decl. & cz & $S_{60}$ & $S_{100}$ & $S_{850}$ & $\sigma_{850}$ & $T_{dust}$ & $\beta$ & Type\\
& (J2000) & (J2000) & (km\,s$^{-1}$) & (Jy) & (Jy) & (Jy) & (Jy) & (K) & & \\
\hline
UGC 148 & 00 15 51.2 &  +16 05 23 & 4213 & 2.21 & 5.04 & 0.055 & 0.012 & 31.6 & 1.4 & 4\\
NGC 99 & 00 23 59.4 &  +15 46 14 & 5322 & 0.81 & 1.49 & 0.063 & 0.015 & 41.8 & 0.4 & 6\\
PGC 3563 & 00 59 40.1 &  +15 19 51 & 5517 & 0.35 & $^{\displaystyle s}$1.05 & 0.027 & 0.008 & 31.0 & 1.0 & 2\\
NGC 786 & 02 01 24.7 &  +15 38 48 & 4520 & 1.09 & 2.46 & 0.066 & 0.019 & 35.2 & 0.8 & 4M\\
NGC 803 & 02 03 44.7 &  +16 01 52 & 2101 & 0.69 & 2.84 & 0.093 & 0.019 & 27.4 & 1.1 & 5\\
UGC 5129 & 09 37 57.9 &  +25 29 41 & 4059 & 0.27 & 0.92 & $<$0.034 & ... & ... & ... & 1\\
NGC 2954 & 09 40 24.0 &  +14 55 22 & 3821 & $<$0.18 & $<$0.59 & $<$0.027 & ... & ... & ... &-5\\
UGC 5342 & 09 56 42.6 &  +15 38 15 & 4560 & 0.85 & 1.66 & 0.032 & 0.008 & 36.4 & 0.9 & 4\\
PGC 29536 & 10 09 12.4 &  +15 00 19 & 9226 & $<$0.18 & $<$0.52 & $<$0.041 & ... & ... & ... & -5\\
NGC 3209 & 10 20 38.4 &  +25 30 18 & 6161 & $<$0.16 & $<$0.65 & $<$0.022 & ... & ... & ... & -5\\
NGC 3270 & 10 31 29.9 &  +24 52 10 & 6264 & 0.59 & 2.39 & 0.059 & 0.014 & 26.8 & 1.3 & 3\\
NGC 3323 & 10 39 39.0 &  +25 19 22 & 5164 & 1.48 & 3.30 & 0.070 & 0.014 & 34.0 & 1.0 & 5\\
NGC 3689 & 11 28 11.0 &  +25 39 40 & 2739 & $^{\displaystyle s}$2.86 & $^{\displaystyle s}$9.70 & 0.101 & 0.017 & 26.8 & 1.7 & 5\\
UGC 6496 & 11 29 51.4 &  +24 56 16 & 6277 & ... & ... & $<$0.018 & ... & ... & ... & -2\\
PGC 35952 & 11 37 01.8 &  +15 34 14 & 3963 & 0.47 & 1.32 & 0.051 & 0.013 & 32.2 & 0.8 & 4\\
NGC 3799$^{\scriptstyle p}$   &  11 40 09.4   &   +15 19 38 & 3312 & U & U & $<$0.268 & ... & ... & ... & 3\\   
NGC 3800$^{\scriptstyle p}$   &  11 40 13.5   &   +15 20 33 & 3312 & U & U & 0.117 & 0.025 & ... & ... & 3\\  
NGC 3812 & 11 41 07.7 &  +24 49 18 & 3632 & $<$0.23 & $<$0.56 & $<$0.038 & ... & ... & ... & -5\\
NGC 3815 & 11 41 39.3 &  +24 48 02 & 3711 & 0.70 & 1.88 & 0.041 & 0.011 & 31.0 & 1.1 & 2\\
NGC 3920 & 11 50 05.9 &  +24 55 12 & 3635 & 0.75 & 1.68 & 0.034 & 0.009 & 34.0 & 1.0 & -2\\
NGC 3987 & 11 57 20.9 &  +25 11 43 & 4502 & 4.78 & 15.06 & 0.186 & 0.030 & 27.4 & 1.6 & 3\\
NGC 3997 & 11 57 48.2 &  +25 16 14 & 4771 & 1.16 & $^{\displaystyle s}$1.95 & $<$0.023 & ... & ... & ... & 3M\\
NGC 4005 & 11 58 10.1 &  +25 07 20 & 4469 & U & U & $<$0.015 & ... & ...  & ... & 3\\
NGC 4015 & 11 58 42.9 &  +25 02 25 & 4341 & 0.25 & $^{\displaystyle s}$0.80 & $<$0.050 & ... & ... & ...& 10M\\
UGC 7115 & 12 08 05.5 &  +25 14 14 & 6789 & $<$0.20 & $<$0.68 & 0.051 & 0.011 & ... & ... & -5\\
UGC 7157 & 12 10 14.6 &  +25 18 32 & 6019 & $<$0.24 & $<$0.63 & $<$0.032 & ... & ... & ... & -2\\
IC 797 & 12 31 54.7 &  +15 07 26 & 2097 & 0.74 & 2.18 & 0.085 & 0.021 & 31.6 & 0.8 & 6\\
IC 800 & 12 33 56.7 &  +15 21 16 & 2330 & 0.38 & 1.10 & 0.076 & 0.019 & 34.6 & 0.4 & 5\\
NGC 4712 & 12 49 34.2 &  +25 28 12 & 4384 & 0.48 & 2.02 & 0.102 & 0.023 & 28.0 & 0.9 & 4\\
PGC 47122 & 13 27 09.9 &  +15 05 42 & 7060 & $<$0.11 & 0.55 & $<$0.035 & ... & ... & ... &-2\\
MRK 1365 & 13 54 31.1 &  +15 02 39 & 5534 & 4.20 & 6.11 & 0.032 & 0.009 & 35.2 & 1.6 & -2\\
UGC 8872 & 13 57 18.9 &  +15 27 30 & 5529 & $<$0.22 & $<$0.45 & $<$0.021 & ... & ... & ... & -2\\
UGC 8883 & 13 58 04.6 &  +15 18 53 & 5587 & 0.45 & 1.19 & $<$0.040 & ... & ... & ... & 4\\
UGC 8902 & 13 59 02.7 &  +15 33 56 & 7667 & 1.23 & 3.32 & 0.067 & 0.018 & 30.4 & 1.2 & 3\\
IC 979 & 14 09 32.3 &  +14 49 54 & 7719 & $^{\displaystyle s}$$^{\ast}$0.19 & $^{\displaystyle s}$$^{\ast}$0.60 & 0.057 & 0.017 & 34.0$^{\ast}$ & 0.3$^{\ast}$ & 2\\
UGC 9110 & 14 14 13.4 &  +15 37 21 & 4644 & U & U & $<$0.046 & ... & ... & ... & 3\\
NGC 5522 & 14 14 50.3 &  +15 08 48 & 4573 & 2.06 & 4.05 & 0.072 & 0.014 & 35.8 & 1.0 & 3\\
NGC 5953$\dag$$^{\scriptstyle p}$   &  15 34 32.4  &   +15 11 38 & 1965 & U & U & 0.184 & 0.024 & ... & ... & 1 \\    
NGC 5954$\dag$$^{\scriptstyle p}$   &  15 34 35.2  &   +15 11 54 & 1959 & U & U & 0.112 & 0.019 & ... & ... & 6\\   
NGC 5980 & 15 41 30.4 &  +15 47 16 & 4092 & 3.45 & 8.37 & 0.253 & 0.043 & 34.0 & 0.8 & 5\\
IC 1174 & 16 05 26.8 &  +15 01 31 & 4706 & $<$0.18 & $<$0.32 & 0.025 & 0.009 & ... & ... & 0\\
UGC 10200 & 16 05 45.8 &  +41 20 41 & 1972 & 1.41 & 1.67 & $<$0.020 & ... & ... & ...& 2M\\
UGC 10205 & 16 06 40.2 &  +30 05 55 & 6556 & 0.39 & 1.54 & 0.058 & 0.015 & 28.0 & 1.0 & 1\\
NGC 6090 & 16 11 40.7 &  +52 27 24 & 8785 & 6.66 & 8.94 & 0.091 & 0.015 & 40.6 & 1.1 & 10M\\
NGC 6103 & 16 15 44.6 &  +31 57 51 & 9420 & 0.64 & 1.67 & 0.052 & 0.012 & 33.4 & 0.8 & 5\\
NGC 6104 & 16 16 30.6 &  +35 42 29 & 8428 & 0.50 & 1.76 & $<$0.033 & ... & ... & ... & 1\\
IC 1211 & 16 16 51.9 &  +53 00 22 & 5618 & $<$0.12 & $<$0.53 & 0.028 & 0.009 & ... & ... & -5\\
UGC 10325$\S$ & 16 17 30.6 &  +46 05 30 & 5691 & 1.57 & 3.72 & 0.041 & 0.009 & 31.0 & 1.4 & 10M\\
NGC 6127 & 16 19 11.5 &  +57 59 03 & 4831 & $<$0.10 & $<$0.30 & 0.086 & 0.020 & ... & ... & -5\\
NGC 6120 & 16 19 48.1 &  +37 46 28 & 9170 & 3.99 & 8.03 & 0.065 & 0.011 & 32.2 & 1.5 & 8\\
NGC 6126 & 16 21 27.9 &  +36 22 36 & 9759 & $<$0.15 & $<$0.43 & 0.023 & 0.008 & ... & ... & -2\\
NGC 6131 & 16 21 52.2 &  +38 55 57 & 5117 & 0.72 & 2.42 & 0.054 & 0.013 & 28.6 & 1.2 & 6\\
NGC 6137 & 16 23 03.1 &  +37 55 21 & 9303 & $<$0.18 & $<$0.53 & 0.029 & 0.010 & ... & ... & -5\\
NGC 6146 & 16 25 10.3 &  +40 53 34 & 8820 & $<$0.12 & $<$0.48 & 0.028 & 0.007 & ... & ... & -5\\
NGC 6154 & 16 25 30.4 &  +49 50 25 & 6015 &  $<$0.15 & $<$0.36 & $<$0.040 & ... & ... & ... & 1\\
NGC 6155 & 16 26 08.3 &  +48 22 01 & 2418 & 1.90 & 5.45 & 0.116 & 0.022 & 29.8 & 1.2 & 6\\
UGC 10407 & 16 28 28.1 &  +41 13 05 & 8446 & 1.62 & 3.12 & 0.026 & 0.009 & 32.8 & 1.5 & 10M\\
NGC 6166 & 16 28 38.4 &  +39 33 06 & 9100 & $^{\displaystyle s}$0.10 & $^{\displaystyle s}$0.63 & 0.073 & 0.017 & 26.2 & 0.6 & -5\\
NGC 6173 & 16 29 44.8 &  +40 48 42 & 8784 & $<$0.17 &  $<$0.23 & $<$0.024 & ... & ... & ... & -5\\
NGC 6189 & 16 31 40.9 &  +59 37 34 & 5638 & 0.75 & 2.57 & 0.072 & 0.019 & 28.6 & 1.1 & 6\\
NGC 6190 & 16 32 06.7 &  +58 26 20 & 3351 & 0.58 & 2.37 & 0.099 & 0.024 & 28.0 & 1.0 & 6\\
\end{tabular}
\end{minipage}
\end{table*}

\begin{table*}
\centering
\begin{minipage}{14.5cm}
\contcaption{}
\begin{tabular}{lllrrrrrllr}
\hline
(1) & (2) & (3) & (4) & (5) & (6) & (7) & (8) & (9) & (10) & (11) \\
Name & R.A. & Decl. & cz & $S_{60}$ & $S_{100}$ & $S_{850}$ & $\sigma_{850}$ & $T_{dust}$ & $\beta$ & Type\\
& (J2000) & (J2000) & (km\,s$^{-1}$) & (Jy) & (Jy) & (Jy) & (Jy) & (K) & &\\
\hline
NGC 6185 & 16 33 17.8 &  +35 20 32 & 10301 & 0.17 & 0.56 & $<$0.030 & ... & ... & ... & 1\\
UGC 10486 & 16 37 34.3 &  +50 20 44 & 6085 & $<$0.19 & $<$0.60 & $<$0.029 & ... & ... & ... & -3\\
NGC 6196 & 16 37 53.9 &  +36 04 23 & 9424 & $<$0.12 & $<$0.44 & $<$0.023 & ... & ... & ... & -3\\
UGC 10500 & 16 38 59.3 &  +57 43 27 & 5218 & $^{\displaystyle s}$$^{\ast}$0.16 & $^{\displaystyle s}$$^{\ast}$0.71 & $<$0.028 & ... & ... & ... & 0\\
IC 5090 & 21 11 30.4 &  $-$02 01 57 & 9340 & 3.04 & 7.39 & 0.118 & 0.017 & 31.6 & 1.2 & 1\\
IC 1368 & 21 14 12.5 &  +02 10 41 & 3912 & 4.03 & 5.80 & 0.047 & 0.011 & 37.6 & 1.3 & 1\\
NGC 7047 & 21 16 27.6 &  $-$00 49 35 & 5626 & 0.43 & 1.65 & 0.055 & 0.013 & 28.0 & 1.1 & 3\\
NGC 7081 & 21 31 24.1 &  +02 29 29 & 3273 & 1.79 & 3.87 & 0.044 & 0.010 & 32.8 & 1.3 & 3\\
NGC 7280 & 22 26 27.5 &  $+$16 08 54 & 1844 & $<$0.12 & $<$0.48 & $<$0.040 & ... & ... & ... & -1\\
NGC 7442 & 22 59 26.5 &  $+$15 32 54 & 7268 & 0.78 & 2.22 & 0.046 & 0.009 & 31.0 & 1.1 & 5\\
NGC 7448$\dag$ & 23 00 03.6 &  $+$15 58 49   & 2194 & 7.23 & 17.43 & 0.193 & 0.032 & 31.0 & 1.4 & 5\\
NGC 7461 & 23 01 48.3 &  $+$15 34 57 & 4272 & $<$0.176 & $<$0.64 & $<$0.022 & ... & ... & ... & -2\\
NGC 7463 & 23 01 51.9 &  +15 58 55 & 2341 & U & U & 0.045 & 0.010 & ... & ... & 3M \\
III ZW 093 & 23 07 21.0 &  +15 51 11 & 14962 & 0.48 & $<$3.16 & $<$0.026 & ... & ... & ... & 10Z\\
III ZW 095 & 23 12 43.3 &  +15 54 12 & 7506 & $<$0.09 & $<$0.80 & $<$0.019 & ... & ... & ... & 10Z\\
UGC 12519 & 23 20 02.7 &  +15 57 10 & 4378 & 0.76 & 2.59 & 0.074 & 0.016 & 29.2 & 1.1 & 5 \\
NGC 7653 & 23 24 49.3 &  +15 16 32 & 4265 & 1.31 & 4.46 & 0.112 & 0.020 & 28.6 & 1.2 & 3\\
NGC 7691 & 23 32 24.4 &  +15 50 52 & 4041 & 0.53 & 1.67 & $<$0.025 & ... & ... & ... & 4\\
NGC 7711 & 23 35 39.3 &  +15 18 07 & 4057 & $<$0.15 & $<$0.50 & $<$0.027 & ... & ... & ... & -2\\
NGC 7722 & 23 38 41.2 &  +15 57 17 & 4026 & 0.78 & 3.03 & 0.061 & 0.015 & 26.8 & 1.4 & 0\\
\hline
\end{tabular}\\
(1) Most commonly used name. \\
(2) Right ascension, J2000 epoch. \\
(3) Declination, J2000 epoch. \\
(4) Recessional velocity taken from NED. [The NASA/IPAC Extragalactic Database (NED) is operated by the Jet Propulsion Laboratory, California Institute of Technology, under contract with the National Aeronautics and Space Administration.] \\
(5) 60\mic flux from the \IRAS\/ Faint Source Catalogue (Moshir et al. 1990); upper limits listed are measured using SCANPI as described in Section~\ref{iras-fluxes}. \\
(6) 100\mic flux from the \IRAS\/ Faint Source Catalogue (Moshir et al. 1990); upper limits listed are measured using SCANPI as described in Section~\ref{iras-fluxes}.\\
(7) 850\mic flux (this work). \\
(8) Error on 850\mic flux, calculated as described in Section~\ref{errors}. \\
(9) Dust temperature derived from a single-component fit to the 60, 100 and 850\mic data points, as described in Section~\ref{sed-fits}. \\
(10) Emissivity index derived from the single-component fit, as described in Section~\ref{sed-fits}. \\
(11) Hubble type (t-type) taken from the LEDA database; we have assigned t=10 to any multiple systems unresolved by \textit{IRAS} or SCUBA (indicated by `10M') and any systems with no type listed in LEDA (indicated by `10Z'; these 2 objects are listed as `compact' sources in NED); all other types marked `M' are listed as multiple systems in LEDA.\\
\smallskip\\
$^{\scriptstyle p}$ Part of a close or interacting pair which was resolved by SCUBA. Fluxes here are the individual galaxy fluxes; fluxes measured for the combined pair are given in Table~\ref{pairstab}.\\
U Unresolved by \textit{IRAS}.\\
$^{\displaystyle s}$ The $\IRAS$ flux is our own SCANPI measurement (see Section~\ref{iras-fluxes}); any individual comments are listed in Section~\ref{maps}.\\
$^{\ast}$ SCANPI measurements and fitted values should be used with caution (see Section~\ref{iras-fluxes}).\\
$\S$ The coordinates of this object refer to one galaxy (NED01) of a the \textit{pair} UGC 10325.\\
$\dag$ Objects are also in the  Paper I \textit{IRAS}-selected sample (DE00).
\end{minipage}
\end{table*}

\section{Observations and Data Reduction}
\label{data-red}

\subsection{The sample}
\label{sample}
This OS sample is taken from the Center for Astrophysics (CfA) optical redshift survey (Huchra et al. 1983), which is a magnitude-limited sample of optically-selected galaxies, complete to \mbox{$m_{B} \leq 14.5$ mag}. It has complete information on magnitude, redshift and morphological-type, and also avoids the Galactic plane. The OS sample consists of all galaxies in the CfA sample lying within three arbitrary strips of sky: (i) all declinations from (B1950.0) \mbox{$16.1<\textrm{RA}<21.5$}, (ii) all RAs from \mbox{$15<\textrm{Dec}<16$} and (iii) RAs from \mbox{$9.6<\textrm{RA}<12.8$} with declinations from \mbox{$25<\textrm{Dec}<26$}. We also imposed a lower velocity limit of \mbox{1900\,km\,s$^{-1}$} to try to ensure that the galaxies did not have an angular diameter larger than the field of view of SCUBA. There are 97 galaxies in the CfA survey meeting these selection criteria and of these we observed 81 (which were at convenient positions given our observing schedule). The OS sample covers an area of \mbox{$\sim$\,570} square degrees and is listed in Table~\ref{fluxtab}. Unlike the IRS sample which contained many interacting pairs (most of which were resolved by SCUBA but not by \textit{IRAS}), the OS sample contains just 2 such pairs.

\subsection{Observations}
\label{obs}
We observed the OS sample galaxies using the SCUBA bolometer array at the 15-m James Clark Maxwell Telescope (JCMT) on Mauna Kea, Hawaii, between December 1997 and January 2001, with a handful of additional observations in February 2003 (due to bad data obtained when observed the first time round; see Section~\ref{red}). Observational methods and techniques were similar to those for the IRS sample described in D00. We give a brief description of these below.

The SCUBA camera has 2 bolometer arrays (850\mic and 450\,\micron, with 37 and 91 bolometers respectively) which operate simultaneously with a field of view of \mbox{$\sim$\,2.3} arcminutes at 850\mic (slightly smaller at 450\,\micron). Beamsizes are measured to be $\sim$15 arcsec at 850\mic and $\sim$8 arcsec at 450\,\micron. Our observations were made in `jiggle-map' mode which, for sources smaller than the field of view, is the most efficient mapping mode. Since the arrangement of the bolometers is such that the sky is instantaneously undersampled, and since we observed using both arrays, the secondary mirror was stepped in a 64-point jiggle pattern in order to fully sample the sky. The cancellation of rapid sky variations is provided by the telescope's chopping secondary mirror, operating at 7.8\,Hz. Linear sky gradients and the gradual increase or decrease in sky brightness are compensated for by nodding the telescope to the `off' position every 16 seconds. We used a chop throw of 120 arcsec in azimuth, except where the galaxy had a nearby companion, in which case we used a chop direction which avoided the companion. 

The zenith opacity $\tau$ was measured by performing regular skydips. The observations were carried out under a wide range of weather conditions, with opacities at 850\mic $\tau_{850}$ ranging from 0.12 to 0.52. This means that some galaxies were observed in excellent conditions ($\tau_{850}<0.2$) while others were observed in far less than ideal conditions. As a result we obtained useful 450\mic data for only a fraction of our galaxies. This is discussed in more detail in Section~\ref{450data}. Our observations were centred on the coordinates taken from the NASA/IPAC Extragalactic Database (NED). We made regular checks on the pointing and found it to be generally good to $\sim$2 arcsec. The integration times depended on source strength and weather conditions. Since most of the OS sample are relatively faint submillimetre sources we typically used $\sim$12 integrations ($\sim$30 mins), although many sources were observed in poorer weather and so required longer integration times.

We calibrated our data by making jiggle maps of Uranus and Mars, or, when these planets were unavailable, of the secondary calibrators CRL 618 and HL Tau. We took the planet fluxes from the JCMT {\small {FLUXES}} program, and CRL 618 and HL Tau were assumed to have fluxes of 4.56 and 2.32 \mbox{Jy beam$^{-1}$} respectively at 850\,\micron.

\begin{table*}
\centering
\begin{minipage}{14.5cm}
\caption{\label{pairstab}\small{Combined SCUBA fluxes for pairs unresolved by \textit{IRAS}. (Notes on individual objects are listed in Section~\ref{maps}).}}
\begin{tabular}{lllrrrrrllr} 
\hline
(1) & (2) & (3) & (4) & (5) & (6) & (7) & (8) & (9) & (10) & (11) \\
Name & R.A. & Decl. & cz & $S_{60}$ & $S_{100}$ & $S_{850}$ & $\sigma_{850}$ & $T_{dust}$ & $\beta$ & Type\\
& (J2000) & (J2000) & (km\,s$^{-1}$) & (Jy) & (Jy) & (Jy) & (Jy) & (K) & &\\
\hline
NGC 3799/3800 & 11 40 11.4 & +15 20 05 & 3312 & 4.81 & 11.85 & 0.135 & 0.035 & 29.8 & 1.5 & 10M\\
NGC 5953/4 & 15 34 33.7 & +15 11 49 & 1966 & 10.04 & 18.97 & 0.273 & 0.034 & 35.2 & 1.1 & 10M\\
\hline
\end{tabular}\\
Note. Columns have the same meanings as in Table~\ref{fluxtab}. \\
\end{minipage}
\end{table*}

\subsection{Data reduction}
\label{red}
The 850\mic and 450\mic data was reduced using the standard SCUBA specific tasks in the {\small {SURF}} package (Jenness \& Lightfoot 1998, 2000; Jenness et al. 2002), where possible via the {\small {XORACDR}} automated data reduction pipeline (Economou et al. 2004). The off-nod position was subtracted from the on-nod in the raw beam-switched data and the data was then flat-fielded and corrected for atmospheric extinction.

In order to correct SCUBA data for atmospheric extinction we must accurately know the value of the zenith sky opacity, $\tau$. Although less crucial at 850\mic if the observation is made in good weather ($\tau_{850}<$0.3) and at low airmass, in worse weather or at 450\mic the measured source flux can be severely affected by an error in $\tau$. $\tau$ is most commonly estimated either by performing a skydip or by extrapolating to the required wavelength (using relations given in the JCMT literature and in Archibald et al. (2002)) from polynomial fits to the continuous measurements of $\tau$ at 225GHz made at the nearby Caltech Submillimetre Observatory. Since skydips are measured relatively infrequently, the polynomial fits to the CSO $\tau_{225}$ data are recommended in the JCMT literature to be the more reliable way of estimating $\tau$ for both SCUBA arrays. As such, for both 850\mic and 450\mic data we have wherever possible (the large majority of observations) used the derived CSO opacity at $225$GHz ($\tau_{cso}$). Where $\tau_{cso}$ values were not available the opacities were derived from 850\mic skydip measurements (at 450\mic using the $\tau_{850}$-to-$\tau_{450}$ relation described in the JCMT literature and Archibald et al. (2002)).

Noisy bolometers were noted but not removed at this stage (it was frequently found to be the case that flagging a noisy bolometer as `bad' creates even worse noise spikes in the final map around the position of the removed bolometer data). Large spikes were removed from the data using standard {\small {SURF}} programs.

The nodding and chopping should remove any noise which is correlated between the different bolometers. In reality, since the data was not observed in the driest and most stable conditions the signal on different bolometers was often highly correlated due to incomplete sky subtraction. In the majority of cases we used the  {\small SURF} task {\small REMSKY}, which takes a set of user-specified bolometers to estimate the sky variation as a function of time. More explicitly, in each time step {\small REMSKY} takes the median signal from the specified sky bolometers and subtracts it from the whole array. To ensure that the sky bolometers specified were looking at sky alone and did not contain any source emission we used a rough SCUBA map together with optical (\textit{Digitised Sky Survey}\footnote{The Digitised Sky Surveys were produced at the Space Telescope Science Institute under U.S. Government grant NAGW-2166. The images of these surveys are based on photographic data obtained using the Oschin Schmidt Telescope on Palomar Mountain and the UK Schmidt Telescope. The plates were processed into the present digital form with the permission of these institutions.} (DSS)) images as a guide when choosing the bolometers, though in this sample there are so few bright sources that in the majority of cases all bolometers could be safely used.

Even after this step, however, due to the relatively poor conditions in which much of the data was observed the residual sky level was sometimes found to vary linearly across the array, giving a `tilted plane' on the array. Moreover, in a number of cases a noisy `striped' sky (due possibly to some short-term instrumentation problem) was found. Though the {\small SURF} task {\small {REMSKY}} was designed to remove the sky noise, it is relatively simplistic and cannot remove such spatially varying `tilted' or `striped' sky backgrounds. In these cases, as for the IRS sample (D00), we used one of our own programs in place of {\small REMSKY}. In a handful of cases the `striped' sky was so severe that it could not be removed, so these objects were re-observed in February 2003.

\begin{table*}
\centering
\begin{minipage}{11cm}
\caption{\label{450tab}\small{450\mic flux densities and two-component SED parameters. (Notes on individual objects are listed in Section~\ref{maps}). }}
\begin{tabular}{lrrrllrcc} 
\hline
(1) & (2) & (3) & (4) & (5) & (6) & (7) & (8) & (9)\\
Name & $S_{450}$ & $\sigma_{450}$ & $\frac{S_{450}}{S_{850}}$ & $T_{w}$ & $T_{c}$ & $\frac{N_{c}}{N_{w}}$ & $M_{d2}$ & $L_{fir}$\\
     & (Jy) & (Jy) & & (K) & (K) & & (log $M_{\odot}$) &  (log $L_{\odot}$)\\
\hline
UGC 148$\dag$ & 0.944 & 0.236 & 17.18 & 34 & 18 & 37 & ... & 10.33\\
NGC 99 & 0.490 & 0.182 & 7.73 & 47 & 17 & 542 & 7.72 & 10.08\\
NGC 803$\ddag$ & 0.631 & 0.196 & 6.79 & 33 & 18 & 92 & 7.02 & 9.46\\
NGC 3689$^{\ast}$ & 1.045 & 0.357 & 10.30 & 59 & 23 & 910 & 7.13 & 10.16\\
PGC 35952 & 0.421 & 0.116 & 8.26 & 58 & 18 & 1859 & 7.31 & 9.77\\
NGC 3987 & 1.110 & 0.319 & 5.98 & 44 & 22 & 279 & 7.85 & 10.78\\
IC 979$\S$ & 0.874 & 0.341 & 15.39 & ... & ... & ... & ... & ...\\
NGC 5953/4 & 2.879 & 0.683 & 10.54 & 54 & 21 & 277 & 7.33 & 10.28\\
NGC 5980 & 1.398 & 0.495 & 5.53 & 43 & 18 & 321 & 8.06 & 10.53\\
NGC 6090 & 0.803 & 0.180 & 8.82 & 55 & 22 & 122 & 8.09 & 11.29\\
NGC 6120 & 0.528 & 0.127 & 8.08 & 45 & 24 & 76 & 7.96 & 11.17\\
NGC 6155$^{\ast}$ & 0.381 & 0.135 & 3.30 & 30 & 20 & 7 & 6.92 & 9.80\\
NGC 6190$^{\ast}$ & 0.880 & 0.308 & 8.89 & 56 & 18 & 2684 & 7.16 & 9.85\\
IC 5090 & 1.018 & 0.240 & 8.66 & 52 & 21 & 346 & 8.28 & 11.19\\
IC 1368 & 0.425 & 0.137 & 9.10 & 55 & 23 & 110 & 7.10 & 10.37\\
NGC 7081 & 0.241 & 0.067 & 5.43 & 32 & 20 & 6 & 6.98 & 9.93\\
NGC 7442 & 0.410 & 0.099 & 9.02 & 54 & 20 & 665 & 7.70 & 10.45\\
UGC 12519 & 0.408 & 0.108 & 5.54 & 28 & 17 & 12 & 7.57 & 10.02\\
NGC 7722 & 0.595 & 0.148 & 9.78 & 54 & 20 & 1224 & 7.33 & 10.04\\
\hline
\end{tabular}\\
(1) Most commonly used name. \\
(2) 450\mic flux (this work). \\
(3) Error on 450\mic flux, calculated as described in Section~\ref{errors}. \\
(4) Ratio of 450- to 850-$\micron$ fluxes. \\
(5) Warm temperature using $\beta=2$. \\
(6) Cold temperature using $\beta=2$. \\
(7) Ratio of cold-to-warm dust. \\
(8) Dust mass calculated using parameters in columns (5)--(7). \\
(9) FIR luminosity (40--1000\micron) integrated under the two-component SED. \\
$\ast$ Some caution is advised (see Section~\ref{maps}).\\
$\S$ The data could not be fitted with a 2-component model (see Section~\ref{maps}).\\
$\dag$ Not well-fitted by two-component model using 850\mic data point; fitted parameters here are from 2-component fit to the 60, 100, 450 and 170\mic (ISO) data points; (see Section~\ref{maps}).\\
$\ddag$The data for NGC 803 are also well-fitted by the parameters: \mbox{$T_{w}$=60\,K}, \mbox{$T_{c}$=19\,K}, \mbox{$\frac{N_{c}}{N_{w}}$=2597}, \mbox{log $M_{d2}$=6.99} and \mbox{log $L_{fir}$=9.48}, (see Section~\ref{maps}).\\
\end{minipage}
\end{table*}

Once the effects of the sky were removed the data was despiked again and the final map produced by re-gridding the data into a pixel grid to form an image on $1$ arcsecond pixels. Where there were multiple data-sets for a given source they were binned together into a co-added final map. In these cases each data set was weighted prior to co-adding using the {\small SURF} task {\small SETBOLWT}, which calculates the standard deviation for each bolometer and then calculates weights relative to the reference bolometer (the central bolometer in the first input map). This method is therefore only suitable if there are no very bright sources present in the central bolometer (if a bright source was present we weighted each dataset using the inverse square of its measured average noise). This step also ensures that noisy bolometers contribute to the final map with their correct statistical weight.

\subsection{850\hbox{\,\boldmath{$\umu$}m} flux measurement}
\label{data-red:flux}
The fluxes were measured from the SCUBA maps by choosing a source aperture over which to integrate the flux, such that the signal-to-noise was maximised. The extent of the galaxy in the optical (DSS) images and the extent of the submillimetre source on the S/N map (see Section~\ref{snmaps}) were used to select an aperture that included as much of the submillimetre flux of the galaxy as possible while minimising the amount of sky included. Note, the optical images in Figure~\ref{egmaps} are shown stretched for optimum contrast -- however, apertures for flux measurement were drawn for a more modest optical extent, as seen at a standard level of contrast.

Conversion of the measured aperture flux in volts to Janskys was carried out by measuring the calibrator flux for that night using the same aperture as for the object. The orientation of the aperture (relative to the chop throw) was also kept the same as for the object, as particularly for more elliptical apertures this has a significant effect.

Objects are said to be detected at $>3\sigma$ if either: (a) the peak S/N in the S/N map was $>3\sigma$ or (b) the flux in the aperture was greater than 3 times the noise in that aperture (where the noise is defined as described in Section~\ref{errors}).

\subsection{450\hbox{\,\boldmath{$\umu$}m} data}
\label{450data}
Due to the increased sensitivity to weather conditions at 450\,\micron, sources emitting at 450\mic will only be detected if they are relatively bright at 850\,\micron. This, together with the wide range of observing conditions for this sample, meant that we found useful 450\mic data for only 19 objects.

Where possible the 450\mic emission was measured in an aperture the same size as used for the 850\mic data. In some cases a smaller aperture had to be used for the 450\mic data, and these individual cases are discussed in Section~\ref{maps}.

\subsection{Error analysis}
\label{errors}
The error on the flux measurement is made up of three components:
\begin{itemize}
\item{A background sky subtraction error $\sigma_{sky}$ due to the uncertainty in the sky level.}
\item{A shot (Poisson) noise term $\sigma_{shot}$ due to pixel-to-pixel variations within the sky aperture. Unlike CCD images, in SCUBA maps the signal in adjoining pixels is correlated; this correlated noise depends on a number of factors, including the method by which the data is binned at the data reduction stage. This has been discussed in some detail by D00, who find that a correction factor is required for each array to account for the fact that pixels are correlated; they find the factor to be 8 at 850\mic and 4.4 at 450\,\micron.}
\item{A calibration error term $\sigma_{cal}$ which for SCUBA observations at 850\mic is typically less than 10\%. We have therefore assumed a conservative calibration factor of 10\% at 850\,\micron. The calibration error at 450\mic was taken to be 15\%, following DE01.}
\end{itemize}

The relationships used to calculate the noise terms are as follows: 
\[
\sigma_{sky}=\sigma_{ms}N_{ap}
\]
and
\[
\sigma_{shot}=8\sigma_{pix}\sqrt{N_{ap}} \qquad \textrm{or} \qquad \sigma_{shot}=4.4\sigma_{pix}\sqrt{N_{ap}}
\]
for 850\mic and 450\mic flux measurements respectively.
The error in the mean sky $\sigma_{ms}=S.D./\sqrt{n}$, where S.D. is the standard deviation of the mean sky values in \textit{n} apertures placed on off-source regions of the map. ${N_{ap}}$ is the number of pixels in the object aperture; $\sigma_{pix}$ is the mean standard deviation of the pixels within the sky apertures. The total error for each flux measurement is then given by
\begin{equation}
\sigma_{tot}=(\sigma_{sky}^{2}+\sigma_{shot}^{2}+\sigma_{cal}^{2})^{1/2} \label{eq1}
\end{equation}
as for the IRS sample. This error analysis is discussed in detail in D00 and DE01.

850\mic fluxes were found to have total errors $\sigma_{tot}$ typically in the range \mbox{15--30\,\%}. 450\mic fluxes were found to have total errors $\sigma_{tot}$ typically in the range \mbox{25--35\,\%}. Note, the $\sigma_{tot}$ used to determine whether a source was detected at the 3$\sigma$ level is defined as in Equation~\ref{eq1} but without the calibration error term.

\subsection{S/N maps}
\label{snmaps}

Unlike the IRS sources the OS sources were not selected on the basis of their dust content. Many of the OS sources, especially the early types, are close to the limit of detection. Also, it is often hard to assess whether a source is detected, or whether some feature of the source is real, due to the variability of the noise across the array. This is due both to an increase in the noise towards the edge of each map, caused by a decrease in the number of bolometers sampling each sky point, and to individual noisy bolometers. For this reason we used the method described in D00 to generate artificial noisemaps, which we used with our real maps to produce signal-to-noise maps. The real maps and the artificial maps were first smoothed (using a 12 pixel FWHM) before creating the S/N map.

We used these S/N maps to aid in choosing the aperture for measuring the 850\mic flux (Section~\ref{data-red:flux}). We have also presented S/N maps of each source (see Section~\ref{results}), as this makes it easier to assess the reality of any features in the maps.

\subsection{\textit{IRAS} fluxes}
\label{iras-fluxes}
\textit{IRAS} 100\mic and 60\mic fluxes, where available, were taken from the \textit{IRAS Faint Source Catalogue} (Moshir et al. 1990; hereafter FSC) via the NED database. Where literature fluxes were unavailable the NASA/IPAC Infrared Science Archive (IRSA) SCANPI (previously ADDSCAN) scan coadd tool was used to measure a flux from the \textit{IRAS} survey data. 

The small number of SCANPI fluxes are indicated by `s' in Table~\ref{fluxtab}, and any special cases are discussed individually in Section~\ref{maps}. We take SCANPI fluxes to be detections if the measurements are formal detections at $>4.5\sigma$  at 100\mic or $>4\sigma$ at 60\,\micron, which Cox et al. (1995) conclude are actually detections at the 98\% confidence level. Otherwise we give a 98\% confidence upper limit (4.5$\sigma$ at 100\mic or 4$\sigma$ at 60\,\micron) using the 1$\sigma$ error found from SCANPI (again following Cox et al. (1995)). If both fluxes are SCANPI measurements we mark the subsequent fitted values by `$\ast$' if there is any doubt as to their viability (for example possible source confusion, confusion with galactic cirrus, or no literature \IRAS\/ fluxes in NED for either band).

\section{Results}
\label{results}
We detected 52 of the 81 galaxies in the OS sample. Table~\ref{fluxtab} lists the 850\mic fluxes and other parameters. For interacting systems resolved by SCUBA but not resolved by \textit{IRAS} the 850\mic fluxes given are for the individual galaxies; the 850\mic fluxes measured for the combined system are listed in Table~\ref{pairstab} along with the \textit{IRAS} fluxes. Table~\ref{450tab} lists the 450\mic fluxes for the 19 galaxies which are also detected at the shorter wavelength. The galaxies detected in the OS sample are shown in Figure~\ref{egmaps}, with our 850\mic SCUBA S/N maps overlaid onto optical (DSS) images. Comments on the individual maps are given in Section~\ref{maps}.

The 850\mic images have several common features. Firstly, we find that many spiral galaxies exhibit two peaks of 850\mic emission, seemingly coincident with the spiral arms. This is most obvious for the more face-on galaxies (for example NGC 99 and NGC 7442), but it is also seen for more edge-on spirals (e.g. NGC 7047 and UGC 12519). This `two-peak' morphology is not seen for all the spirals, however. Some, for example NGC 3689, are core-dominated and exhibit a single central peak of submillimetre emission, while others (NGC 6131 and NGC 6189 are clear examples) exhibit a combination of these features, with both a bright nucleus and peaks coincident with the spiral arms. In a number of cases the 850\mic peaks clearly follow a prominent dust lane (e.g. NGC 3987 and NGC 7722). These results are consistent with the results of numerous mm/submm studies. For example, Sievers et al. (1994) observe 3 distinct peaks in NGC 3627 and note that the two outer peaks are coincident with the transition region between the central bulge and the spiral arms -- they also observe dust emission tracing the dust lanes of the spiral arms; Gu\'elin et al. (1995), Bianchi et al. (2000), Hippelein et al. (2003) and Meijerink et al. (2005) observe a bright nucleus together with extended dust emission tracing the spiral arms. Many of the features seen in our OS sample 850\mic maps are also found by Stevens et al. (2005) in their SCUBA observations of nearby spirals.

Secondly, we find that a number of galaxies appear to be extended at 850\mic compared to the optical emission seen in the DSS images. In many cases this extended 850\mic emission appears to correspond to very faint optical features, as can be seen for NGC 7081 and NGC 7442 in Figure~\ref{egmaps}. In order to investigate this further we have already carried out follow-up optical imaging for $\sim$ half the sample detected at 850\,\micron, to obtain deeper images than available from the DSS. The results and discussion of this deeper optical data will be the subject of a separate paper (Vlahakis et al., in preparation).

\begin{figure*}
 \begin{center}
 \includegraphics[angle=0, width=8cm]{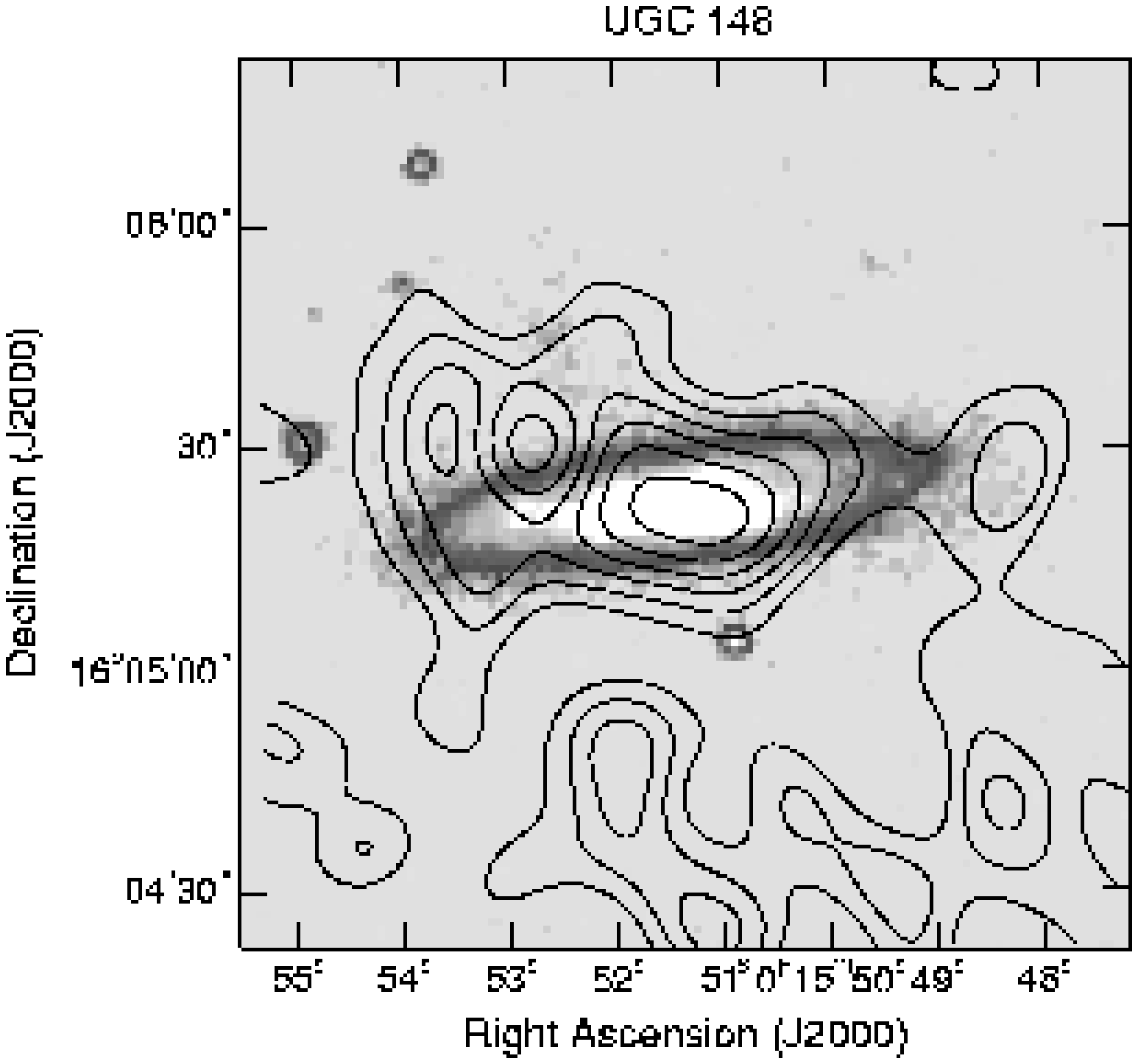}
 \hfill
 \includegraphics[angle=0, width=8cm]{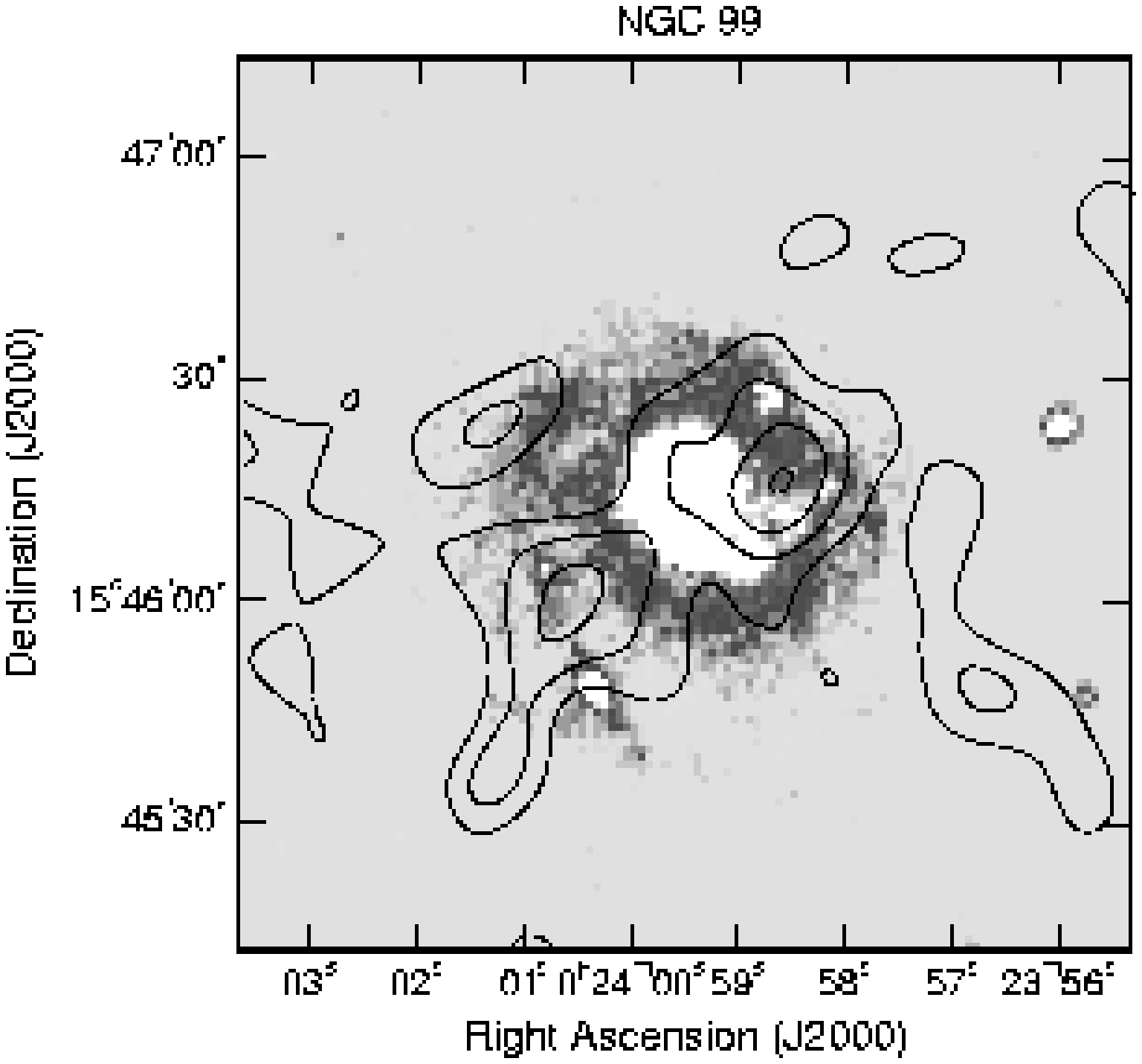}\\[-1ex]
 \hfill
 \vfill
 \includegraphics[angle=0, width=8cm]{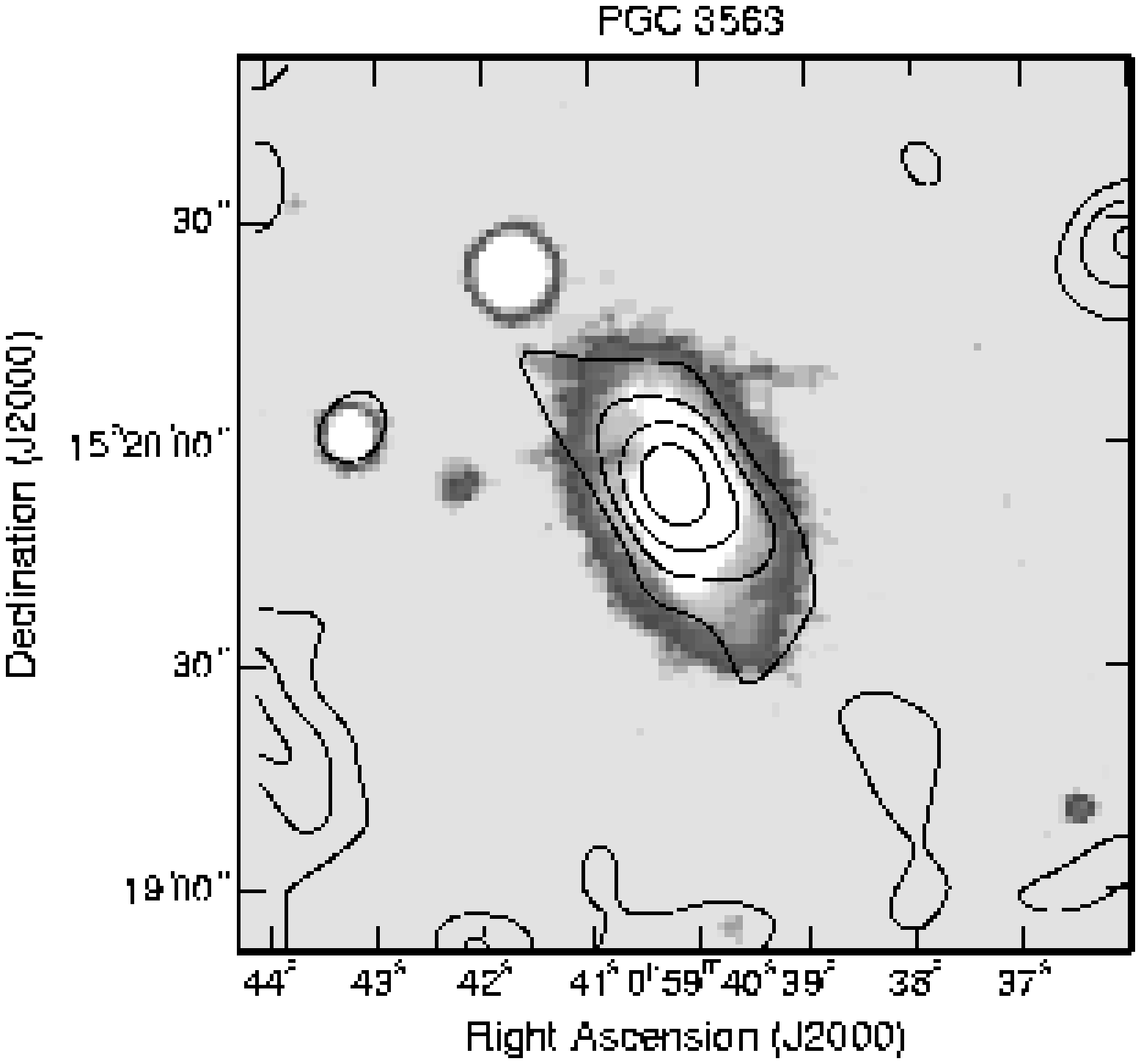}
 \hfill
 \includegraphics[angle=0, width=8cm]{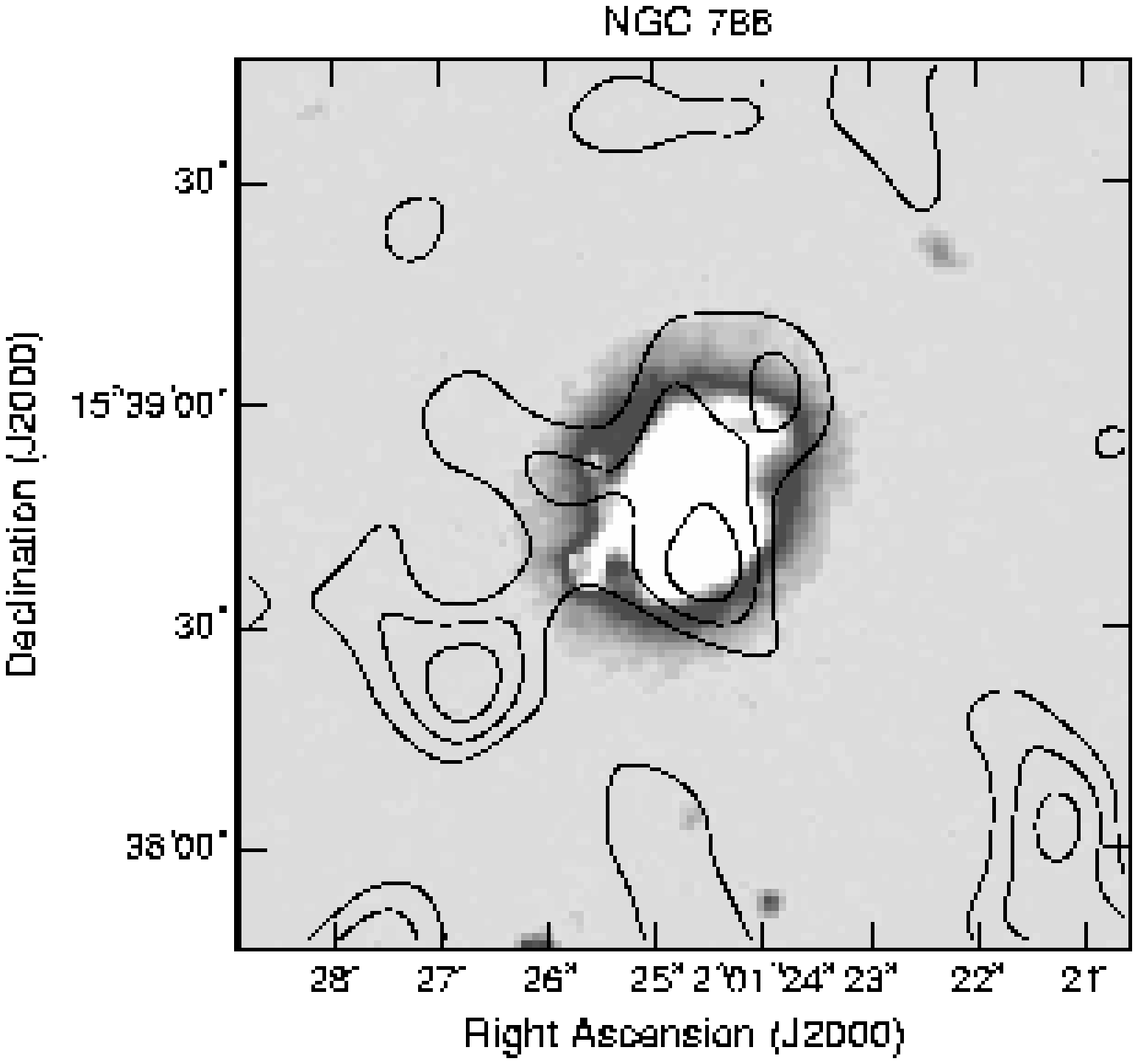}\\[-1ex]
 \hfill
 \vfill
 \includegraphics[angle=0, width=8cm]{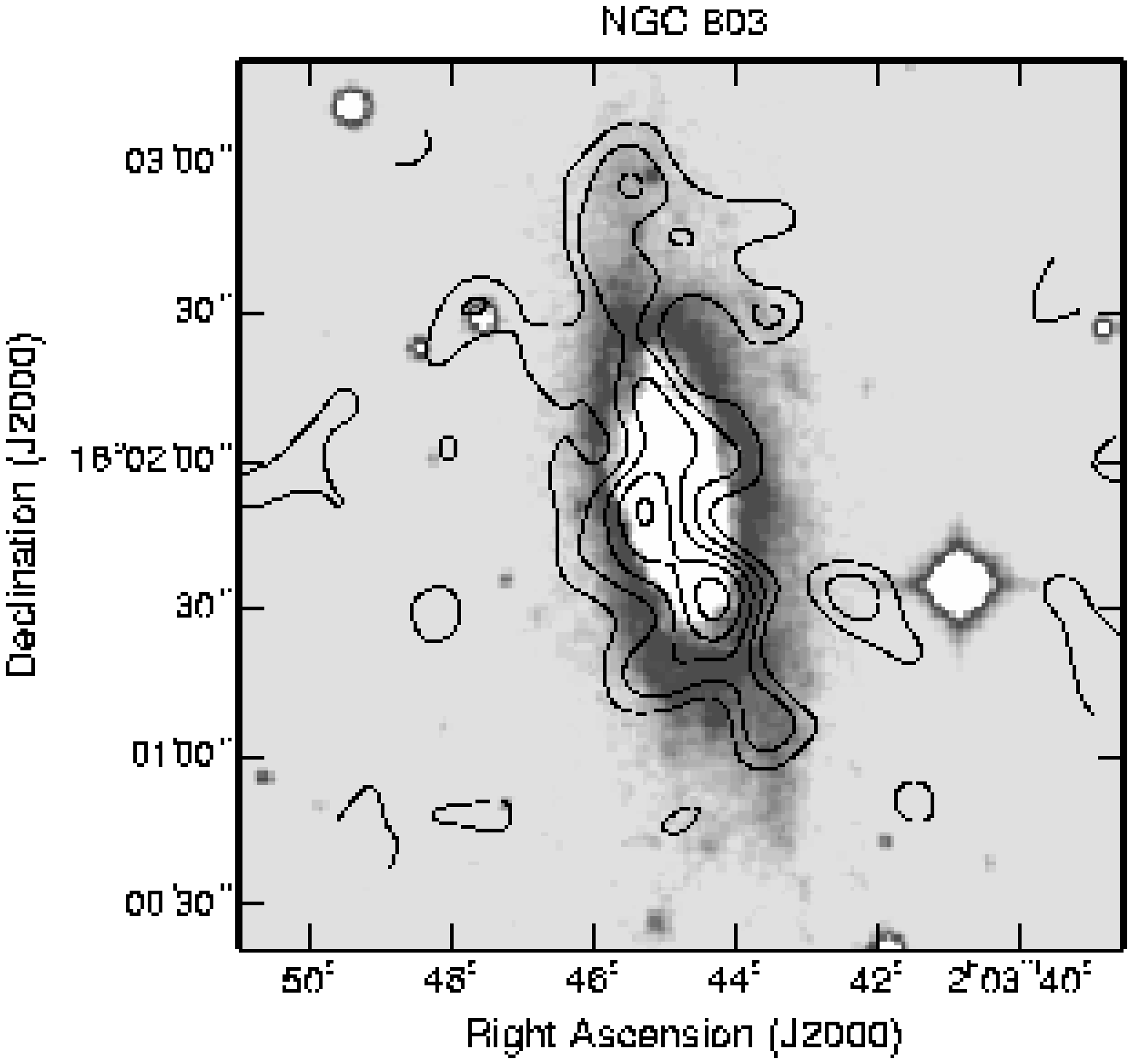}
 \hfill
 \includegraphics[angle=0, width=8cm]{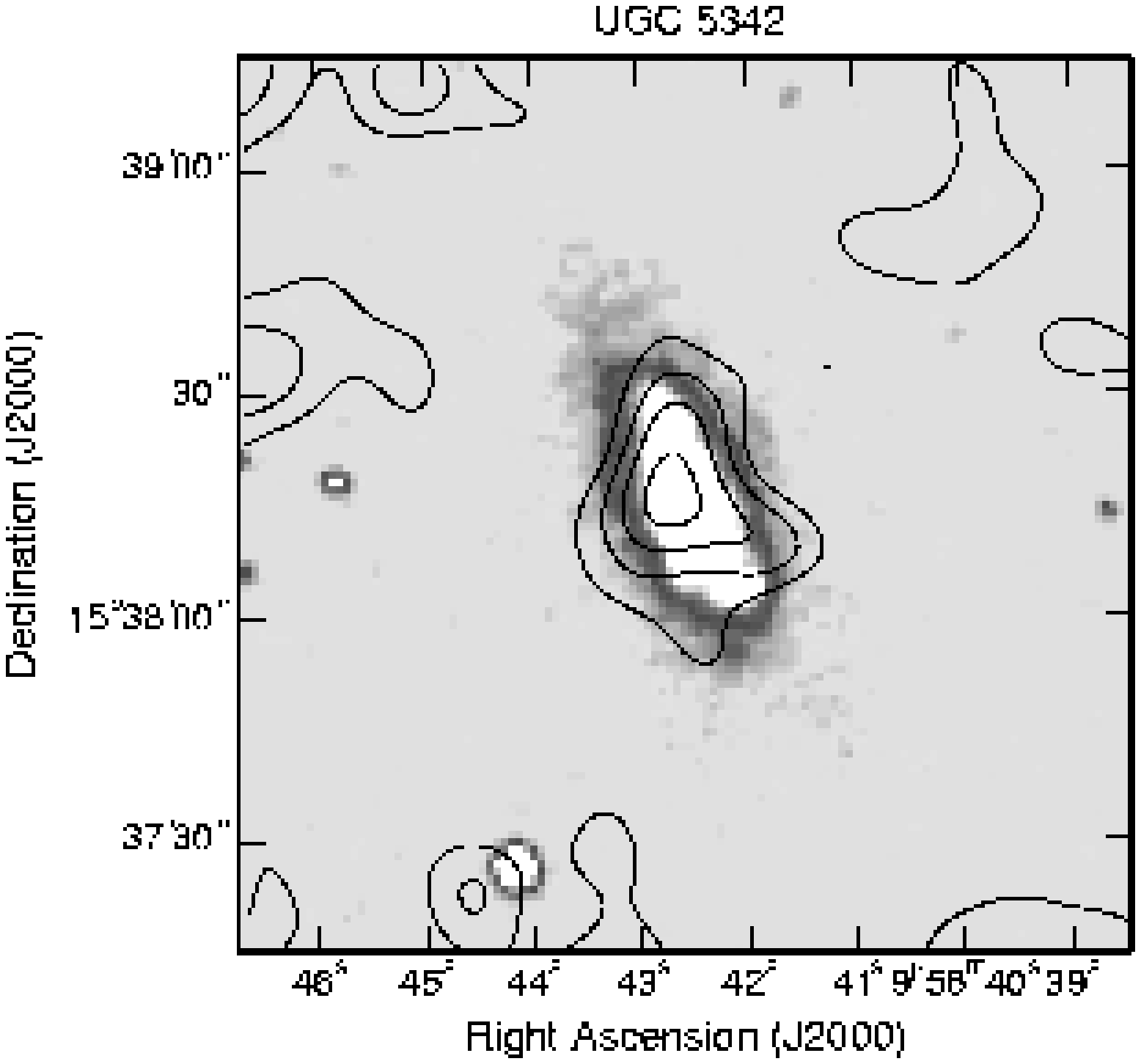}
 \hfill
 \caption{\label{egmaps}{The optically-selected SLUGS: 850\mic SCUBA S/N maps (produced as described in Section~\ref{snmaps}; 1$\sigma$ contours) overlaid onto DSS optical images ($2\arcmin\!\times\!2\arcmin$, except for NGC 803 and NGC 6155 which are $3\arcmin\!\times\!3\arcmin$). (Optical images are shown here with a contrast that optimises the optical features, however when used as a guide for drawing SCUBA flux measurement apertures a more conservative stretch was applied).}}
 \end{center}
\end{figure*}

\begin{figure*}
 \begin{center}
 \includegraphics[angle=0, width=8cm]{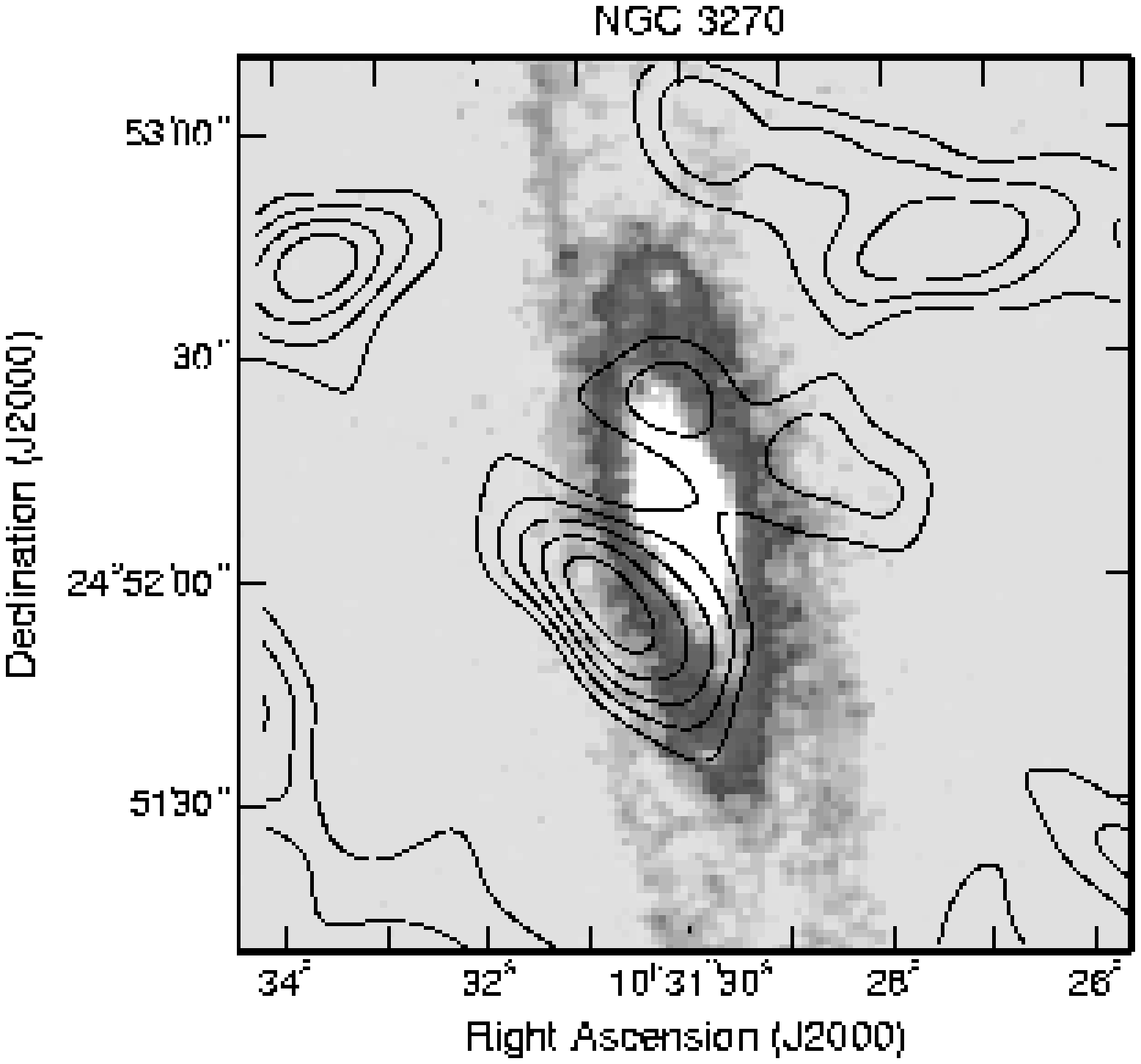}
 \hfill
 \includegraphics[angle=0, width=8cm]{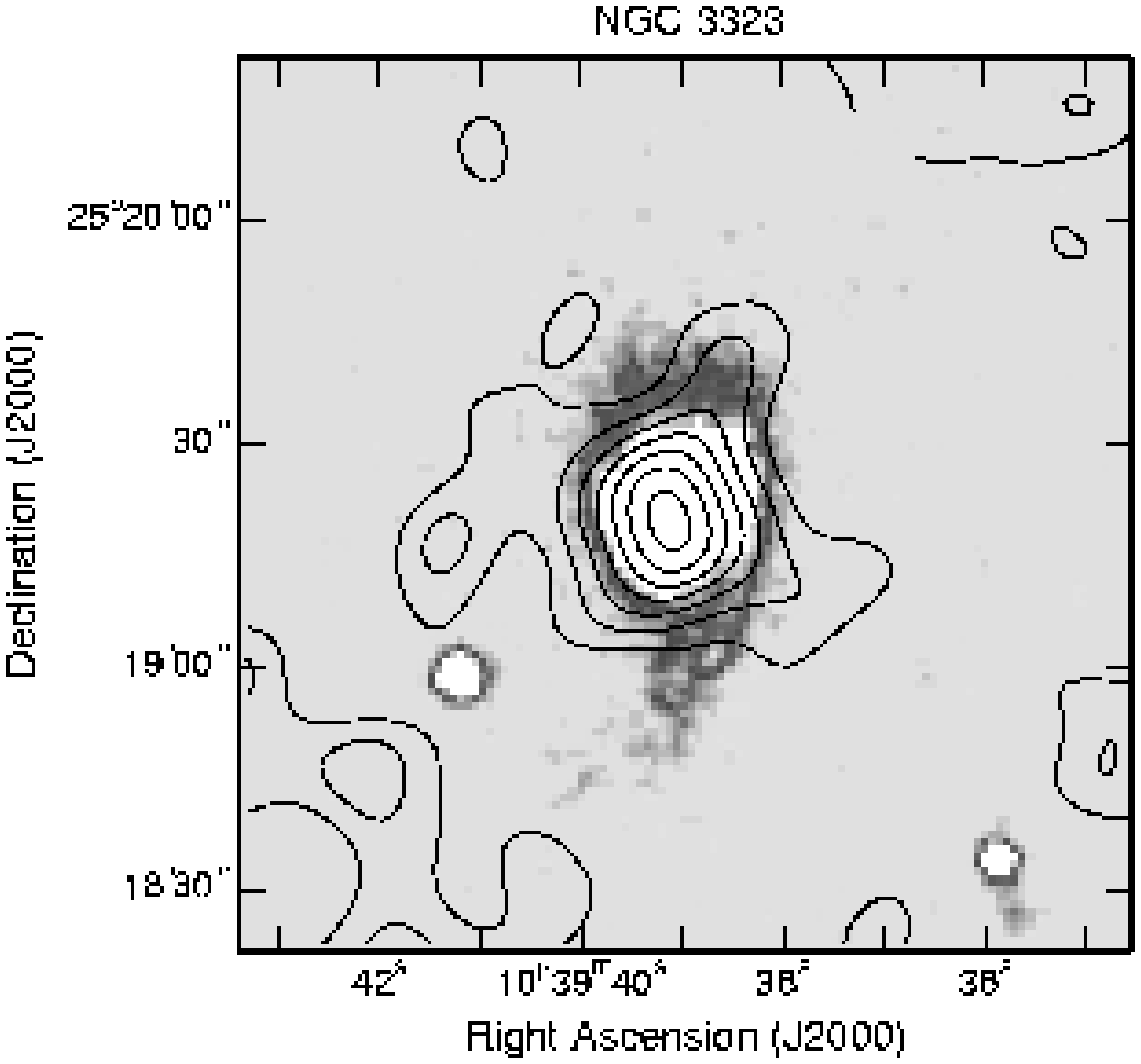}
 \hfill
 \includegraphics[angle=0, width=8cm]{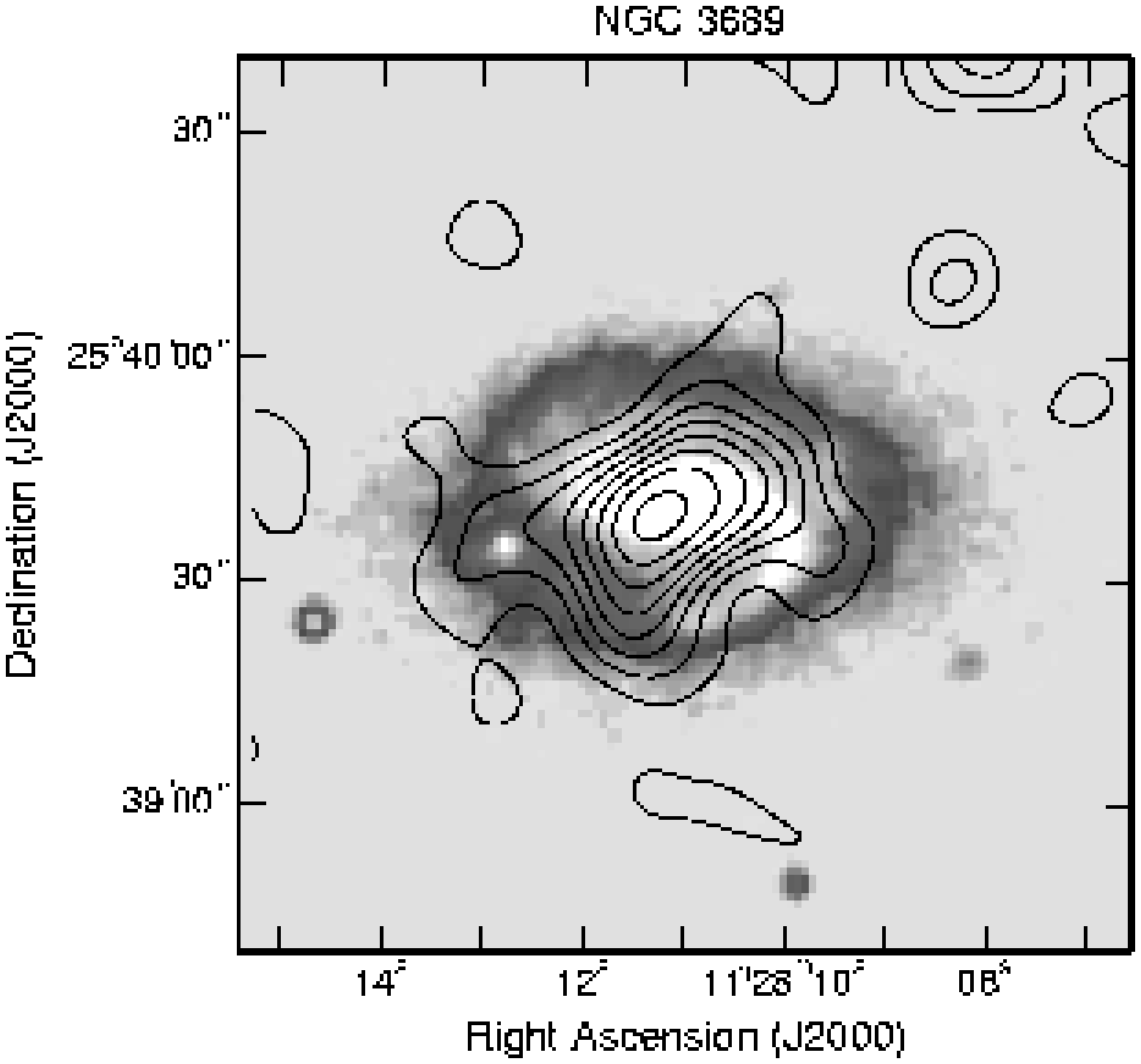}
 \hfill
 \includegraphics[angle=0, width=8cm]{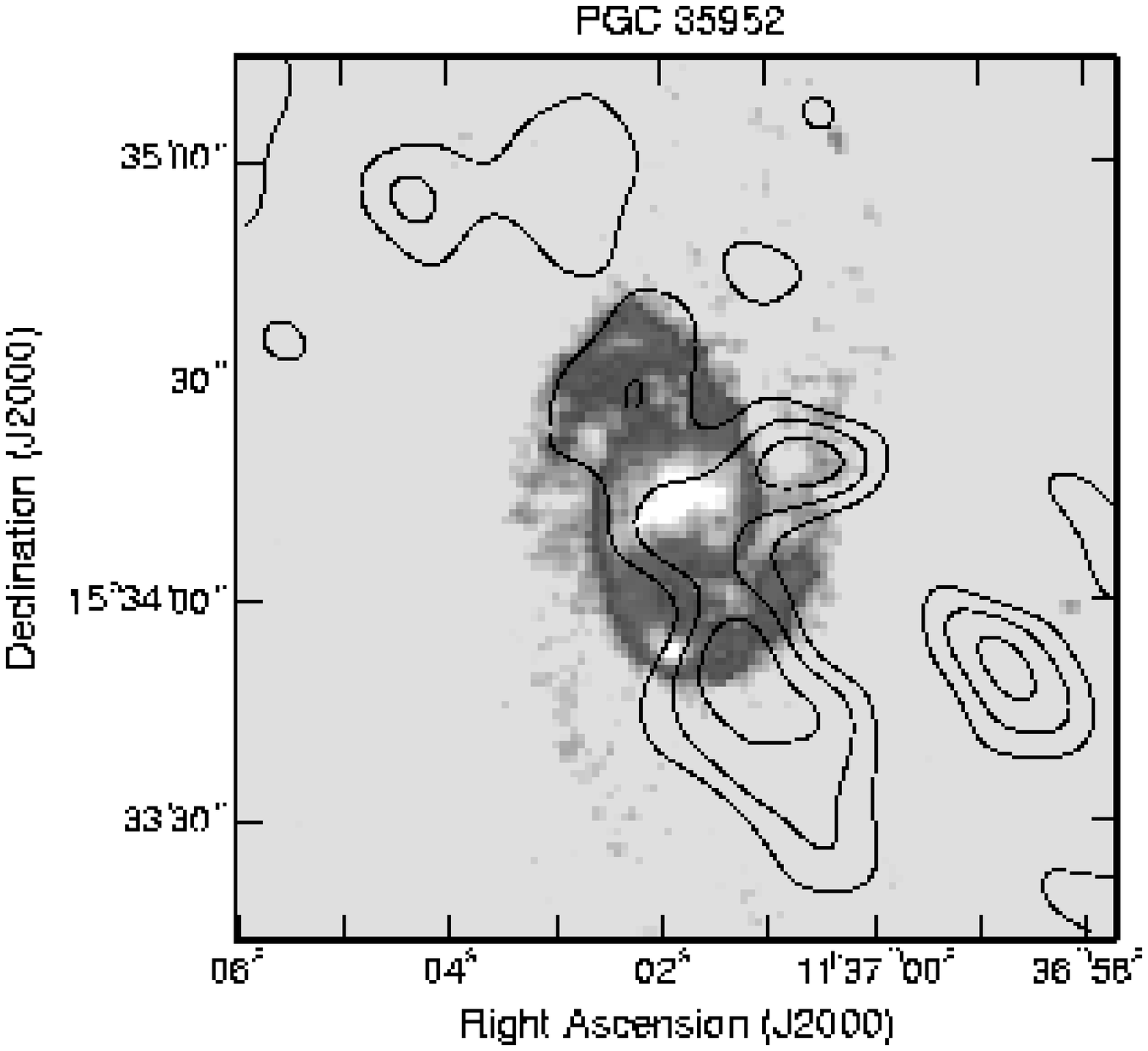}
 \hfill
 \includegraphics[angle=0, width=8cm]{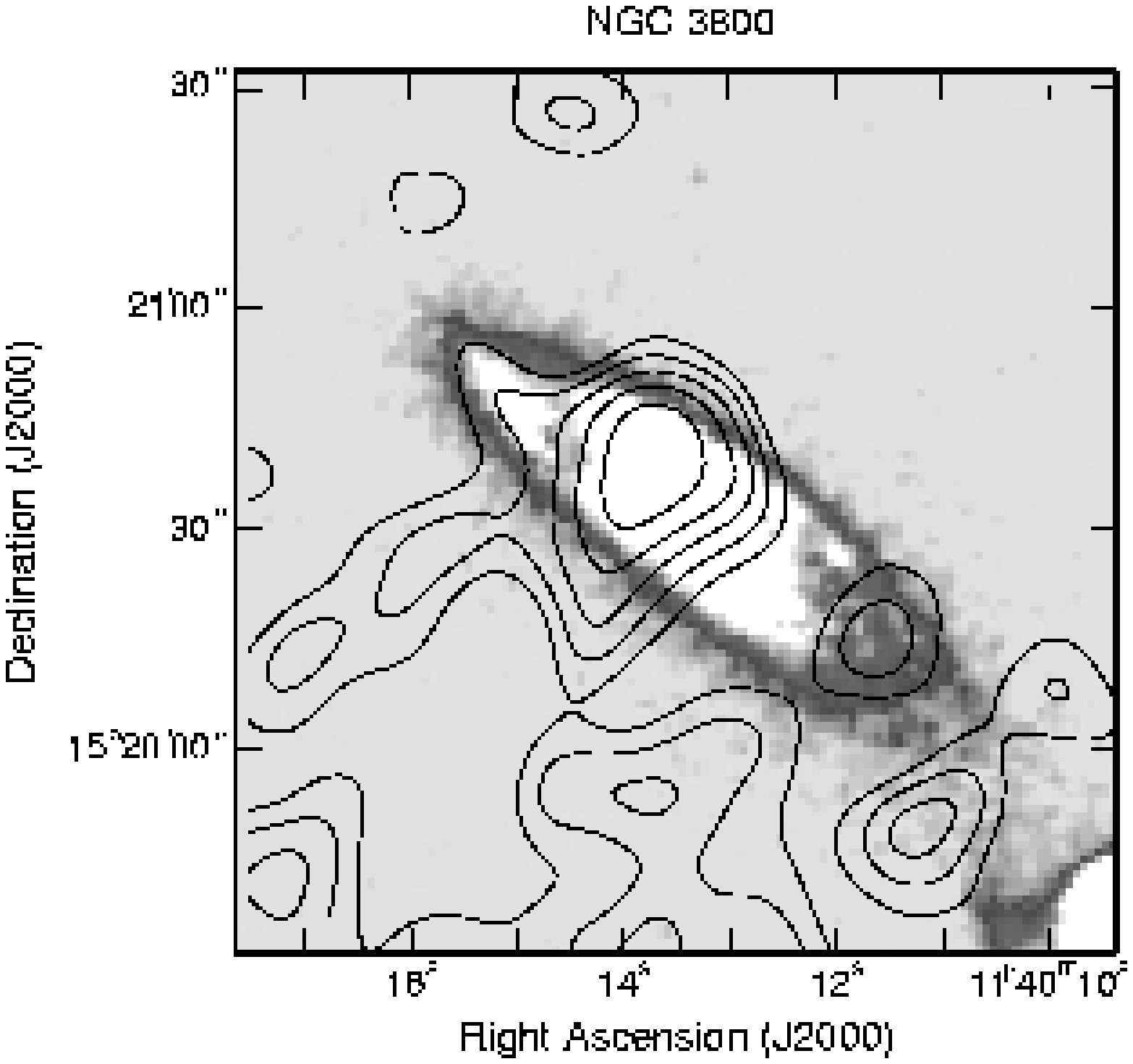}
 \hfill
 \includegraphics[angle=0, width=8cm]{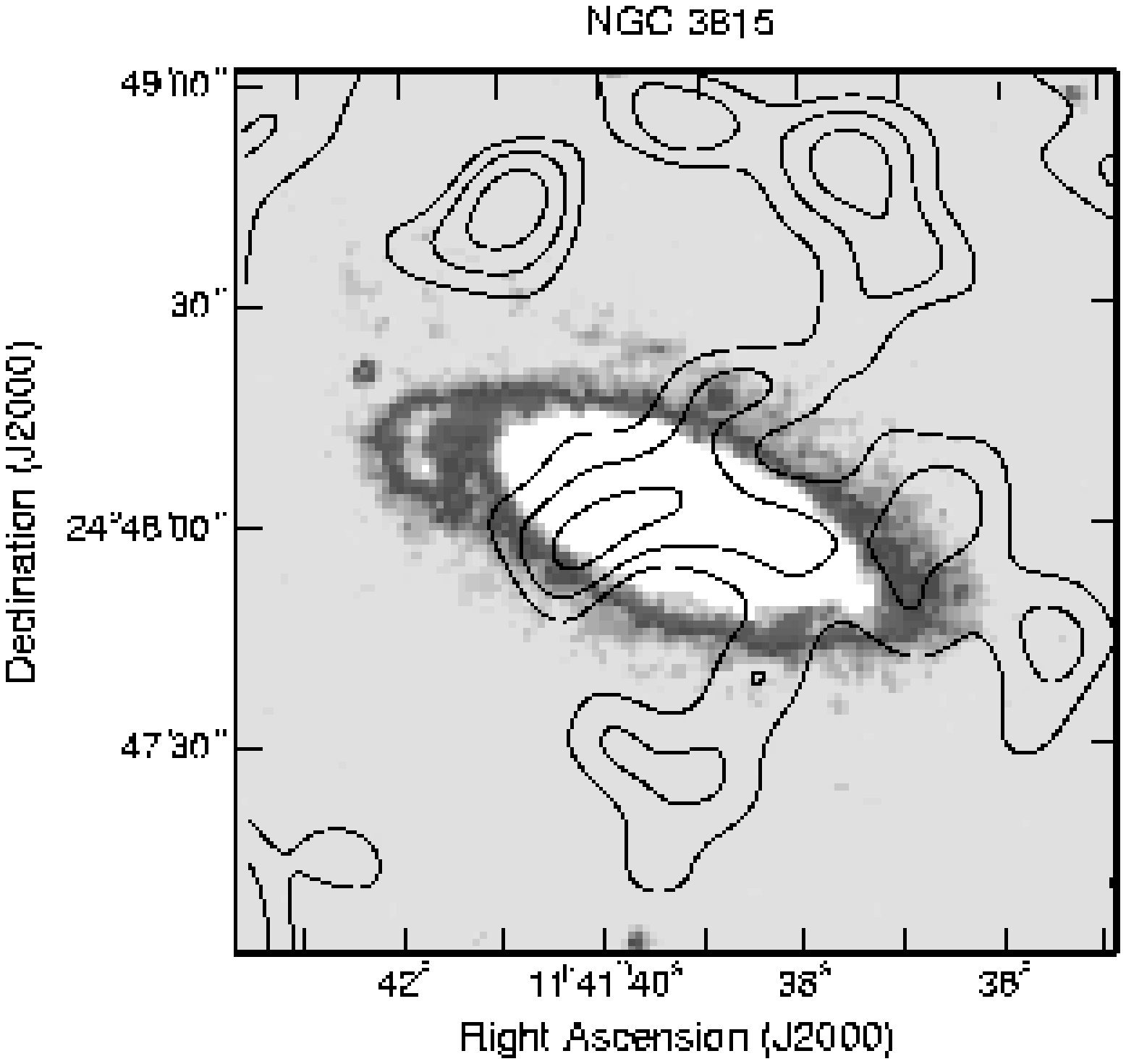}
 \hfill
 \contcaption{}
 \end{center}
\end{figure*} 

\begin{figure*}
 \begin{center}
 \includegraphics[angle=0, width=8cm]{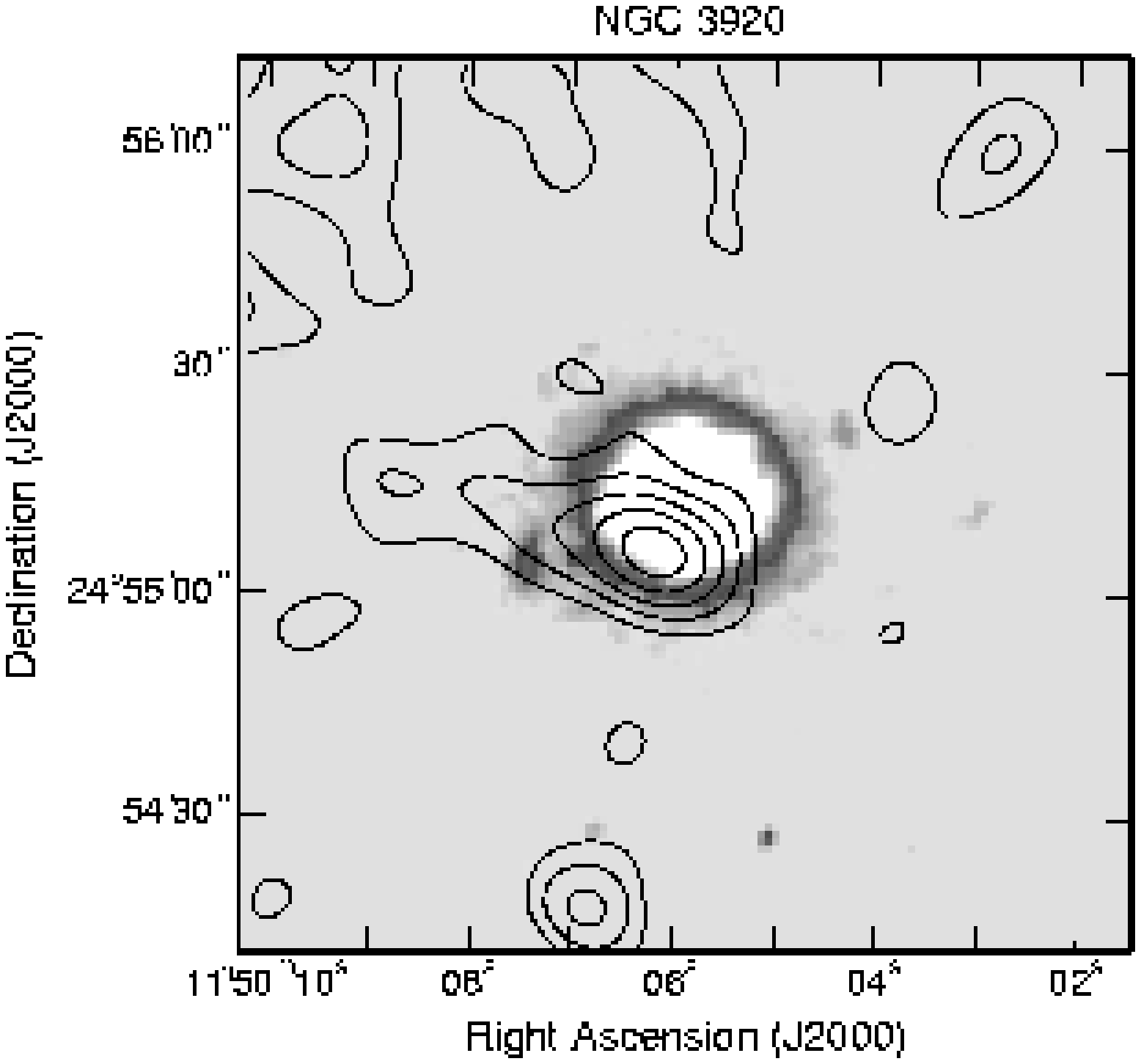}
 \hfill
 \includegraphics[angle=0, width=8cm]{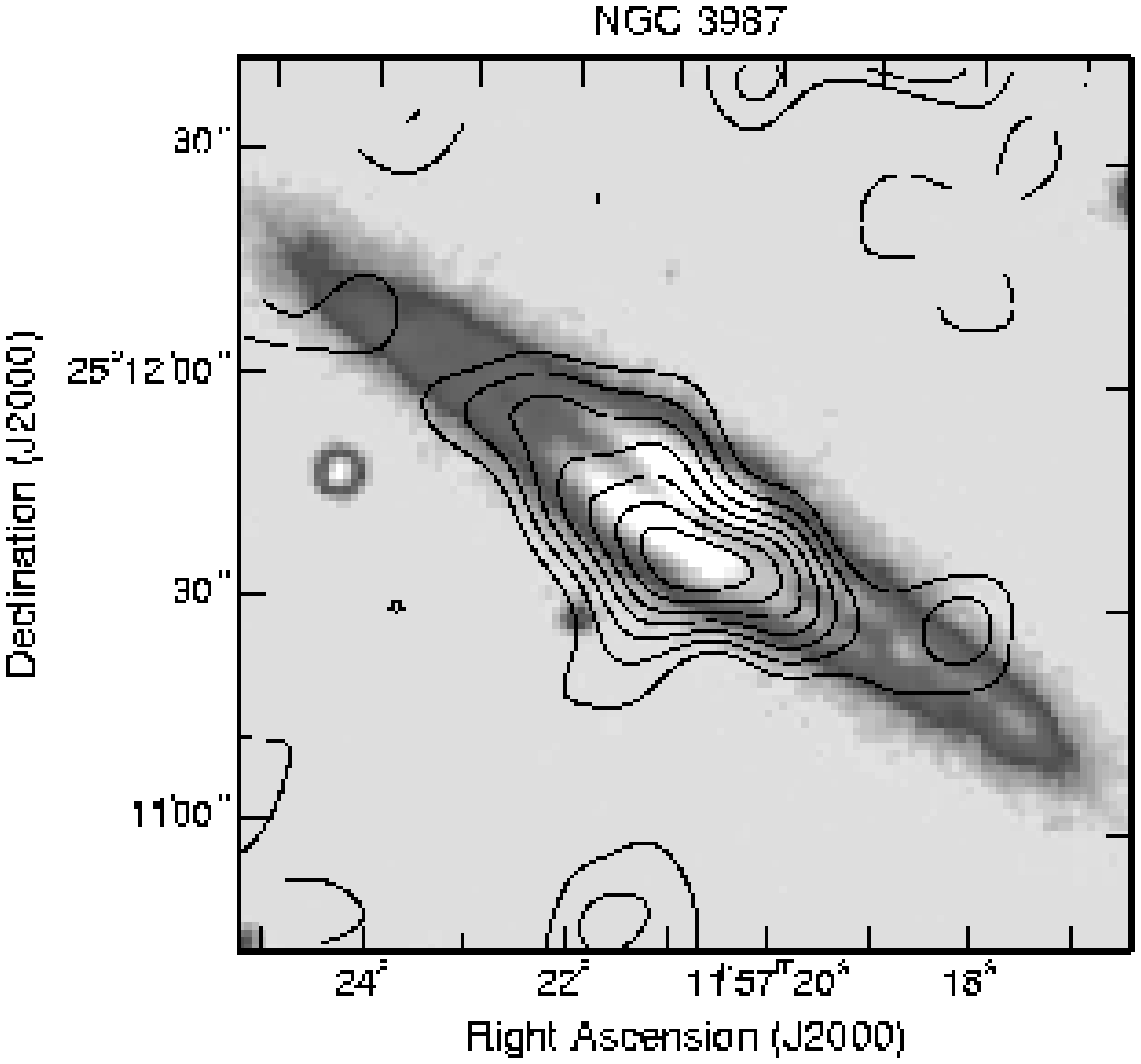}
 \hfill
 \includegraphics[angle=0, width=8cm]{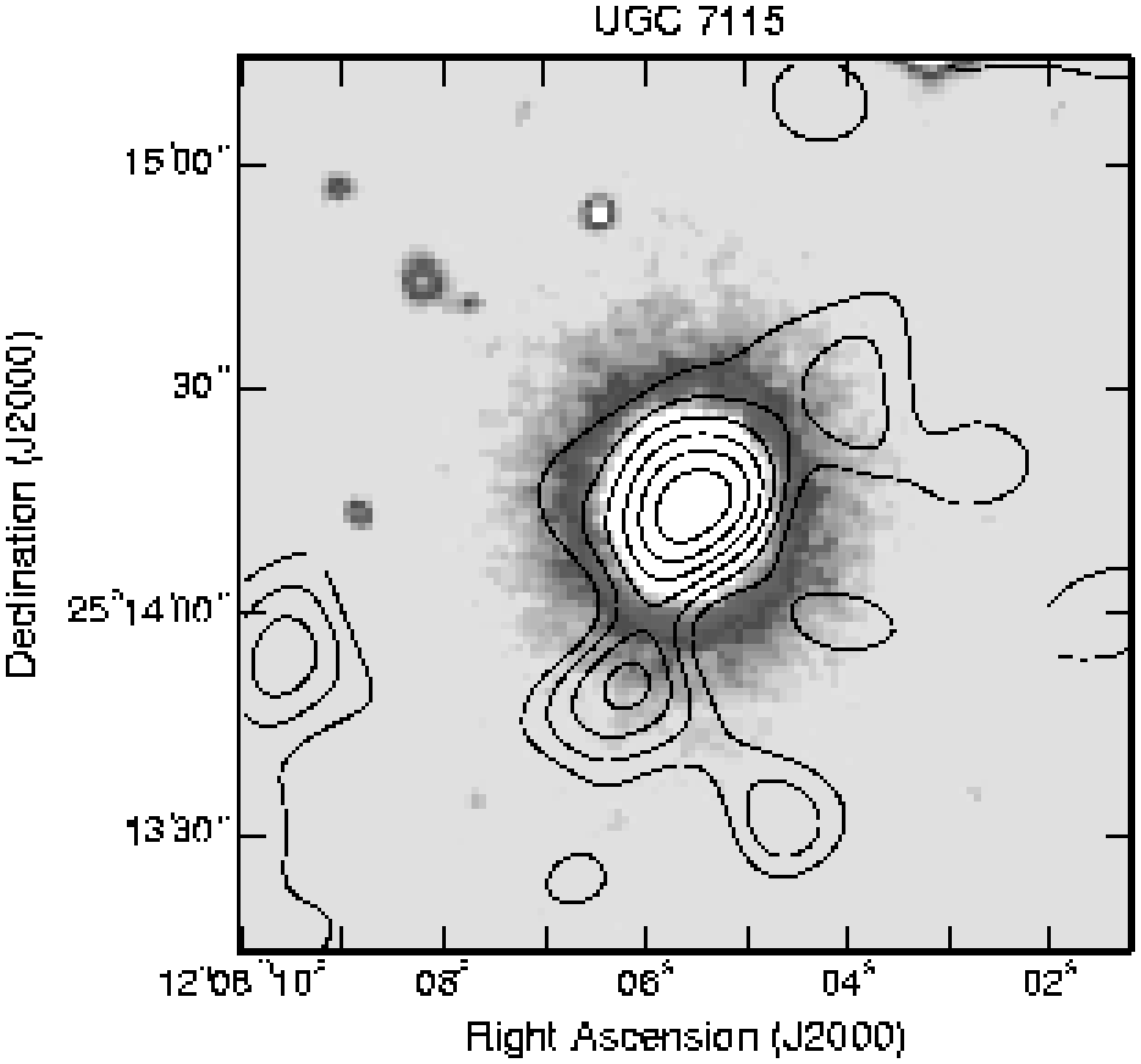}
 \hfill
 \includegraphics[angle=0, width=8cm]{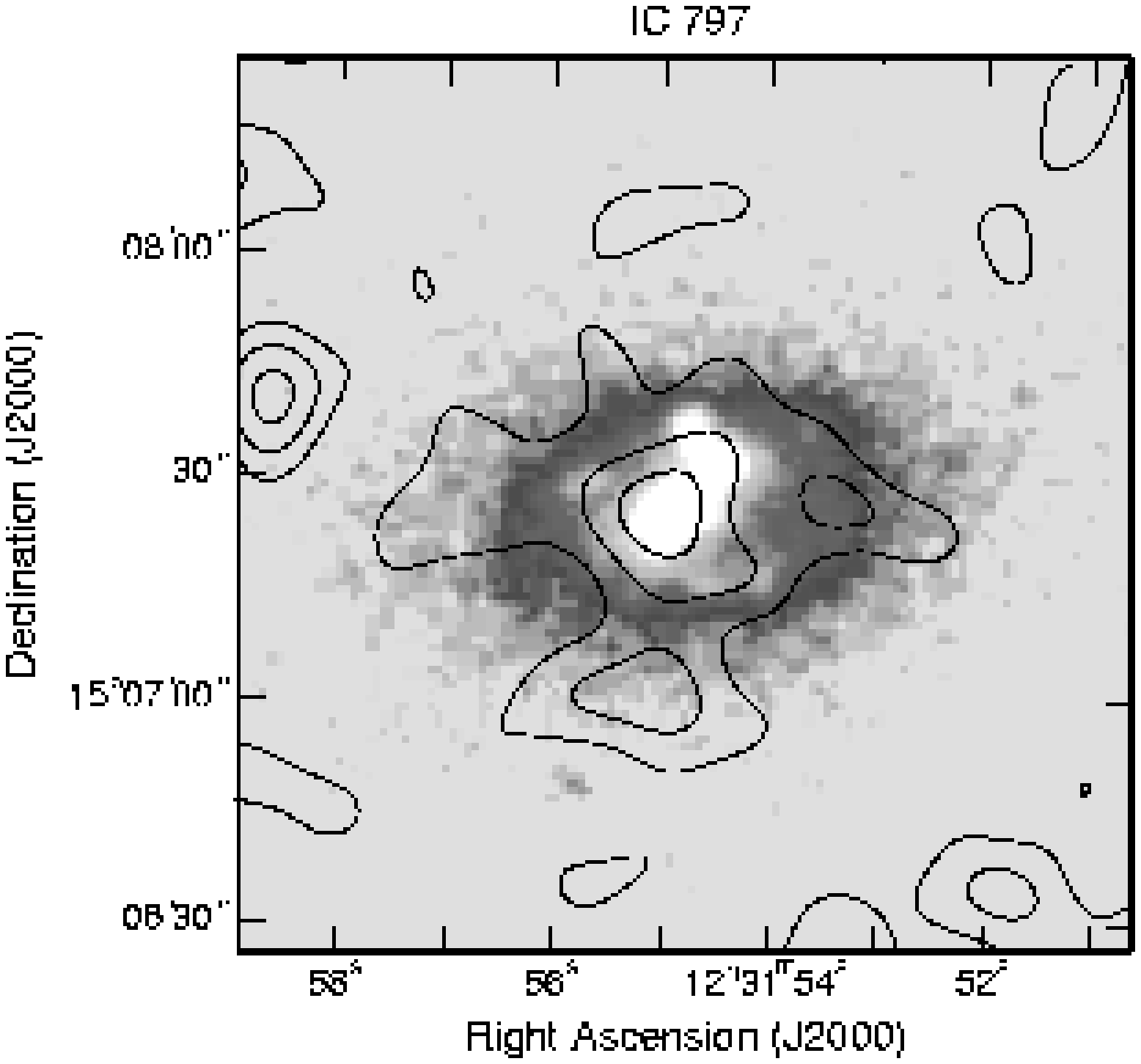}
 \hfill
 \includegraphics[angle=0, width=8cm]{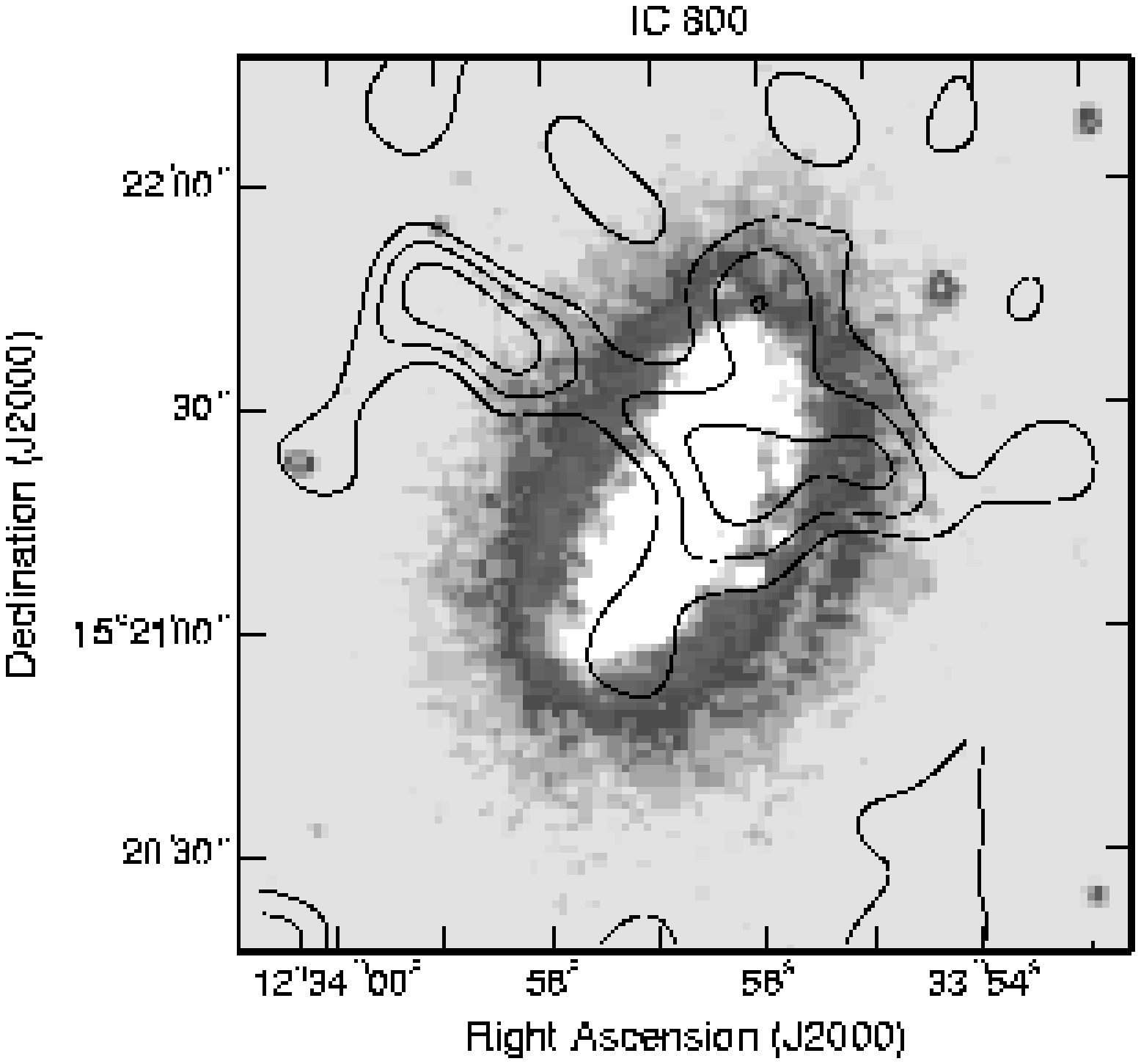}
 \hfill
 \includegraphics[angle=0, width=8cm]{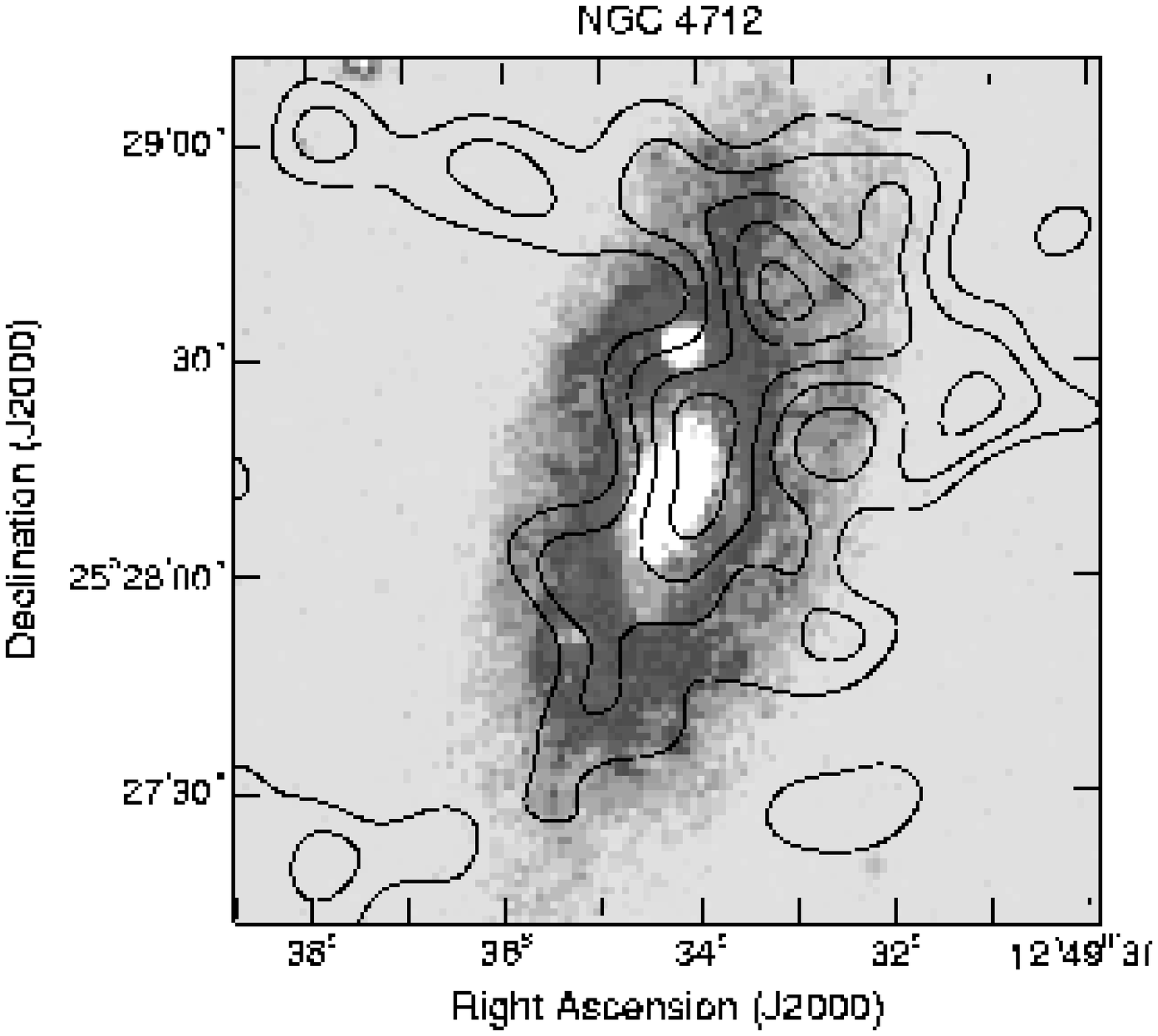}
 \hfill
 \contcaption{}
 \end{center}
\end{figure*} 

\begin{figure*}
 \begin{center}
 \includegraphics[angle=0, width=8cm]{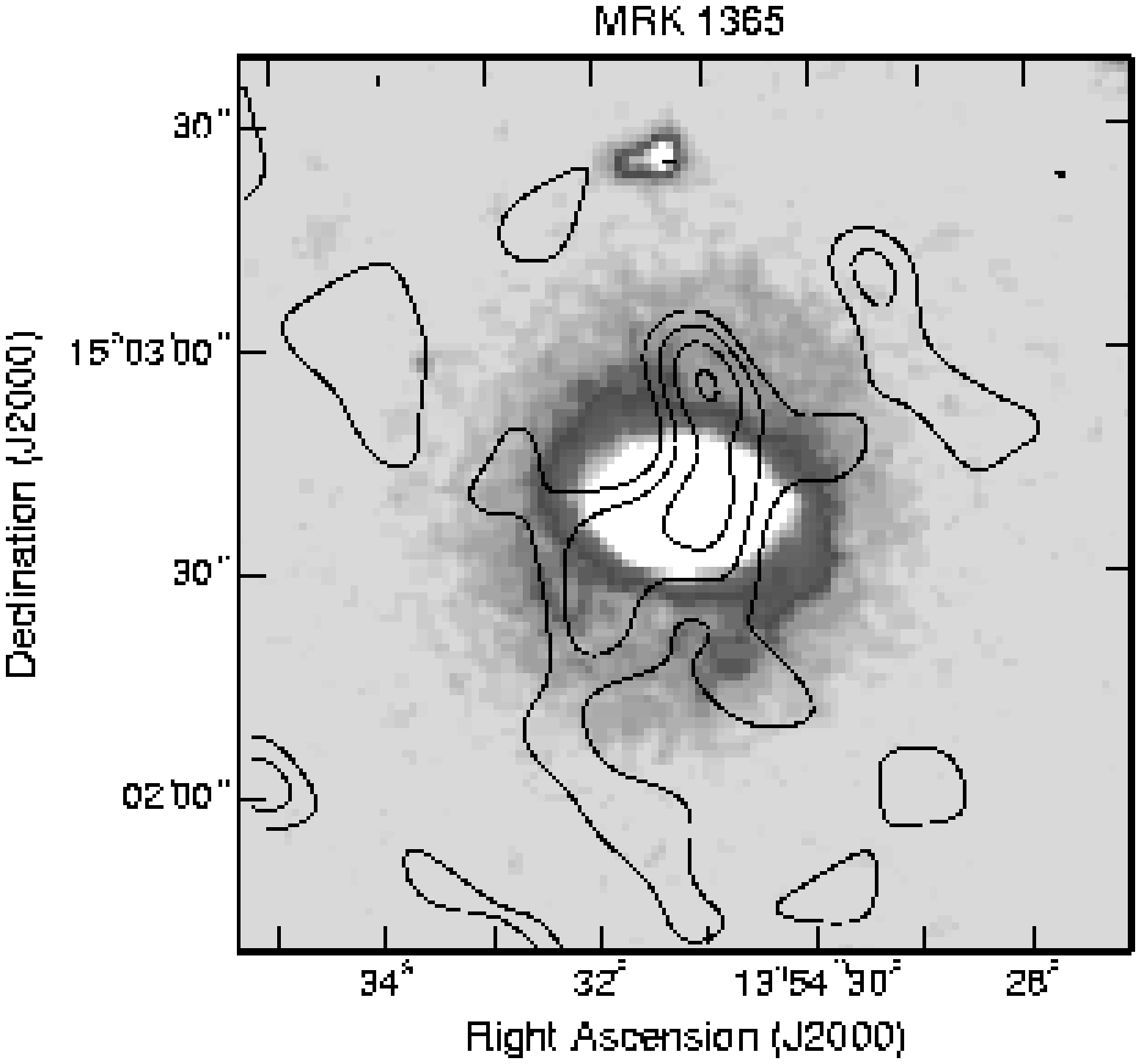}
 \hfill
 \includegraphics[angle=0, width=8cm]{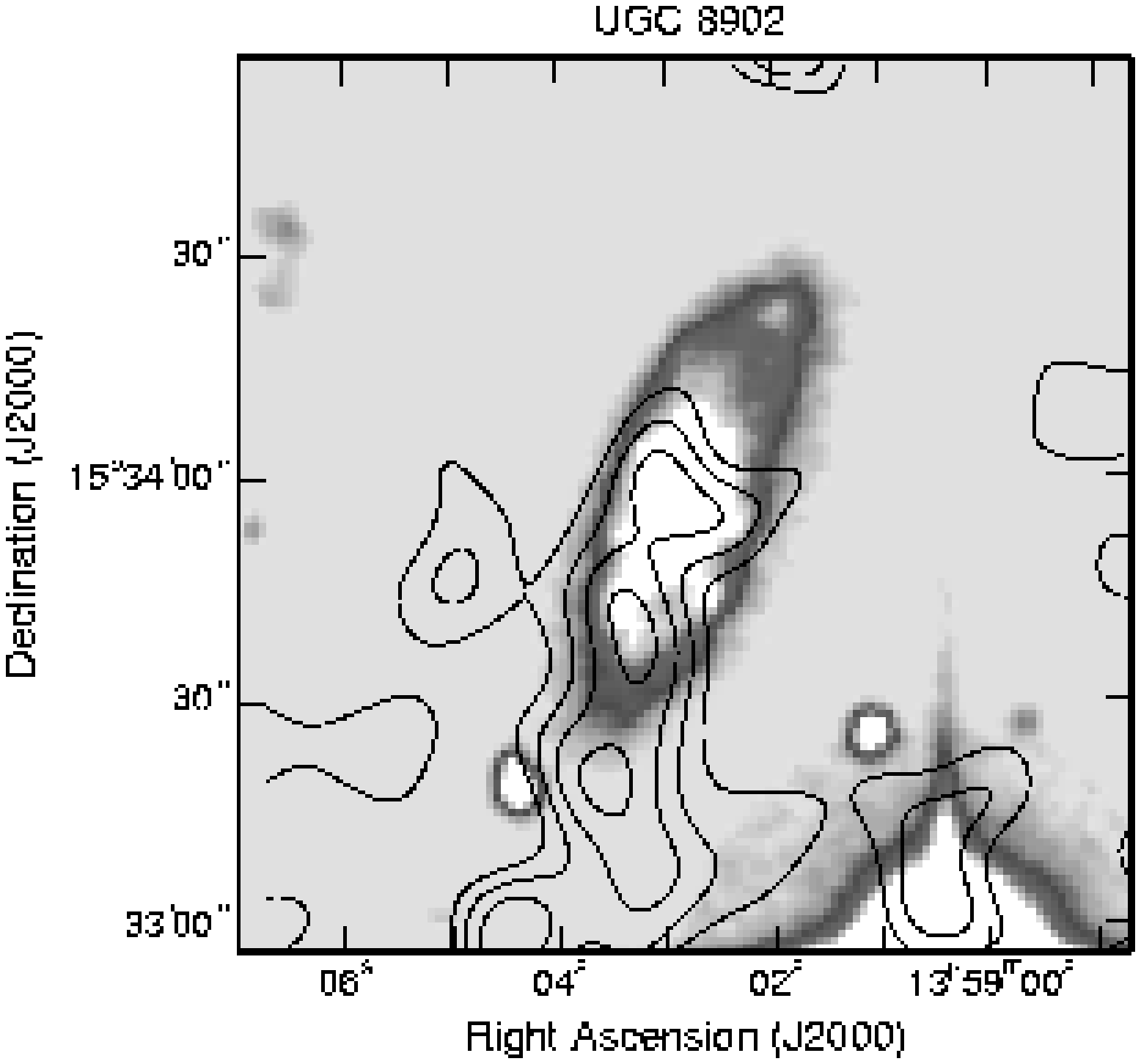}
 \hfill
 \includegraphics[angle=0, width=8cm]{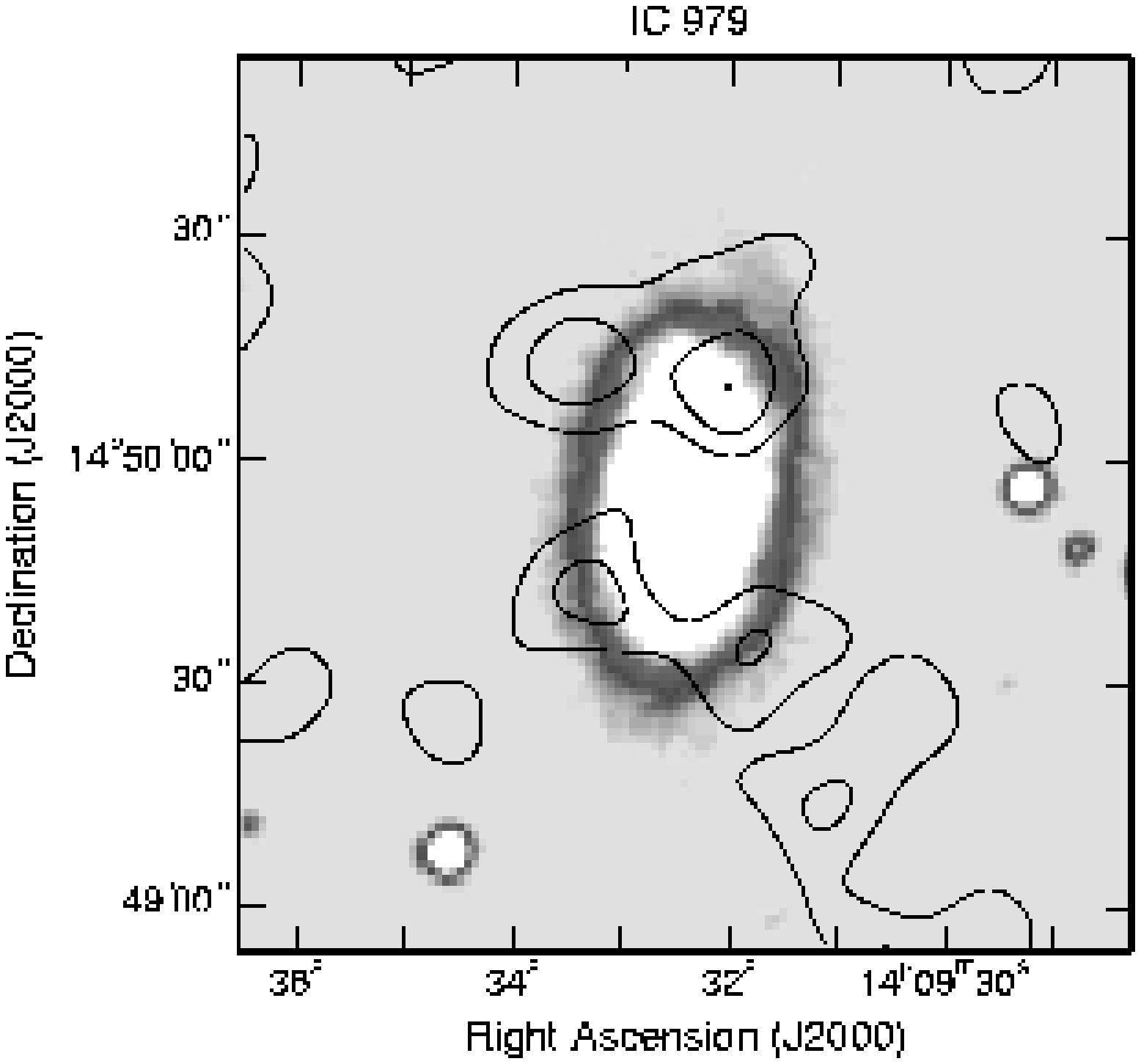}
 \hfill
 \includegraphics[angle=0, width=8cm]{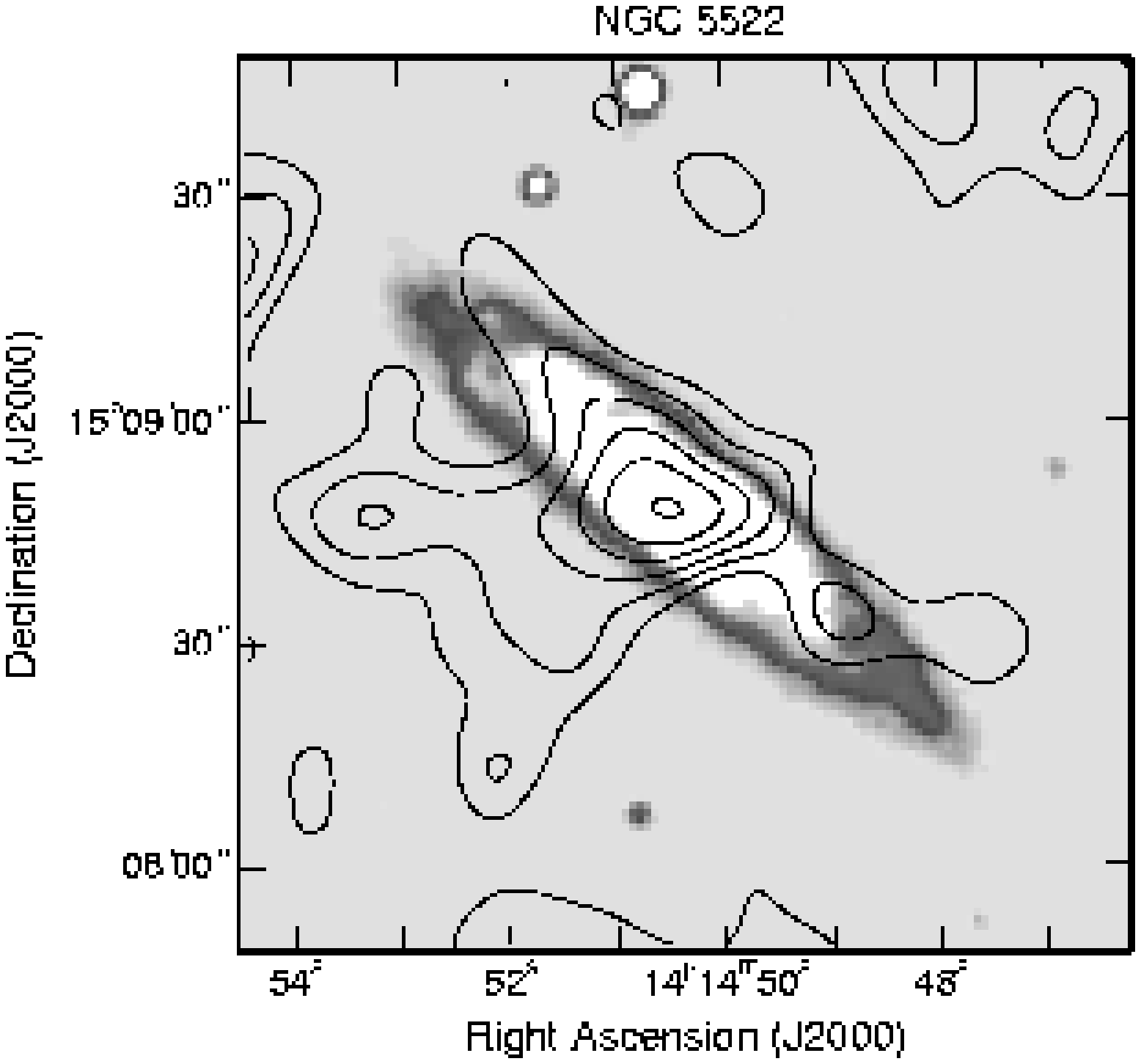}
 \hfill
 \includegraphics[angle=0, width=8cm]{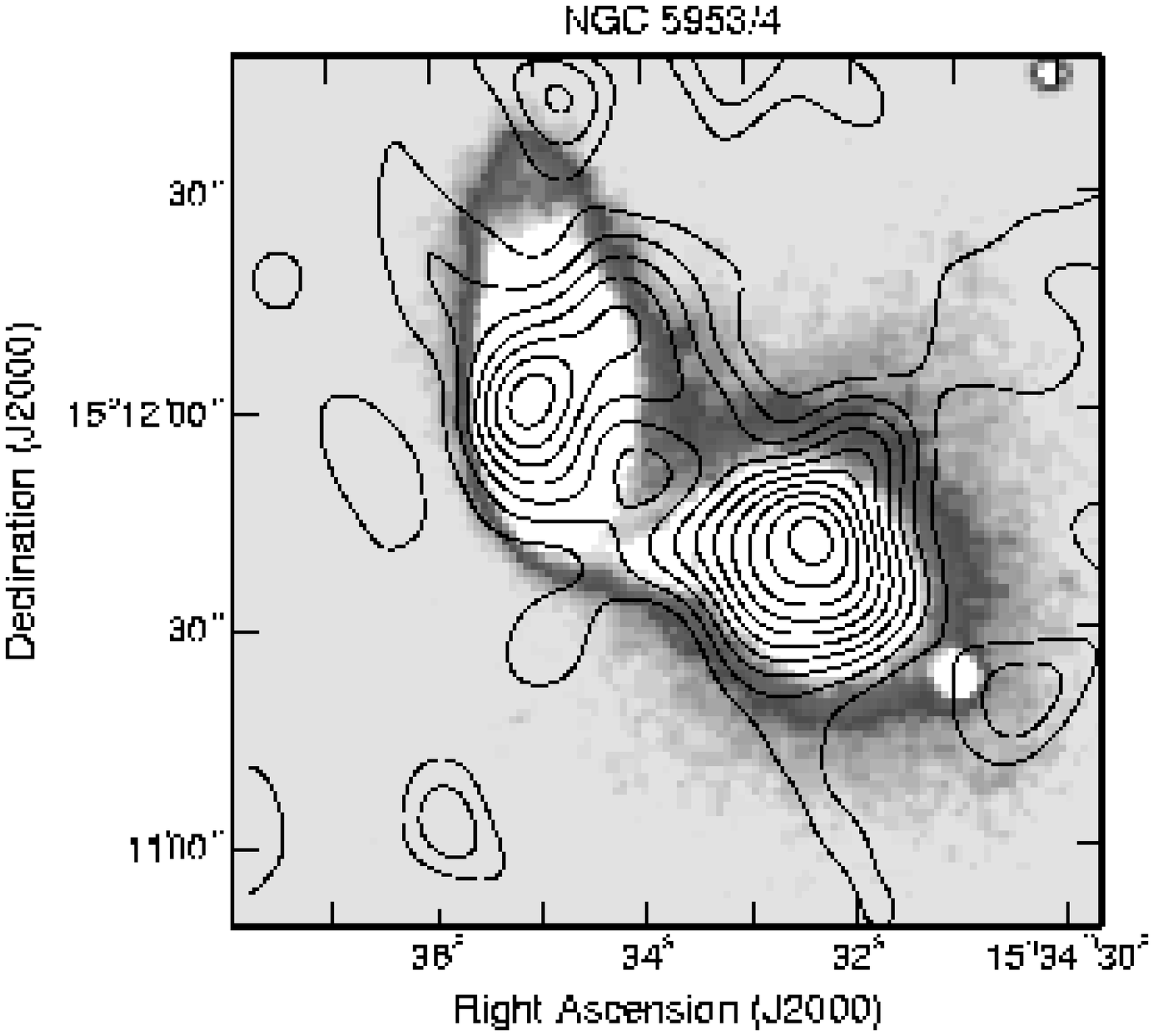}
 \hfill
 \includegraphics[angle=0, width=8cm]{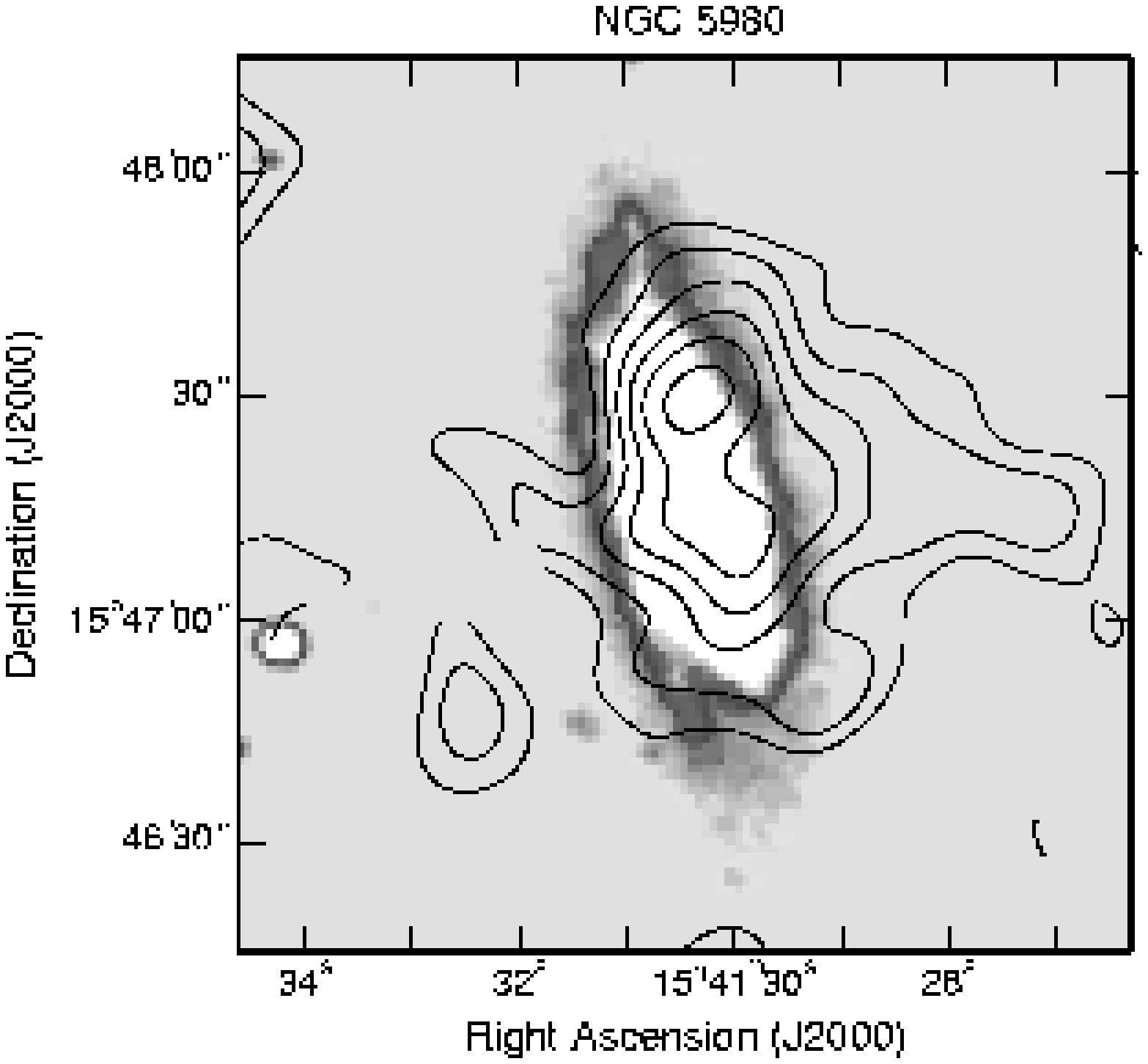}
 \hfill
 \contcaption{}
 \end{center}
\end{figure*}

\begin{figure*}
 \begin{center}
 \includegraphics[angle=0, width=8cm]{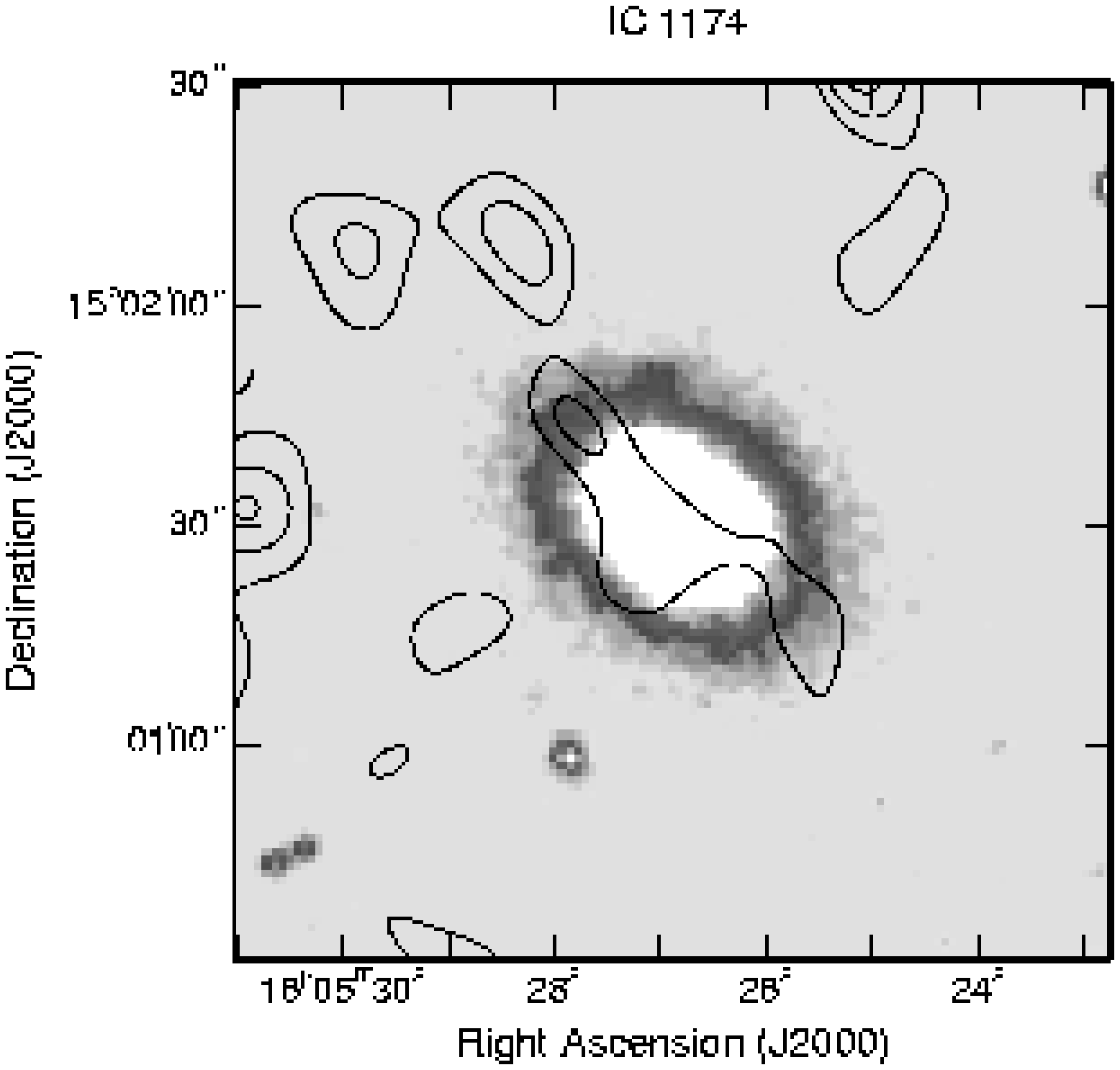}
 \hfill
 \includegraphics[angle=0, width=8cm]{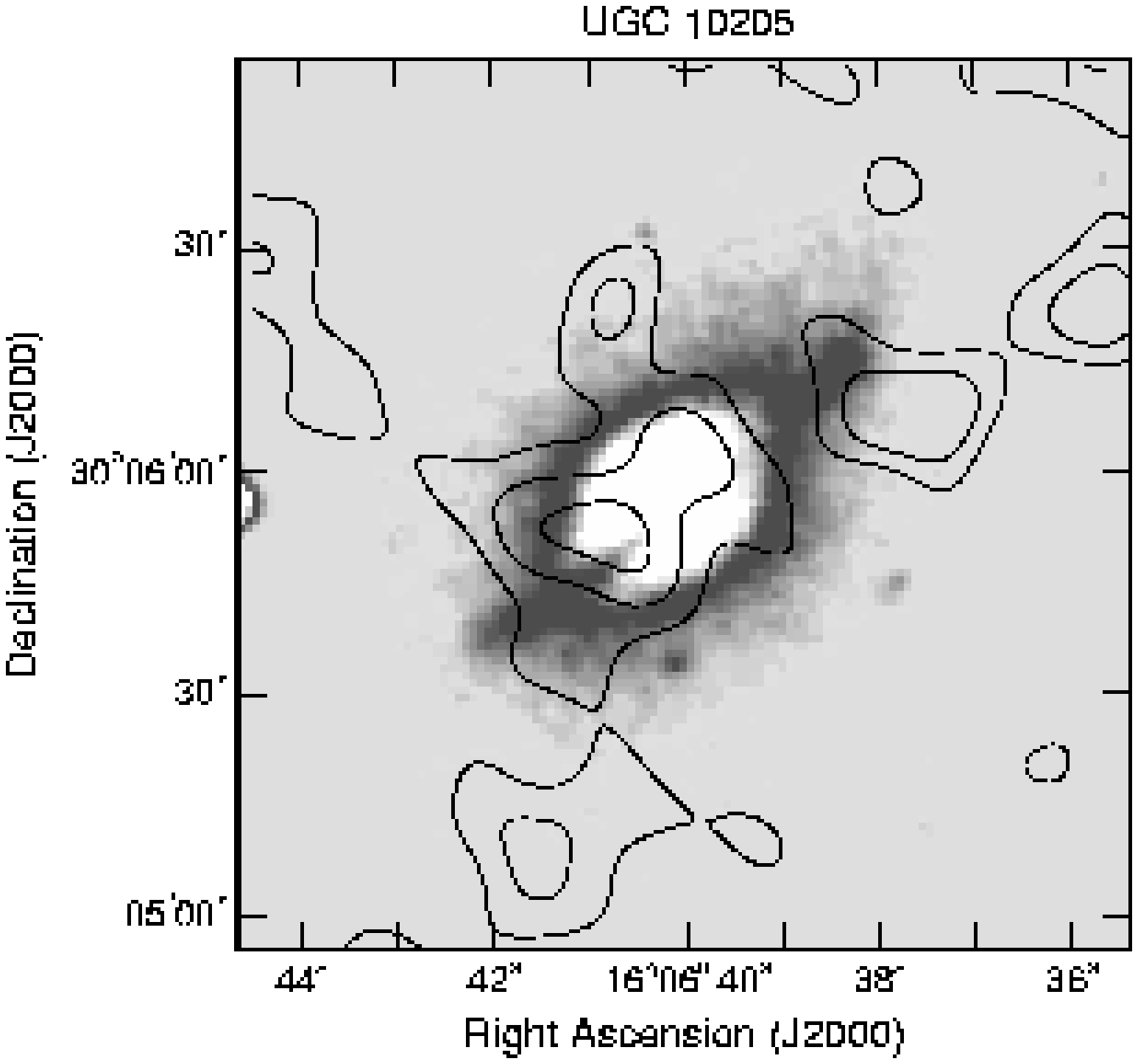}
 \hfill
 \includegraphics[angle=0, width=8cm]{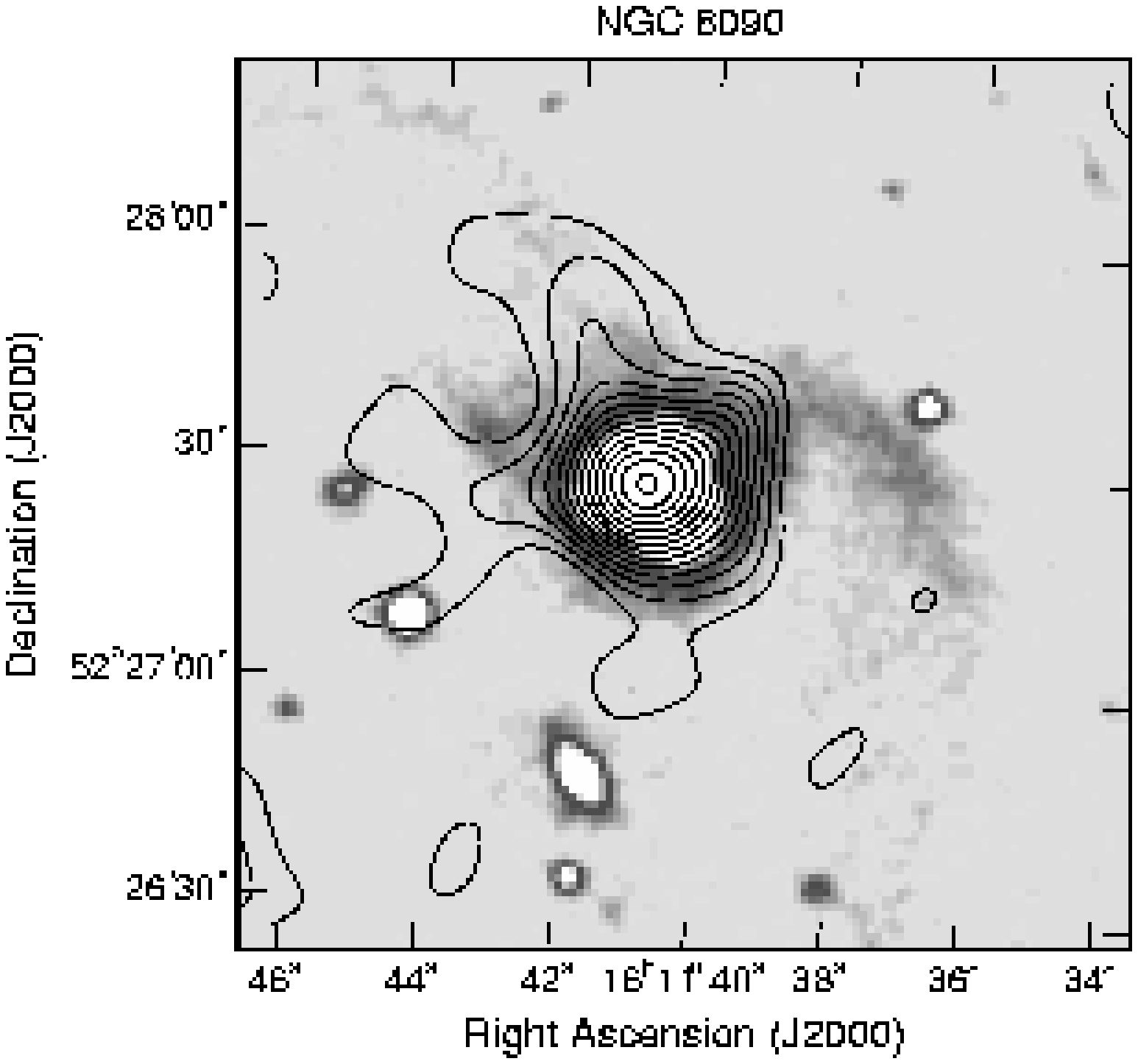}
 \hfill
 \includegraphics[angle=0, width=8cm]{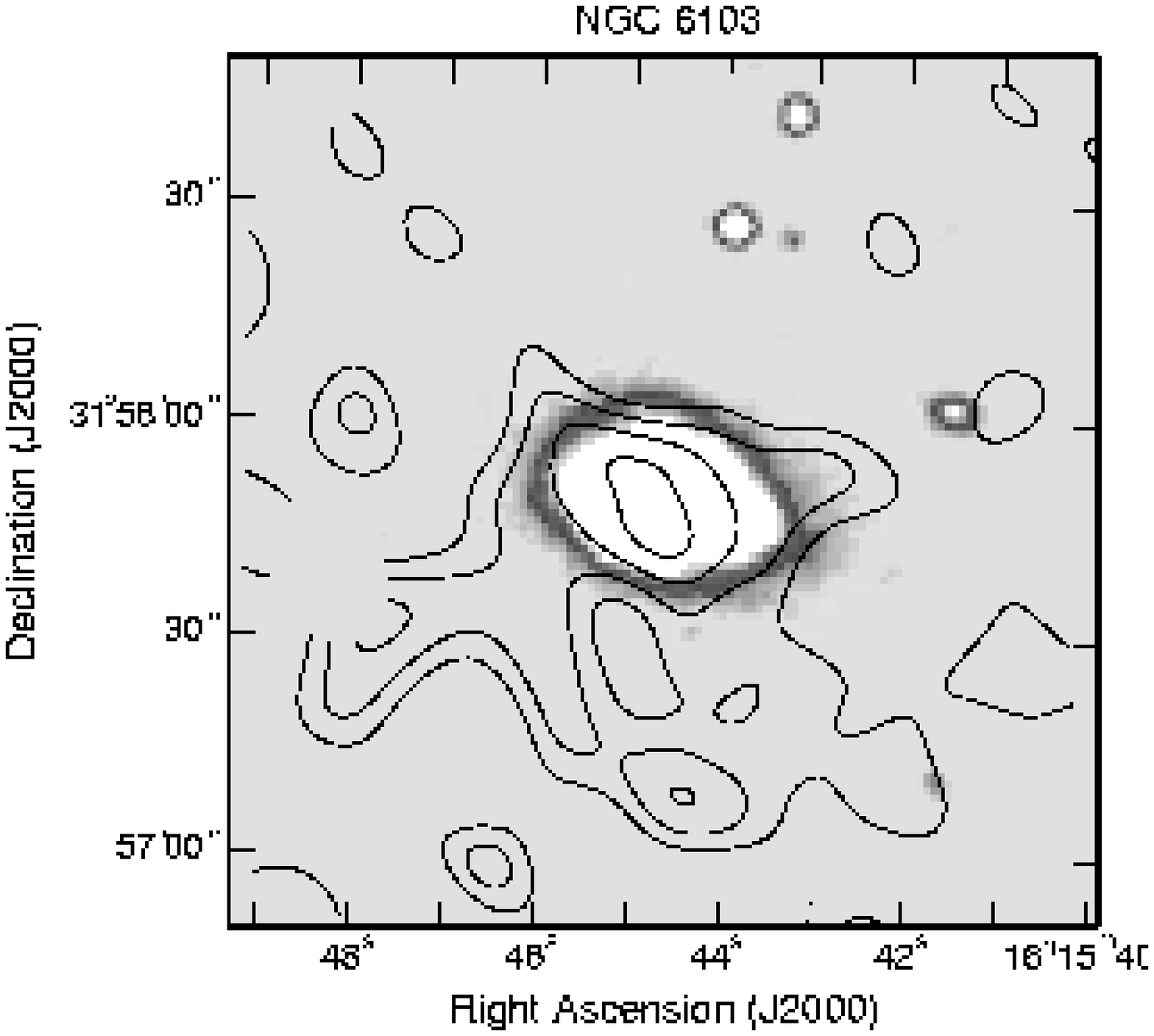}
 \hfill
 \includegraphics[angle=0, width=8cm]{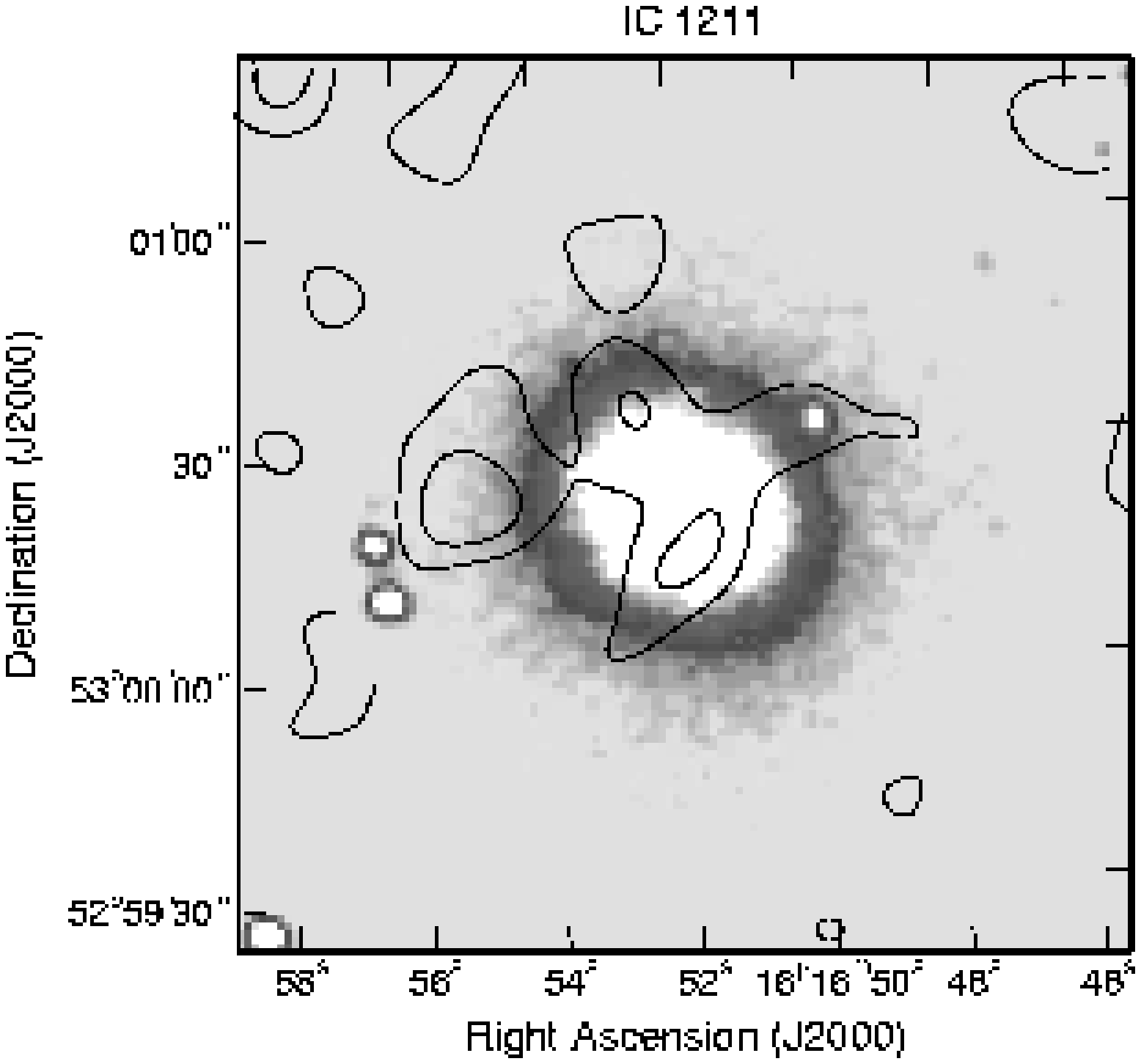}
 \hfill 
 \includegraphics[angle=0, width=8cm]{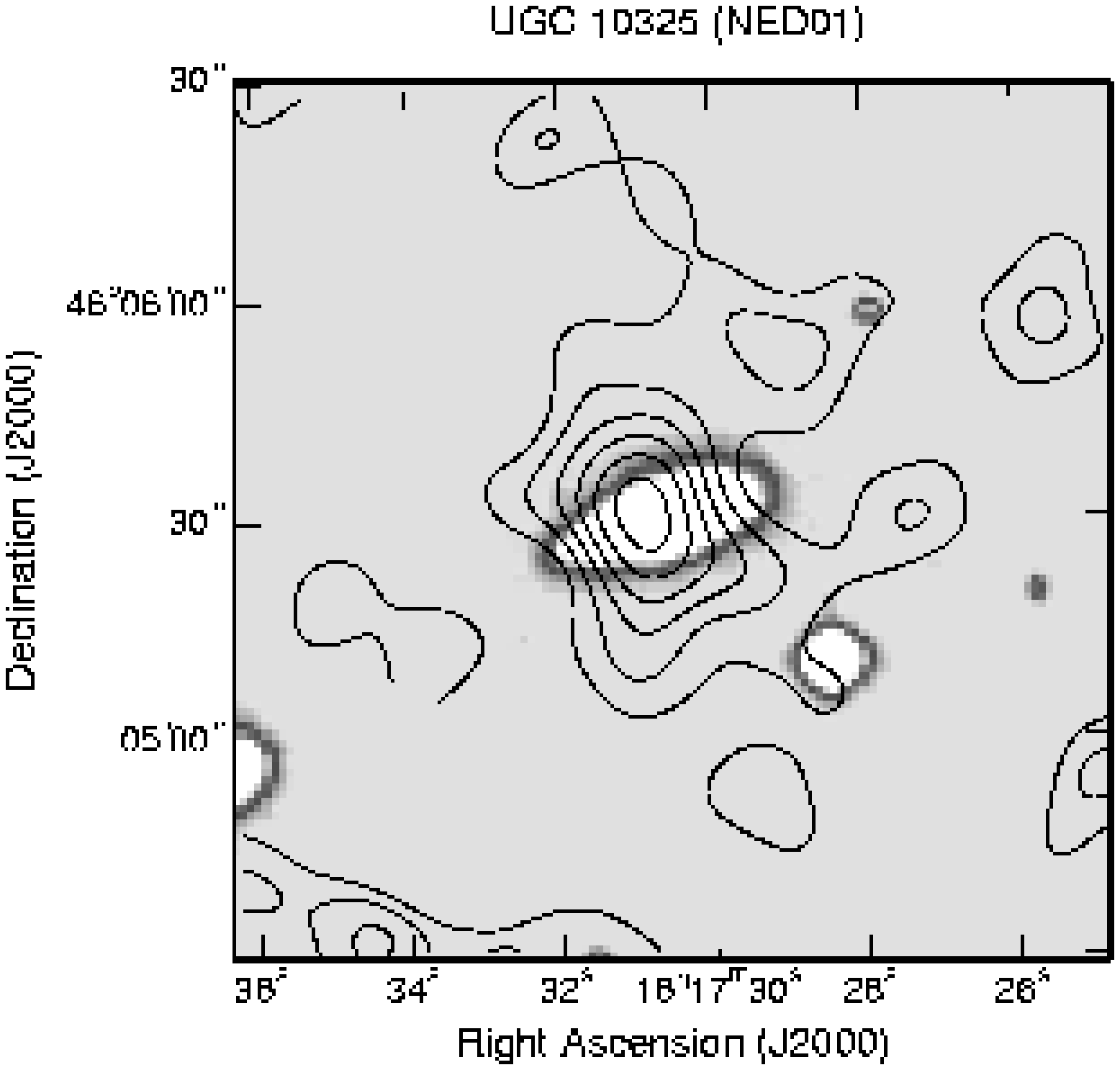}
 \hfill
 \contcaption{}
 \end{center}
\end{figure*}

\begin{figure*}
 \begin{center}
 \includegraphics[angle=0, width=8cm]{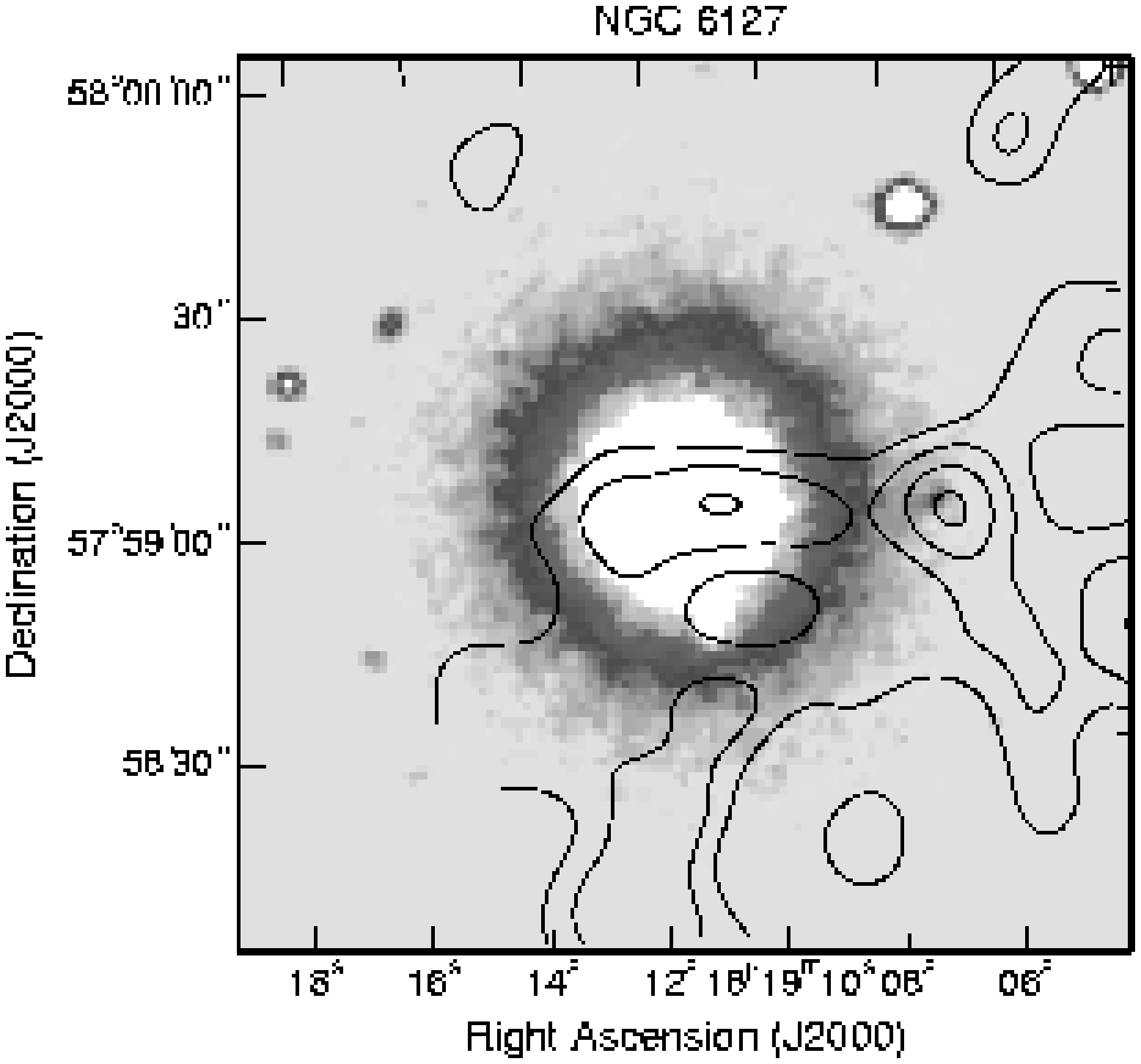}
 \hfill
 \includegraphics[angle=0, width=8cm]{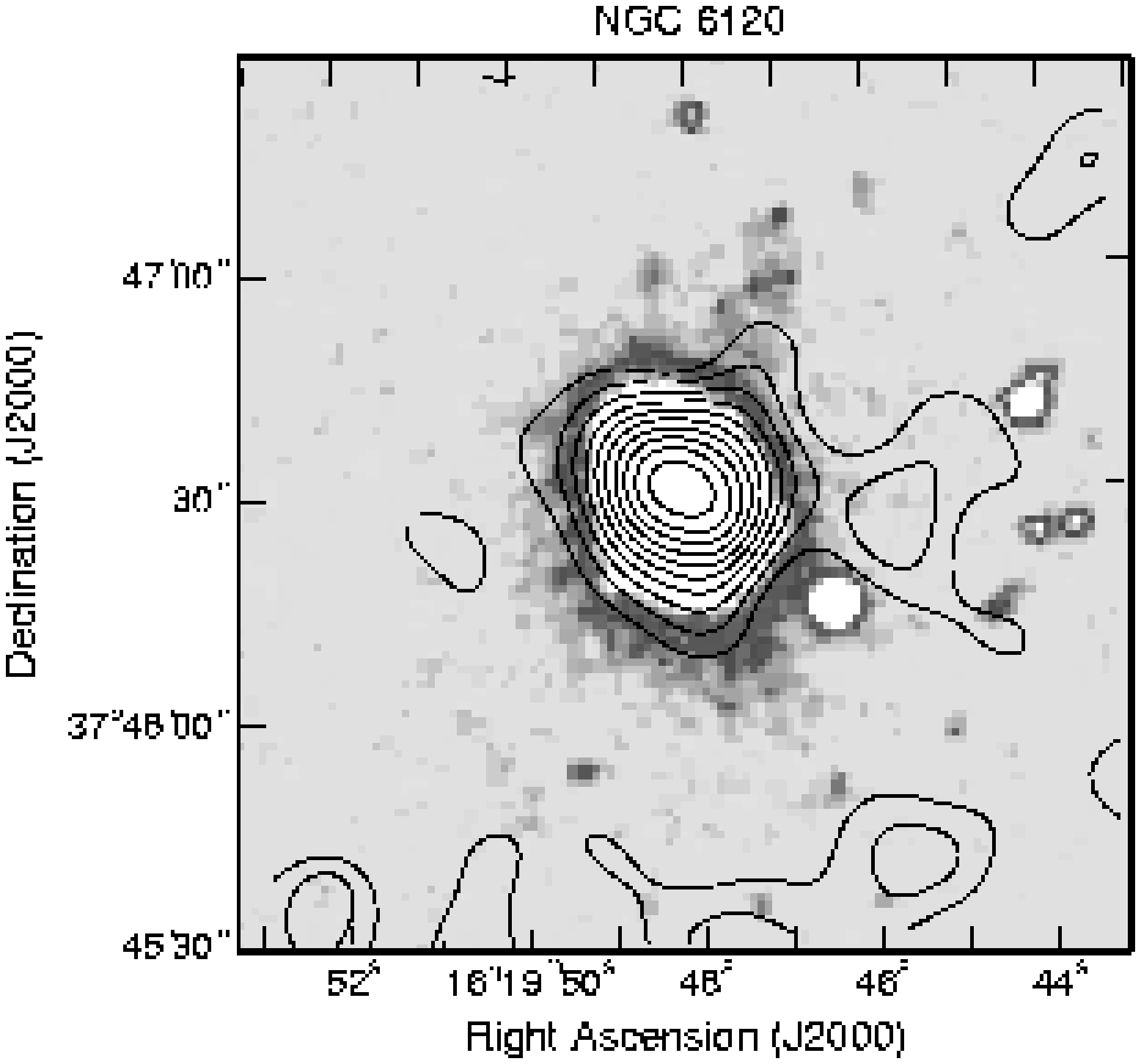}
 \hfill
 \includegraphics[angle=0, width=8cm]{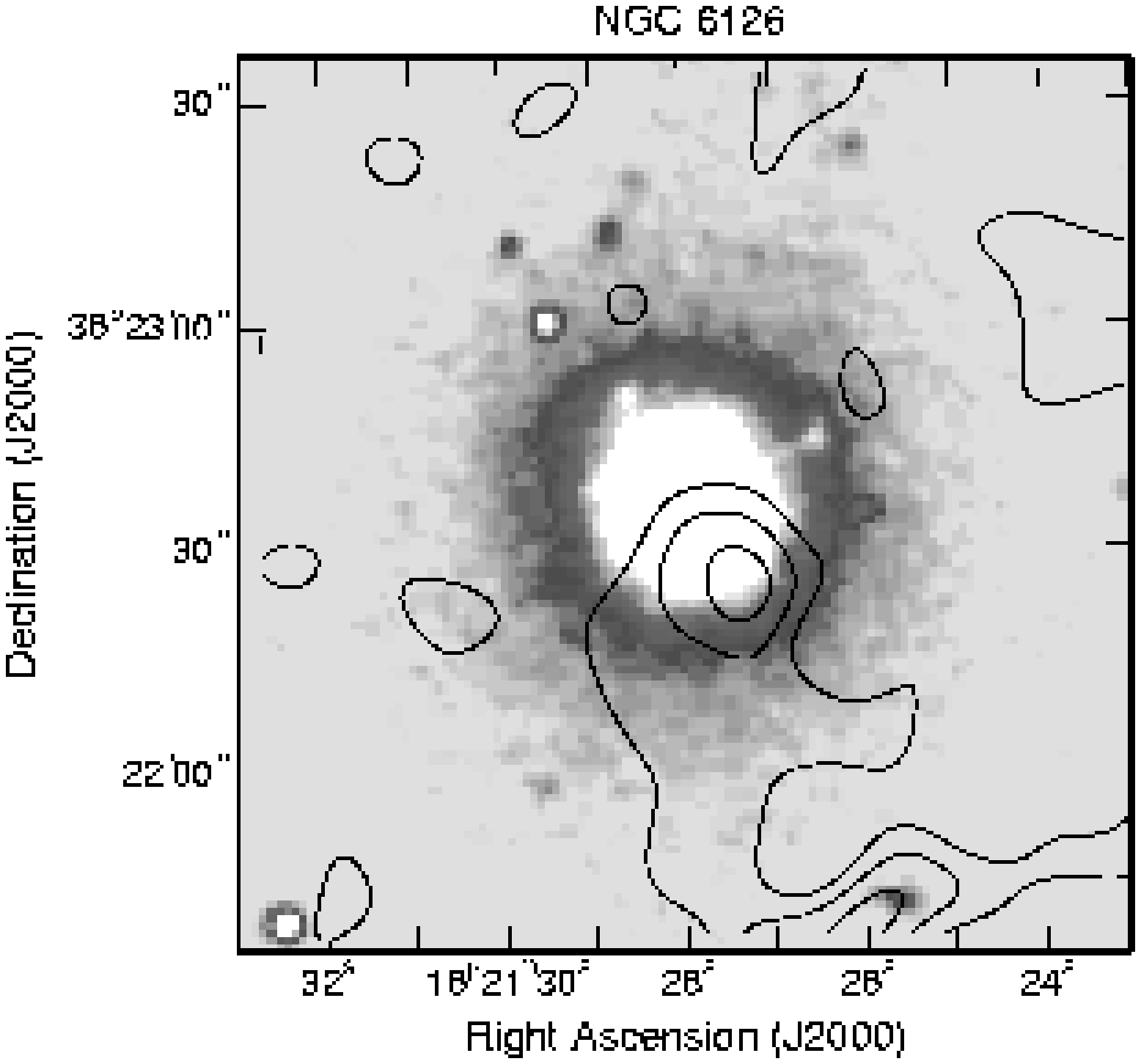}
 \hfill
 \includegraphics[angle=0, width=8cm]{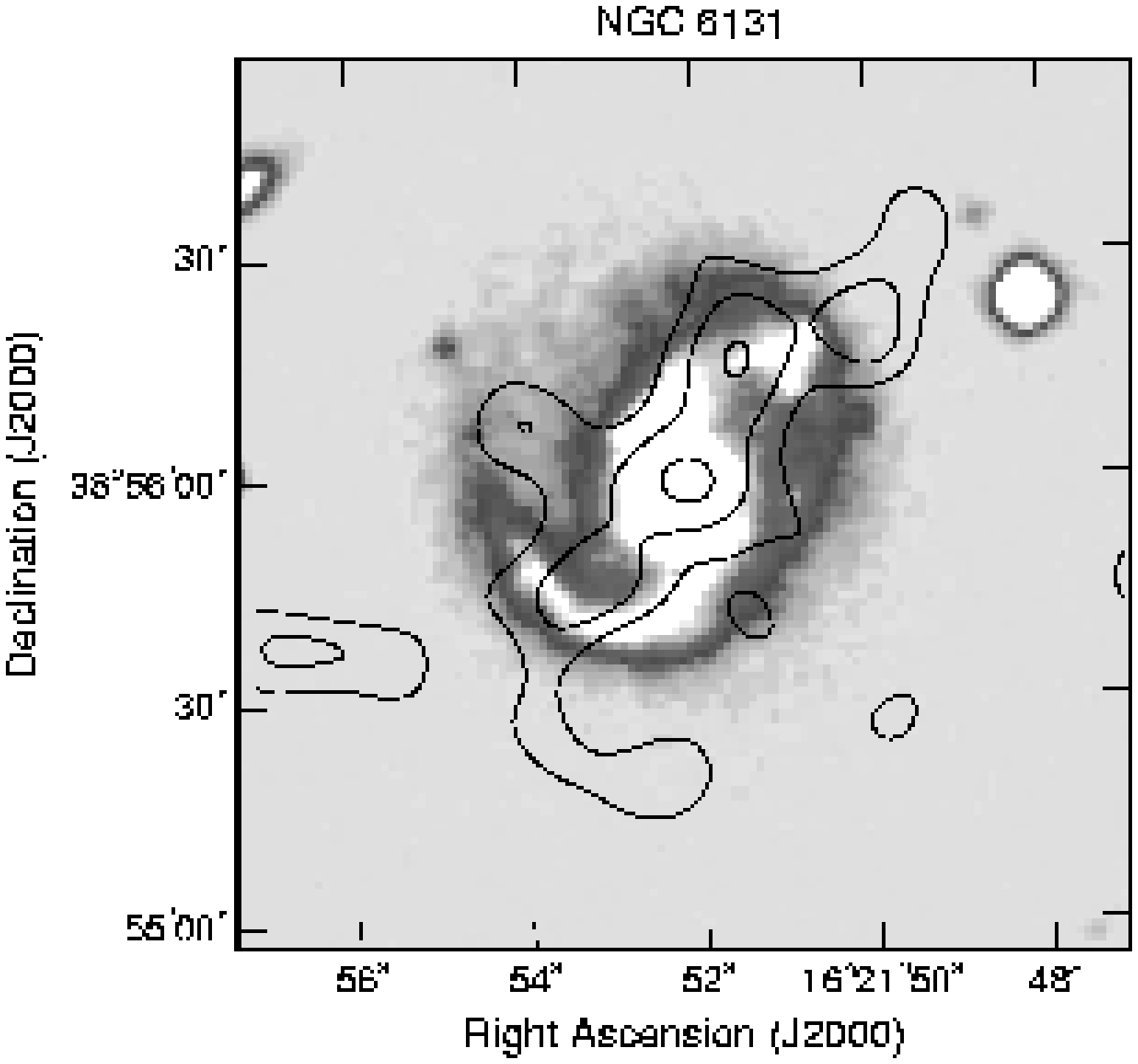}
 \hfill 
 \includegraphics[angle=0, width=8cm]{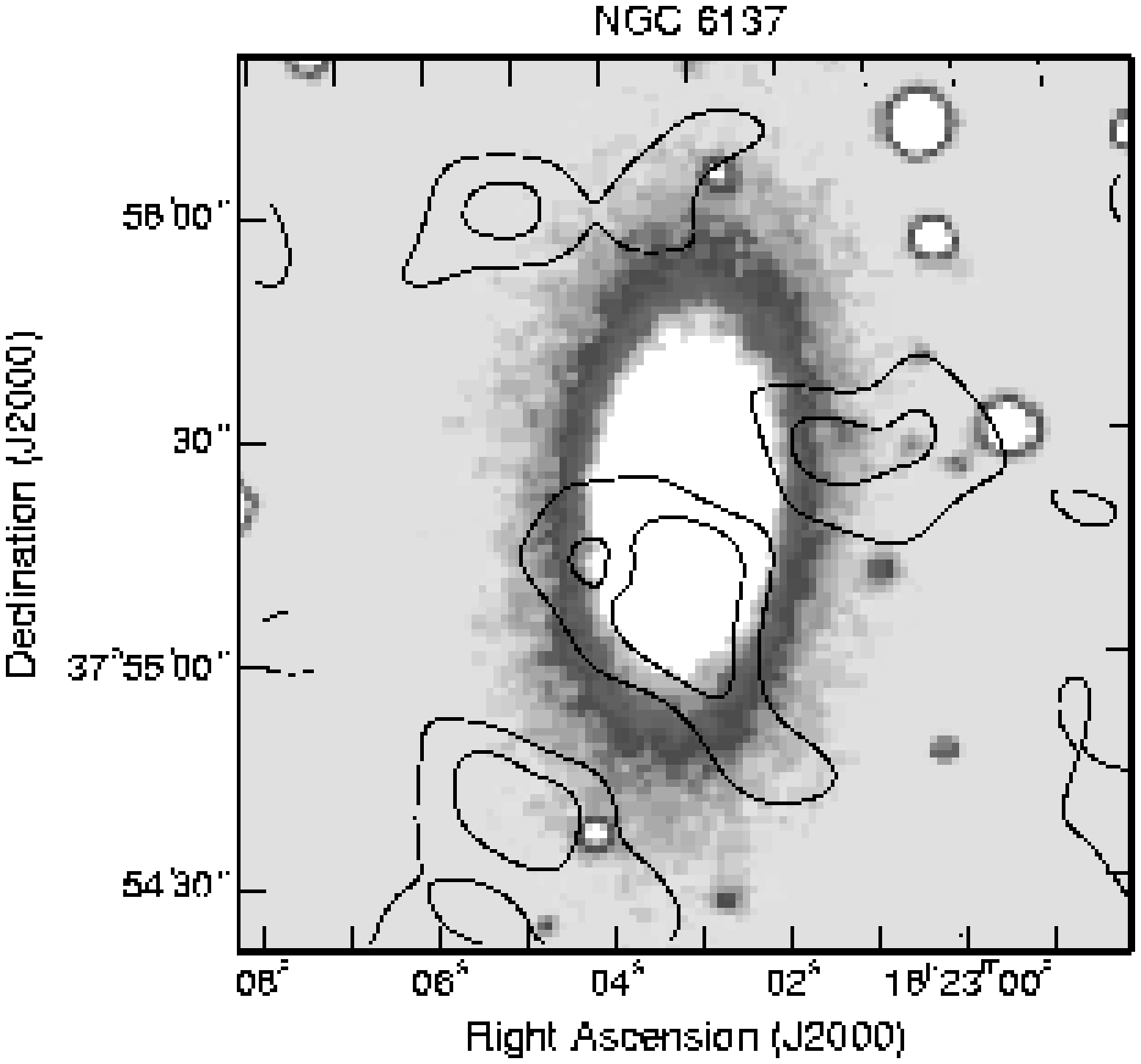}
 \hfill
 \includegraphics[angle=0, width=8cm]{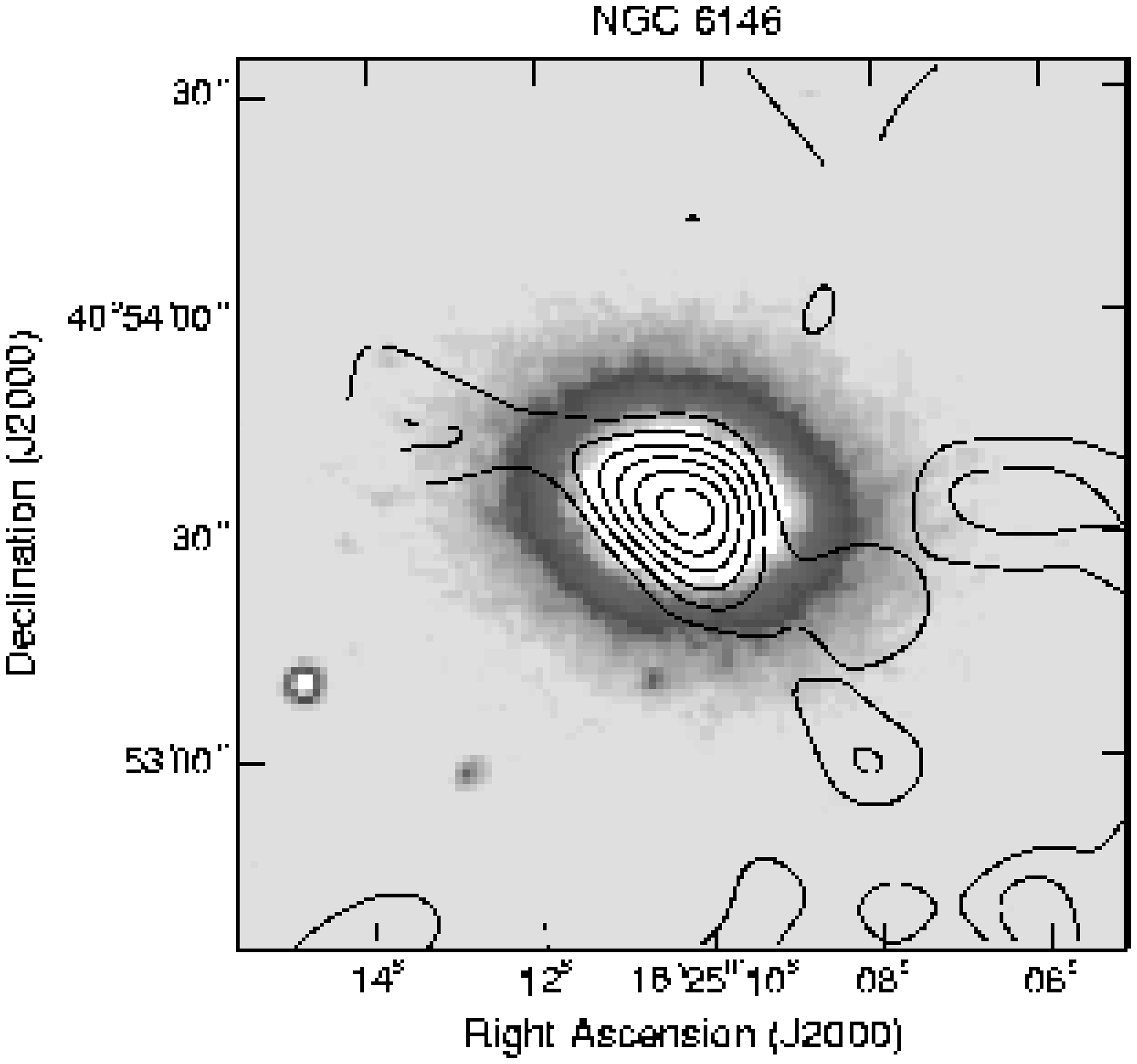}
 \hfill
 \contcaption{}
 \end{center}
\end{figure*} 

\begin{figure*}
 \begin{center}
 \includegraphics[angle=0, width=8cm]{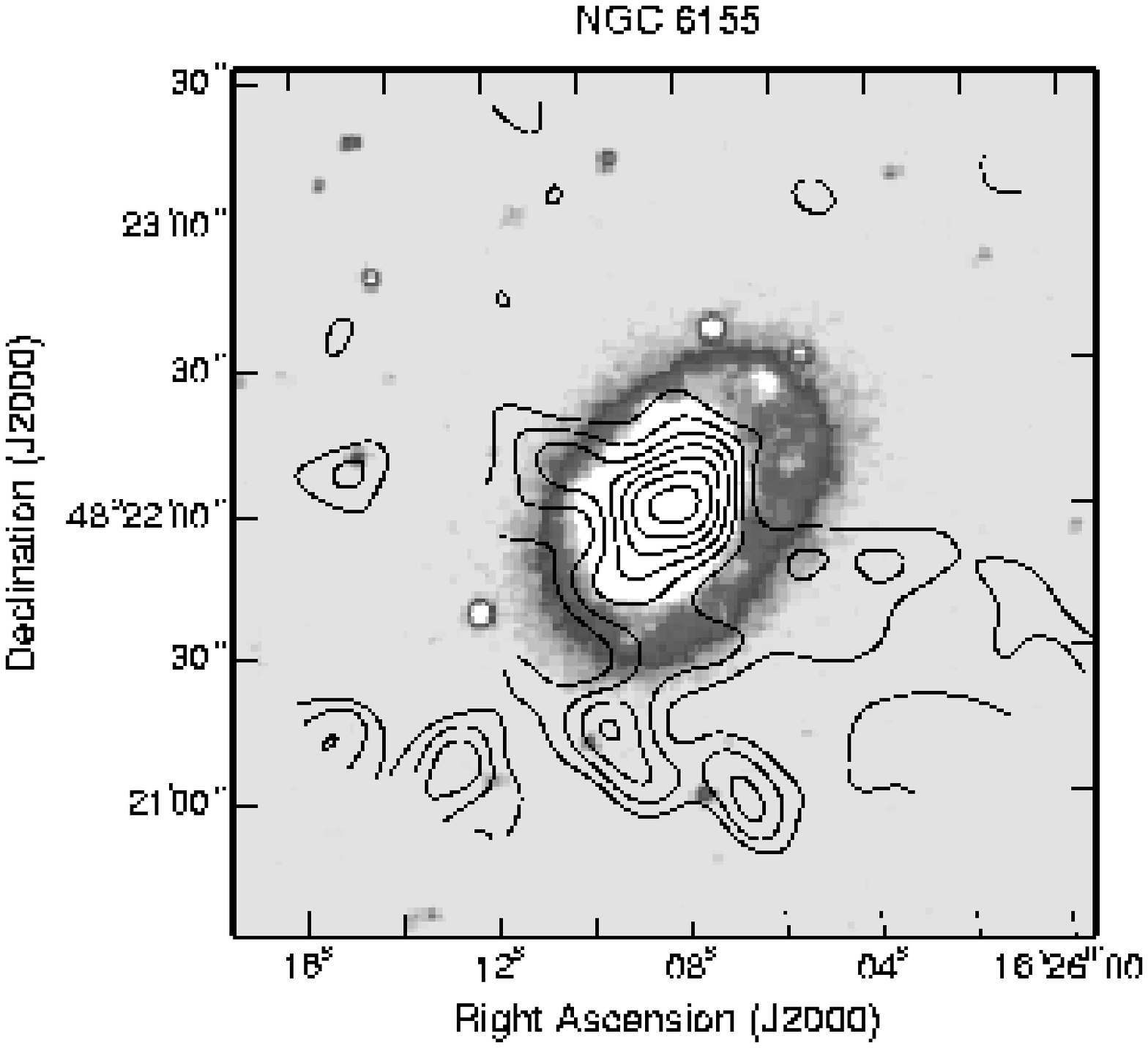}
 \hfill
 \includegraphics[angle=0, width=8cm]{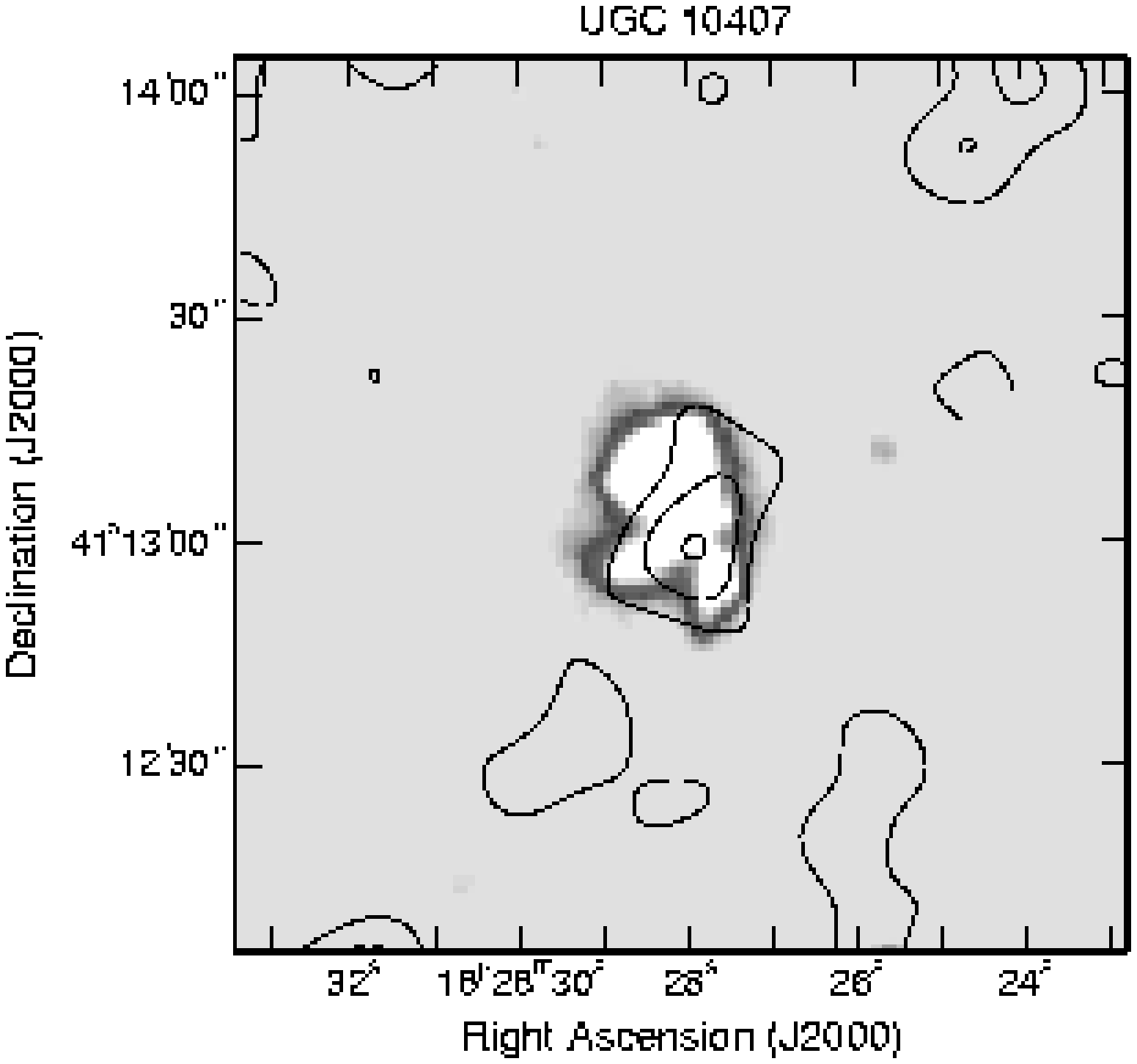}
 \hfill
 \includegraphics[angle=0, width=8cm]{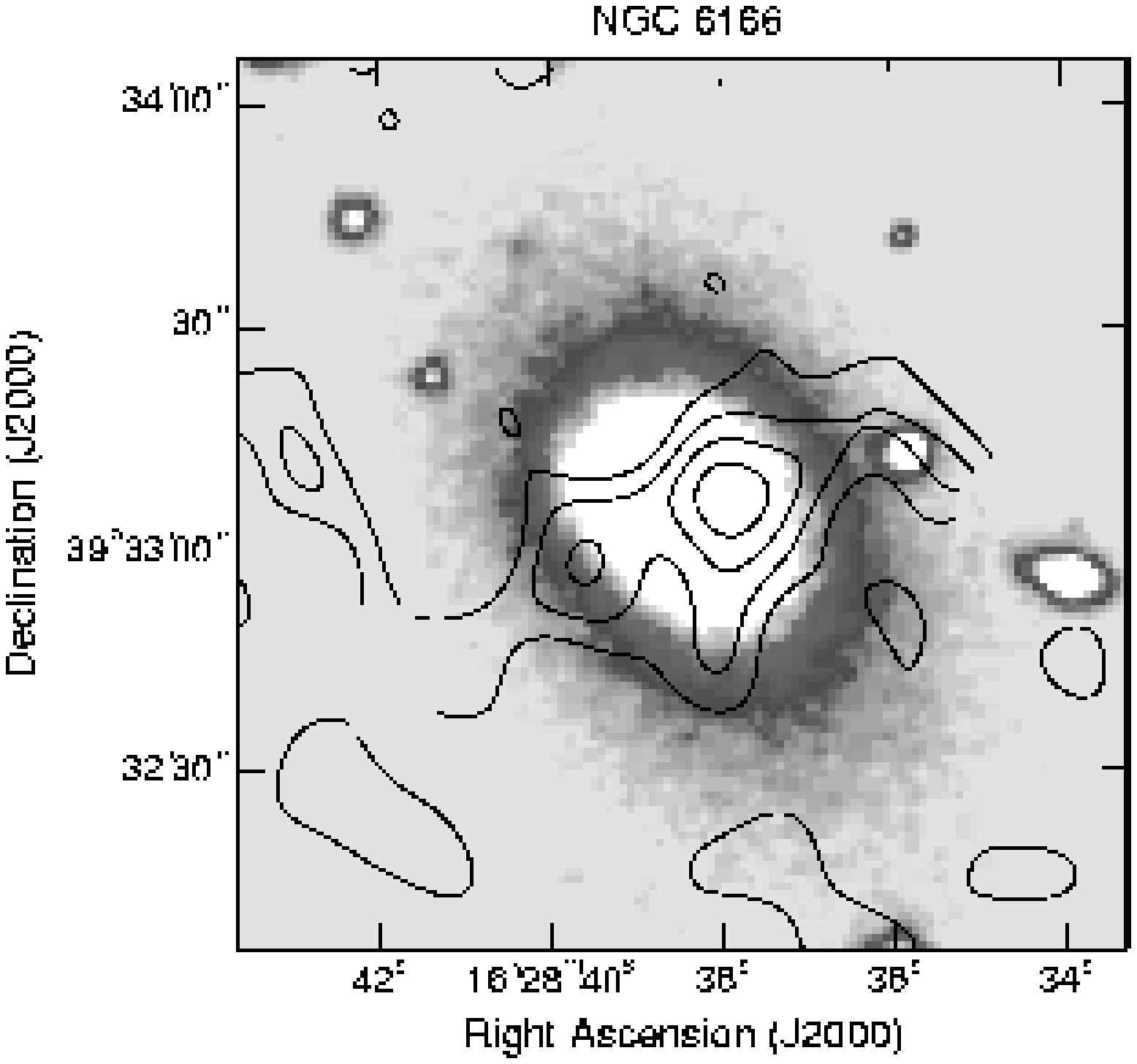}
 \hfill
 \includegraphics[angle=0, width=8cm]{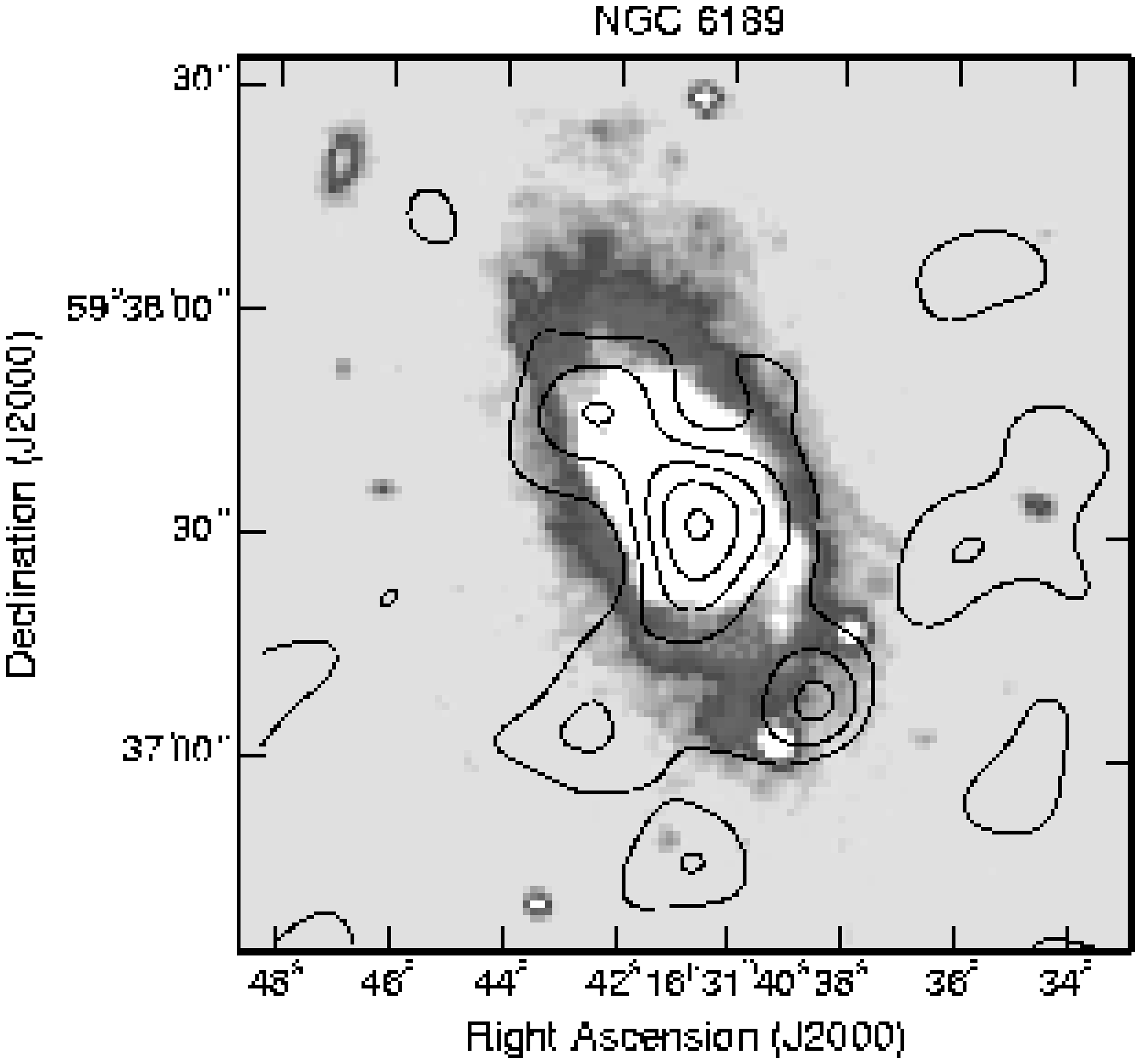}
 \hfill
 \includegraphics[angle=0, width=8cm]{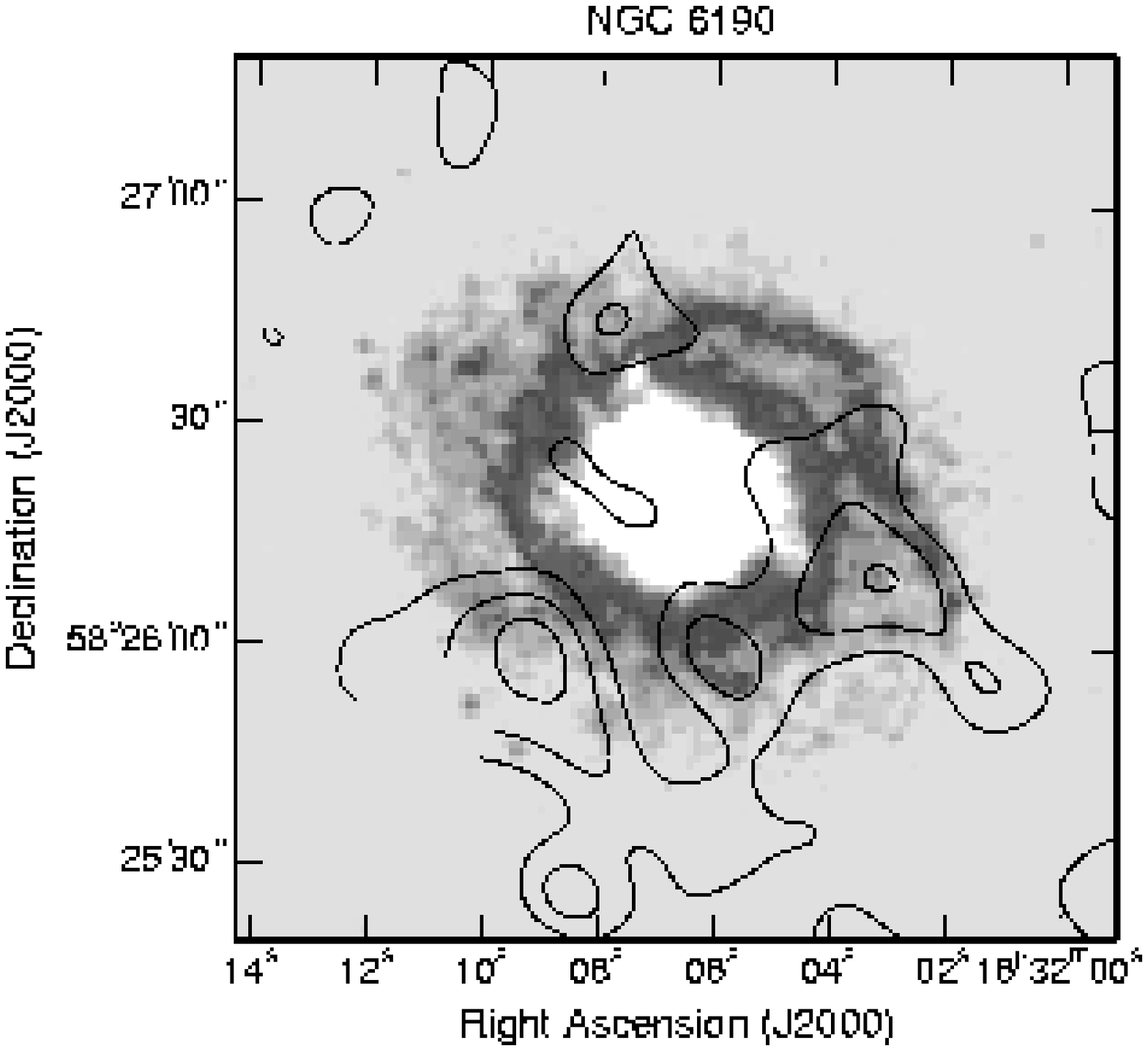}
 \hfill 
 \includegraphics[angle=0, width=8cm]{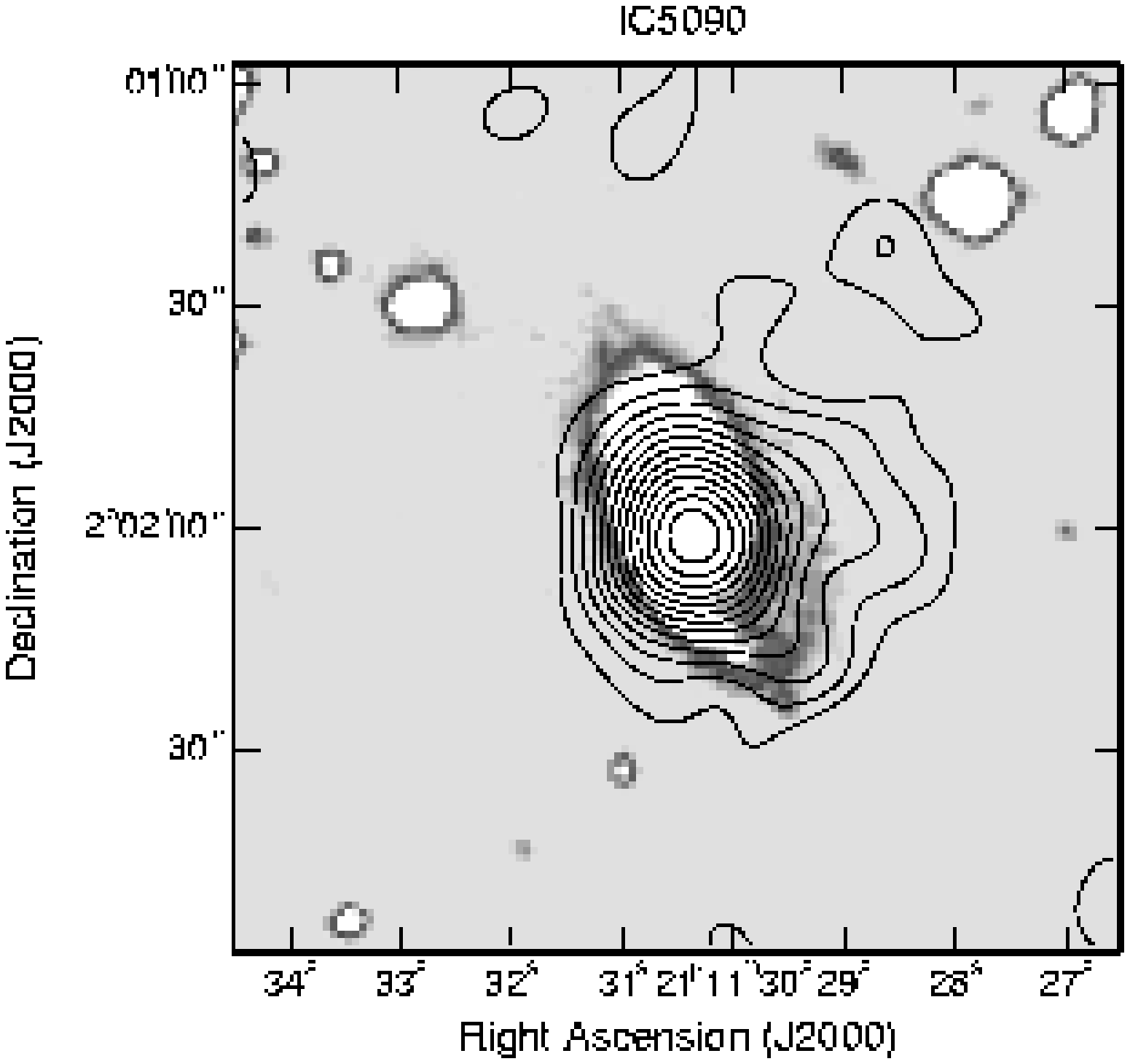}
 \hfill
 \contcaption{}
 \end{center}
\end{figure*} 

\begin{figure*}
 \begin{center}
 \includegraphics[angle=0, width=8cm]{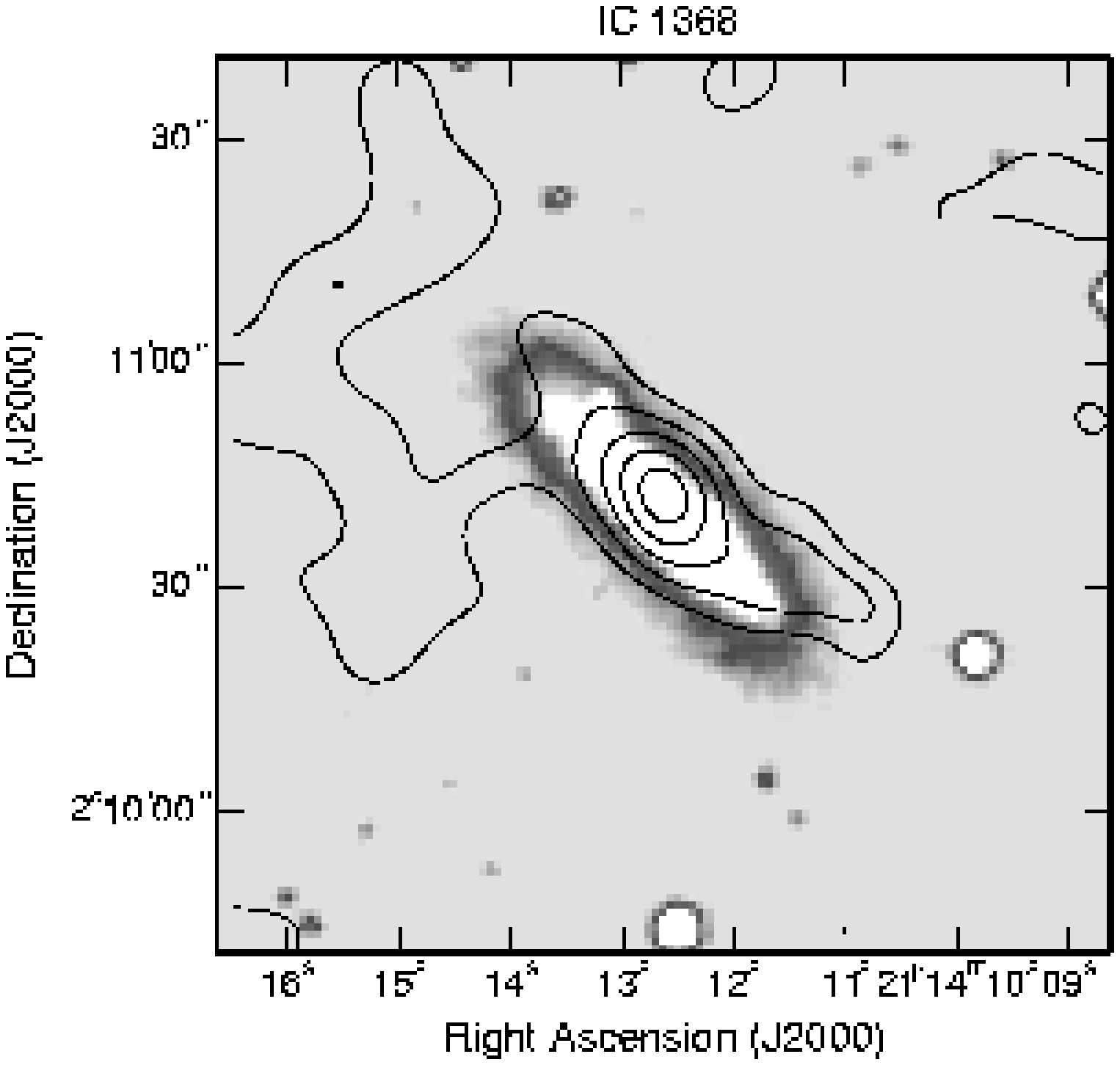}
 \hfill
 \includegraphics[angle=0, width=8cm]{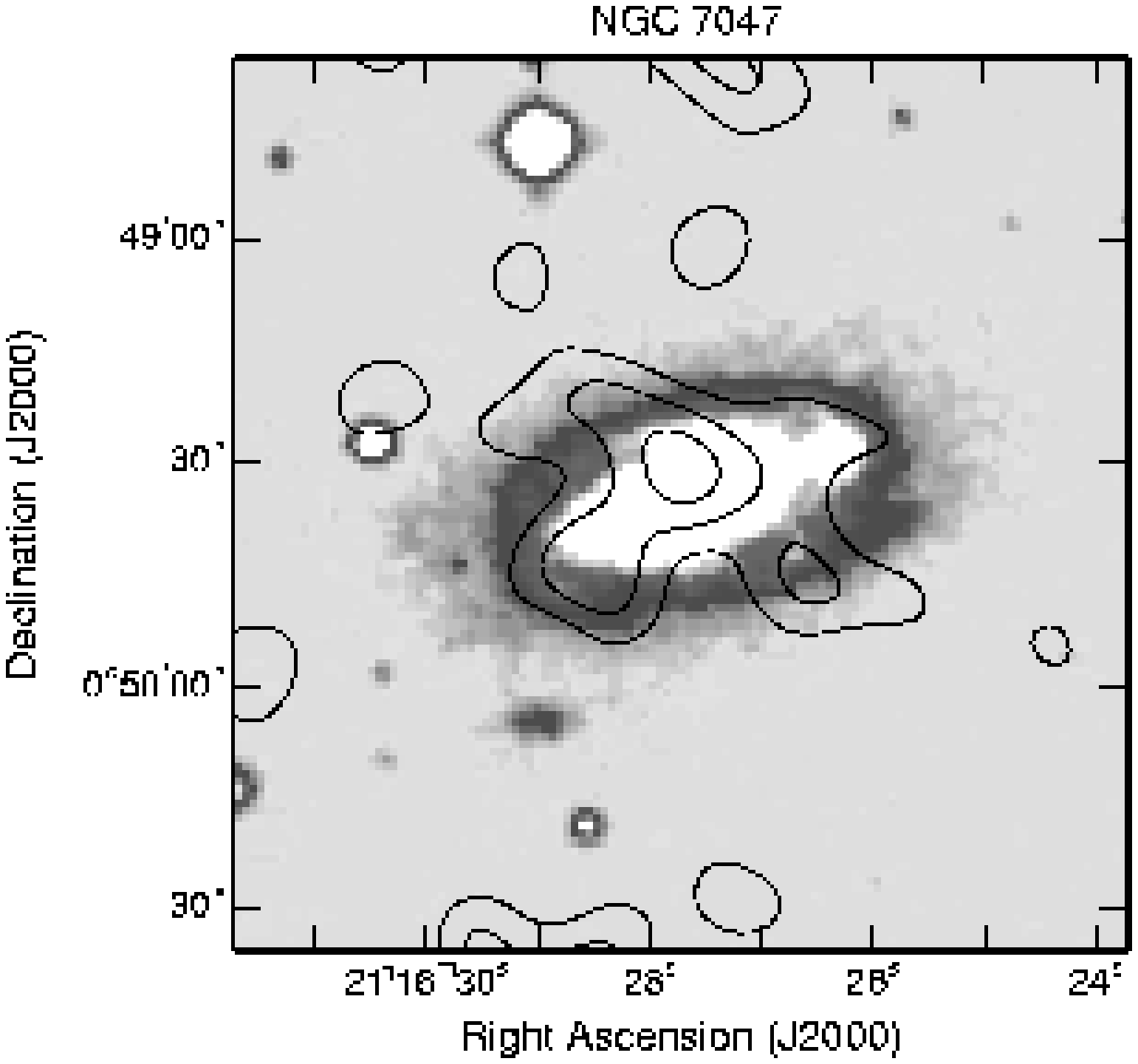}
 \hfill
 \includegraphics[angle=0, width=8cm]{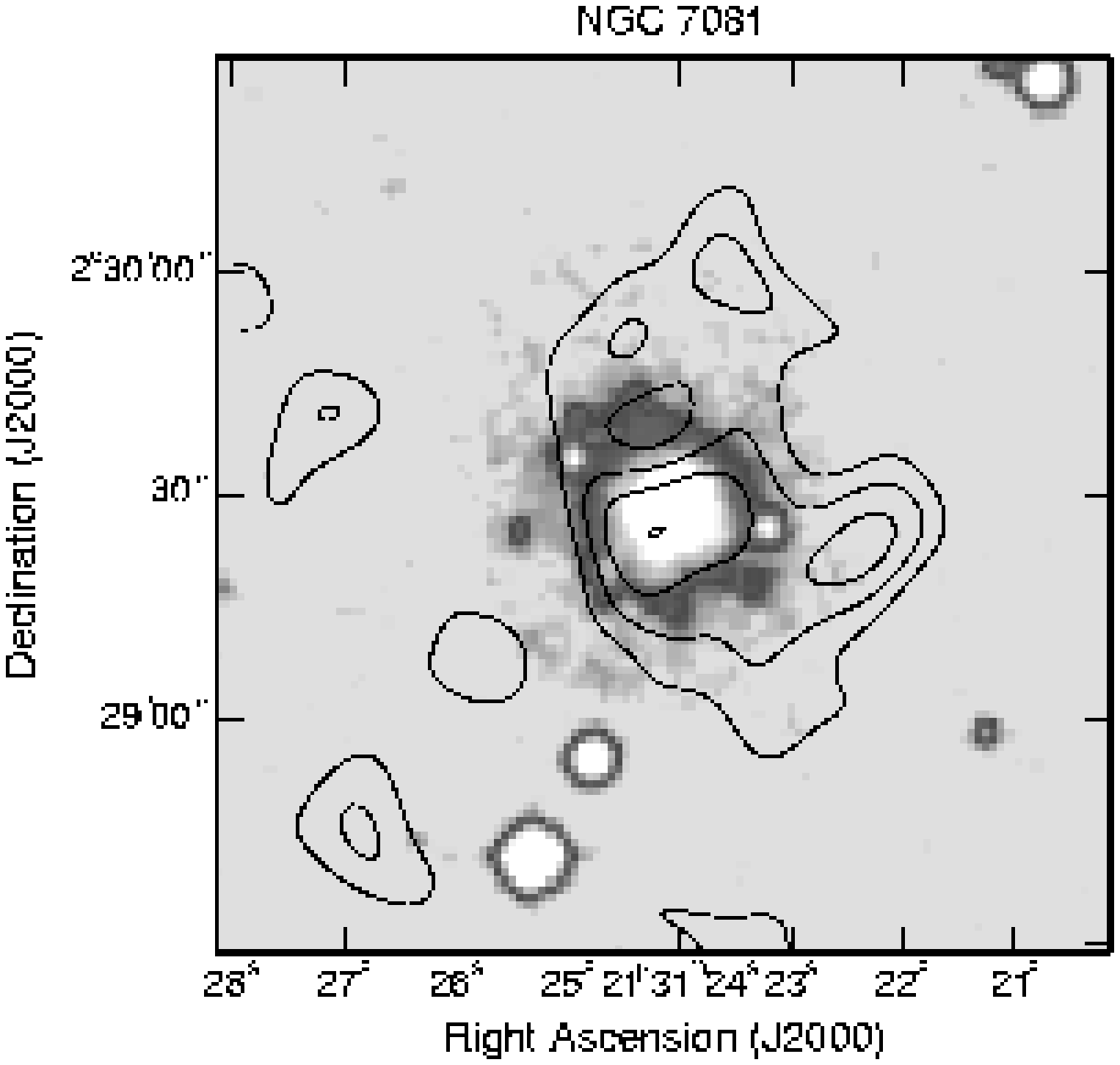}
 \hfill
 \includegraphics[angle=0, width=8cm]{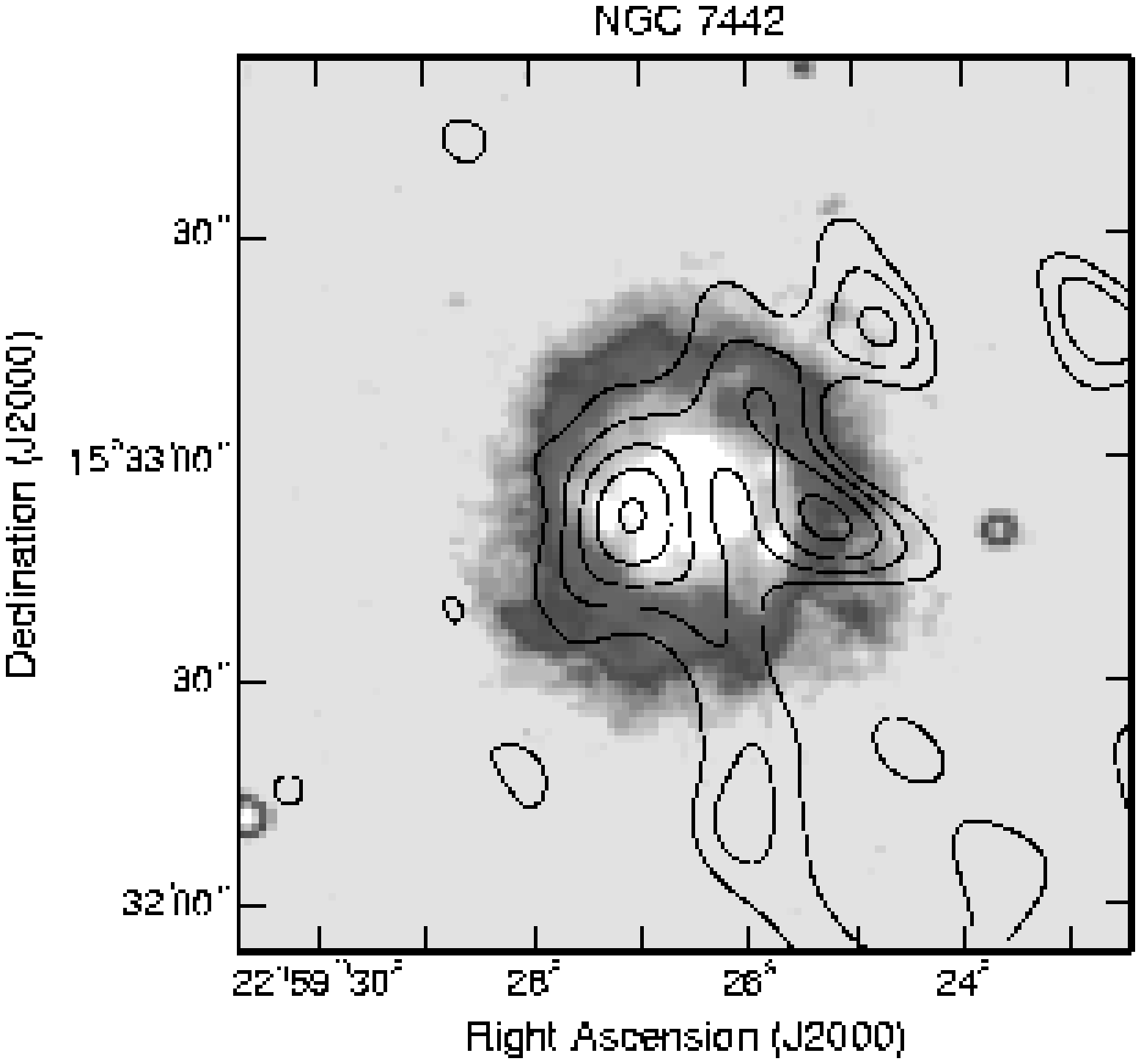}
 \hfill
 \includegraphics[angle=0, width=8cm]{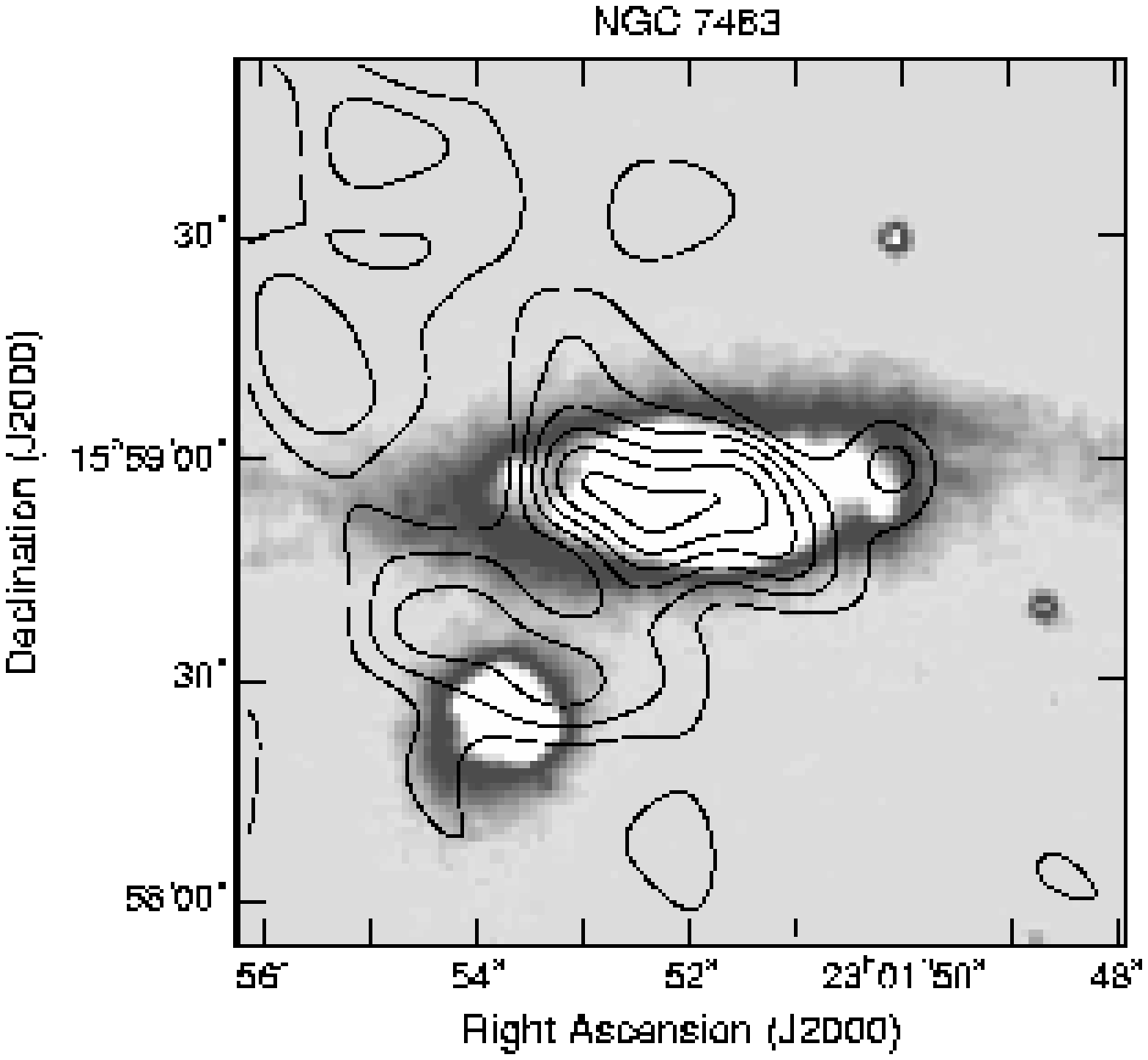}
 \hfill
 \includegraphics[angle=0, width=8cm]{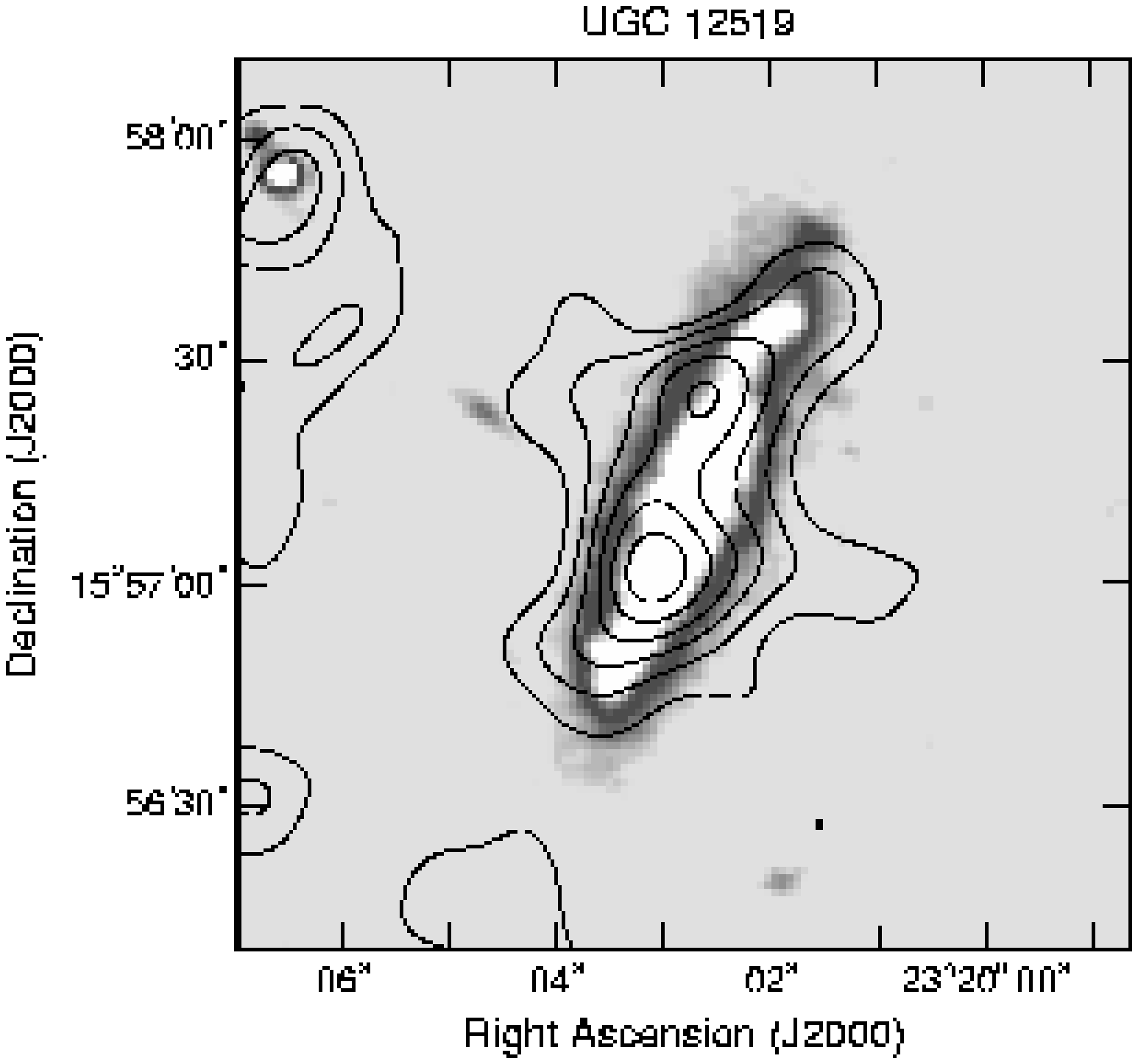}
 \vfill
 \contcaption{}
 \end{center}
\end{figure*} 

\begin{figure*}
 \begin{center}
 \includegraphics[angle=0, width=8cm]{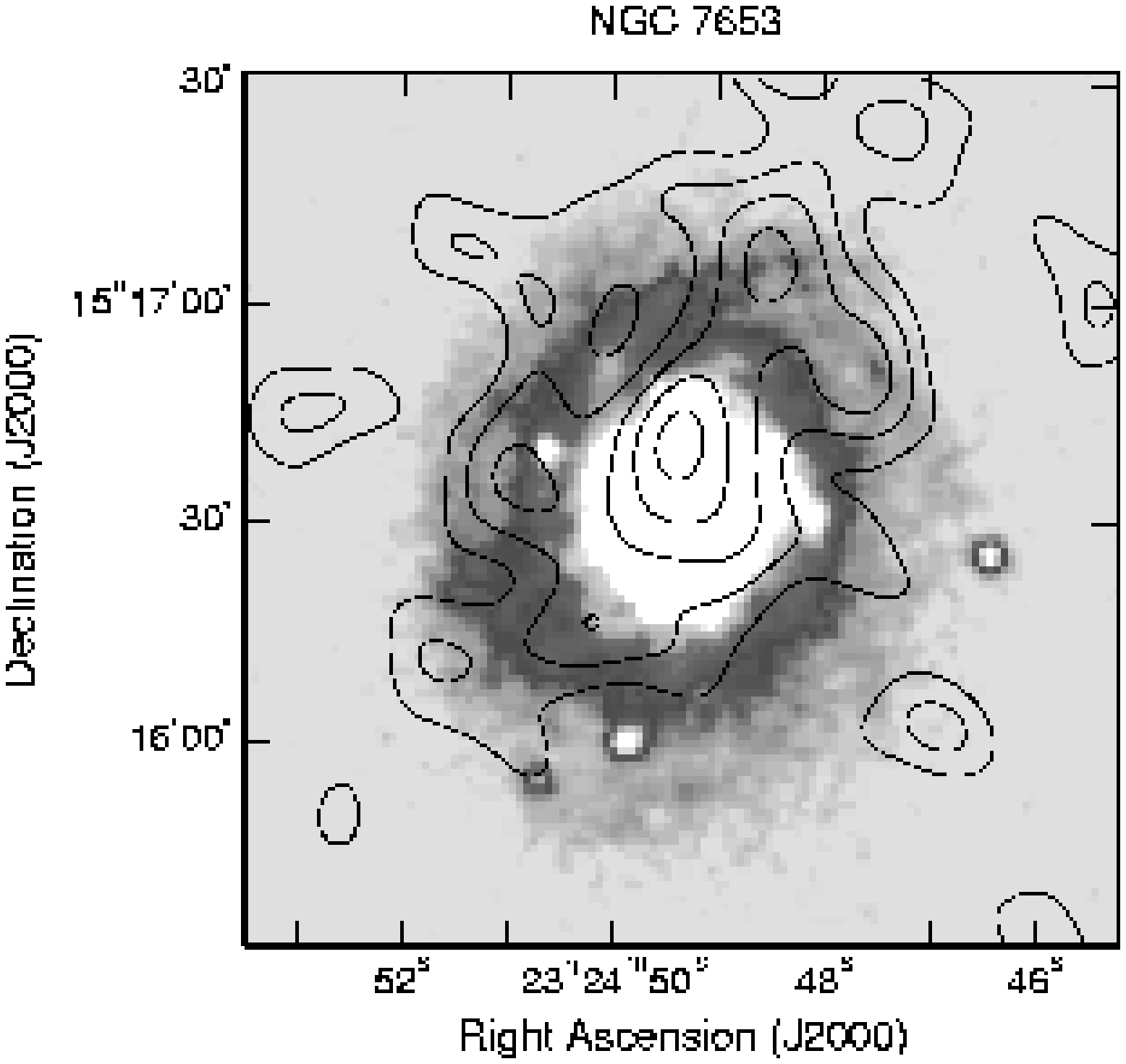}
 \hfill
 \includegraphics[angle=0, width=8cm]{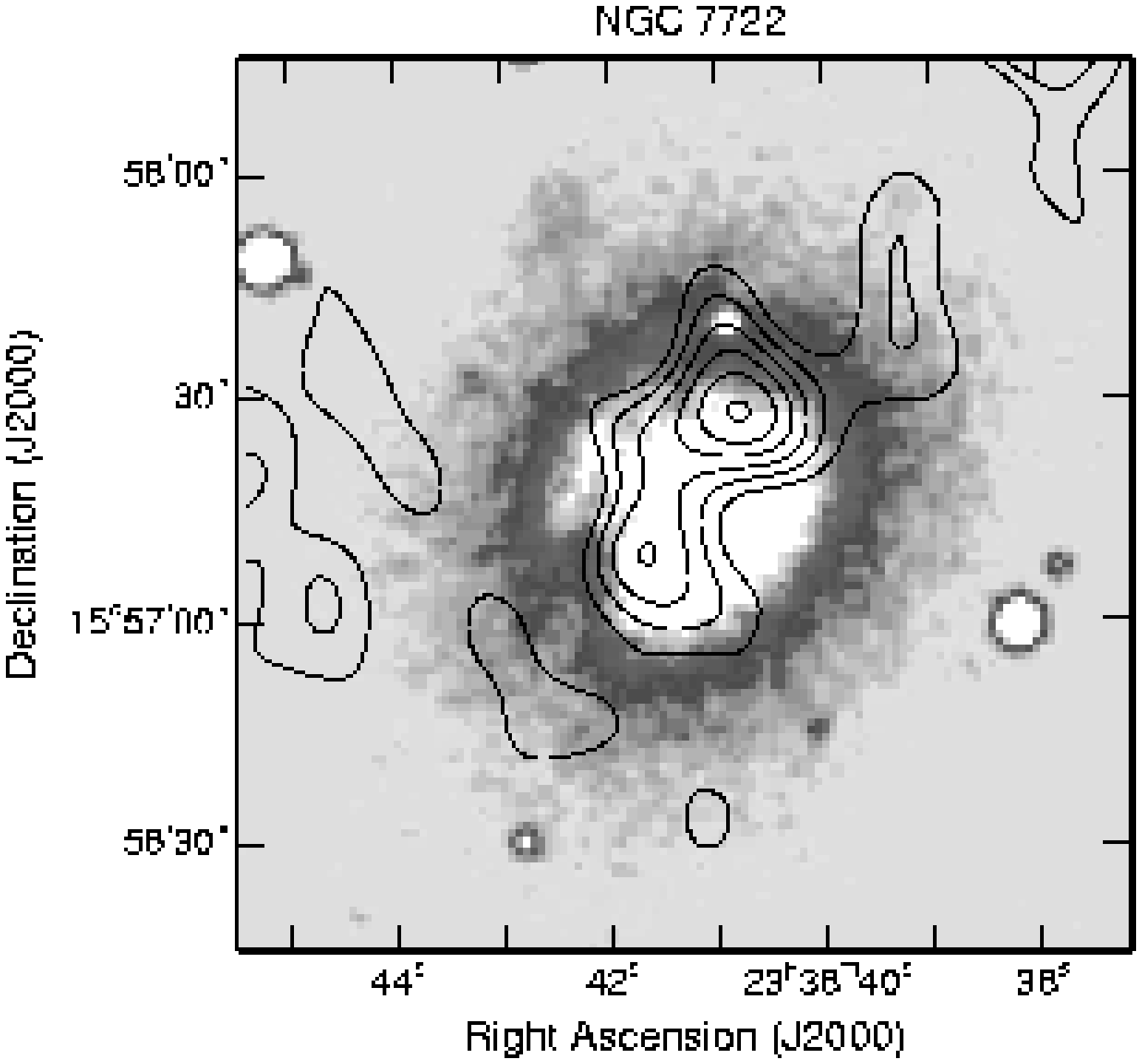}
 \hfill
 \contcaption{}
 \end{center}
\end{figure*}

\subsection{Notes on individual objects}
\label{maps}

In the following discussion of individual objects we note that since the number of bolometers sampling each sky point decreases towards the edges of the  submillimetre maps the noise increases towards the edge of the maps. Although the S/N maps in Figure~\ref{egmaps} were produced using artificial noisemaps (Section~\ref{snmaps}) which should normally account for this effect there are certain circumstances, such as a `tilted-sky' (see Section~\ref{red}) or the very noisiest bolometers, where residual `noisy' features may remain in the S/N maps. This means that any submillimetre emission in Figure~\ref{egmaps} seen beyond the main optical extent (and away from the centre of the map) should be regarded with some caution. However, in order to aid distinction between probable residual noisy features in the S/N map and potential extended submillimetre emission we have made a thorough investigation of each individual map. In the following discussion, unless otherwise stated we have found all \mbox{$\ge$\,2$\sigma$} submillimetre peaks away from the main optical galaxy to be associated with noisy bolometers or a tilted sky.

\textbf{UGC 148}. Data points for this object are not well-fitted by the two-component dust model (Section~\ref{sed-fits}), probably due to the 850\mic flux having been underestimated -- the 850\mic S/N contours shown in Figure~\ref{egmaps} for this object show evidence of a residual tilted sky plane (the sky is more positive on one side of the map than the other), suggesting that sky removal techniques may have been inadequate in this case and that therefore the source flux may have been under- (or over-) estimated. Also, since the E-NE part of the galaxy is coincident with noisy bolometers in the 850\mic map the flux-measurement aperture was drawn to avoid this region, so the 850\mic flux may be underestimated. However, an additional data point at 170\mic (ISO) is available from the literature (Stickel et al. 2000, 2004). Using the 60, 100, 170 and 450\mic fluxes we find that the data points are well-fitted by the two-component dust model (we take an average of all 170\mic fluxes available, see Section~\ref{sed-fits}), and these results are listed in Table~\ref{450tab} and the SED is shown in Figure~\ref{2compSEDfig}.

\textbf{NGC 99}. The submillimetre emission follows the spiral arm structure. The 2$\sigma$ peak to the NE of the galaxy is not associated with any noisy bolometers.

\textbf{NGC 786}. The submillimetre emission to the SE of the galaxy is not associated with any noisy bolometers but we note that this object was observed in \textit{very} poor weather.

\textbf{NGC 803}. None of the submillimetre peaks are associated with noisy bolometers. Due to the high ratio of $S_{25}/S_{60}$ good two-component SED fits (Section~\ref{sed-fits}) to the 4 data points (60, 100, 450 and 850\,\micron) can be achieved for two quite different values of the warm component temperature. In addition to the parameters listed in Table~\ref{450tab} a good fit is also found with the following parameters: $T_{w}$=60\,K, $T_{c}$=19\,K, \mbox{$\frac{N_{c}}{N_{w}}$=2597}, \mbox{log $M_{d2}$=6.99} and \mbox{log $L_{fir}$=9.48}.
\\[-2.2ex]

\textbf{UGC 5342}. This observation had a tilted sky.

\textbf{NGC 3270}. All the submillimetre emission away from the main optical extent, and the emission to the N of the galaxy, is associated with noisy bolometers. This observation also suffered from a tilted sky, which may explain much of the submillimetre emission in the N part of the map. However, the emission towards the centre of the map, coinciding with the main optical galaxy, is not associated with any noisy bolometers. This emission seems to occur where the galaxy bulge ends and the inter-arm region begins, as was found for NGC 3627 by Sievers et al. (1994).

\textbf{NGC 3689}. Very few scans were available via SCANPI, so $\IRAS$ fluxes (and corresponding fitted parameters) for this object should be used with caution.

\textbf{PGC 35952}. The submillimetre emission to the S and SW is not associated with any noisy bolometers. The submillimetre peaks coincident with the main optical extent of the galaxy appear to follow the spiral arm structure, as for NGC 6131, and it is possible the extended submillimetre emission to the S relates to the very extended faint spiral arms seen in the optical.

\textbf{NGC 3799/3800}. NGC 3799 and NGC 3800 were observed in separate maps. NGC 3799 individually is not detected at the 3$\sigma$ level (although we measure flux at the 2$\sigma$ level). For NGC 3800 (shown in Figure~\ref{egmaps}) most of the integrations for the bolometers to the S of the map were unusable, and consequently this part of the map is very much noisier. Generally this observation is very noisy, and especially bad at 450\,\micron; thus no 450\mic flux is available. In fact only the main region of submillimetre emission at the centre of the map is in an area free from noisy bolometers, and it is this region over which we have measured the submillimetre flux of the galaxy.

The S$_{850}$ listed for NGC 3799/3800 in Table~\ref{pairstab} is a conservative measurement of the 850\mic emission from the system, the sum of the separately measured fluxes from each of the two component galaxies. In coadding the two maps there appears to be a `bridge' of 850\mic emission between the two galaxies, consistent with emission seen in the optical (NGC 3799 is to the SW of NGC 3800 in Figure~\ref{egmaps}). However, since this region of the map has several noisy bolometers we only measure fluxes for the main optical extent of the galaxies.

\textbf{NGC 3815}. The submillimetre emission to the NE and W of the galaxy is associated with noisy bolometers. However, the arm-like submillimetre structures seen extending from the galaxy to the N and S are \textit{not}. Both of these `arms' extend in the direction of faint optical features seen in the DSS and 2MASS images. The optical images also show evidence of extended optical emission around the main optical extent (just visible to the NE in Figure~\ref{egmaps}), but 2MASS (JHK) images show a band of emission stretching E-W between two nearby galaxies on either side of NGC 3815. It seems clear that some kind of interaction is taking place in this system, and therefore it is perhaps not unlikely that there might also be significantly extended submillimetre emission.

\textbf{NGC 3920}. The submillimetre emission to the E and S is not associated with any noisy bolometers.

\textbf{NGC 3987}. This edge-on galaxy has a prominent dust lane in the optical. Though the submillimetre emission follows the dust lane it is seen slightly offset. A similar result was found for another edge-on spiral by Stevens et al. (2005), who conclude this effect is simply an effect of the inclination of the galaxy on the sky.

\textbf{NGC 3997}. The FSC gives an upper limit \mbox{$<$3.101\,Jy} for S$_{100}$, likely due to possible source confusion with NGC 3993. The S$_{100}$ we measure with SCANPI should therefore be used with some caution.

\textbf{NGC 4005}. This object is detected at 850\mic at only the 2.5$\sigma$ level. It is unresolved by $\IRAS$ and may be confused with $\IRAS$ source IRASF11554+2524 (NGC 4000).

\textbf{UGC 7115}. With the exception of the submillimetre emission to the SE, none of the submillimetre emission in this map is associated with noisy bolometers. However, we found this SCUBA observation to have a tilted sky. We estimate that as much as 80\% of the 850\mic flux from this elliptical may be due to synchrotron contamination from a radio source associated with the galaxy (Vlahakis et al., in prep.).

\textbf{IC 800}. The submillimetre emission to the E of this galaxy corresponds to a region of the map which is only slightly noisy, making it unlikely any residual features of this noise remains in the S/N map (Section~\ref{snmaps}). However, we also note that this object was observed in poor weather.

\textbf{NGC 4712}. None of the submillimetre emission is associated with noisy bolometers, with the exception of the 2$\sigma$ peak closest to the galaxy in the arm-like structure extending to the E. However, we note that this arm-like feature, though faint, is also seen in the optical (though not reproduced in the optical image in Figure~\ref{egmaps}); it appears to originate from the main galaxy extent, where there appears to be a significant amount of dust obscuration.

\textbf{UGC 8902}. The region of submillimetre emission to the E, far SW and far S are all associated with noisy bolometers. However, the 4$\sigma$ submillimetre peak lying to the S/SE beyond the main optical extent, at a similar declination to the small galaxy to the SE, is not associated with any noisy bolometers. This submillimetre emission is consistent with the fact that the overall emission associated with the galaxy is offset to the S/SE with respect to the optical. We also note that this region in the optical contains a number of faint condensations in the direction of the small galaxy to the SE.

\textbf{IC 979}. Although this galaxy is detected with relatively low S/N at 850\mic it is also detected at 450\,\micron. We allocate a higher 450\mic calibration error (25\%) for this source, since there were no good 450\mic calibrator observations that night (calibration was achieved by taking the mean results from a number of calibrators observed that and the previous night). Note also that no two-component fit could be made to the data since the 450\mic data point is higher than the 100\mic value, possibly due to the problems with calibrating the 450\mic data but more likely due to an underestimate of the 100\mic flux. We note this object was observed in poor weather.

\textbf{UGC 9110}. There appears to be flux present at both 850\mic and 450\mic at the 2$\sigma$ level, but the maps are very noisy and several integrations unusable, most likely due to unstable and deteriorating weather conditions during the observation. This object is unresolved by \IRAS: \IRAS\/ source IRASF14119+1551 (FSC fluxes S$_{100}$=2.341 and \mbox{S$_{60}$=0.802\,Jy}) is likely a blend of UGC9110 and companion CGCG103-124 (Condon, Cotton \& Broderick 2002).

\textbf{NGC 5522}. The 850\mic emission to the SE of NGC 5522 is associated with a region of the map which is slightly noisy and where there are a number of spikes in the data. We note that this observation was carried out in poor weather.

\begin{figure*}
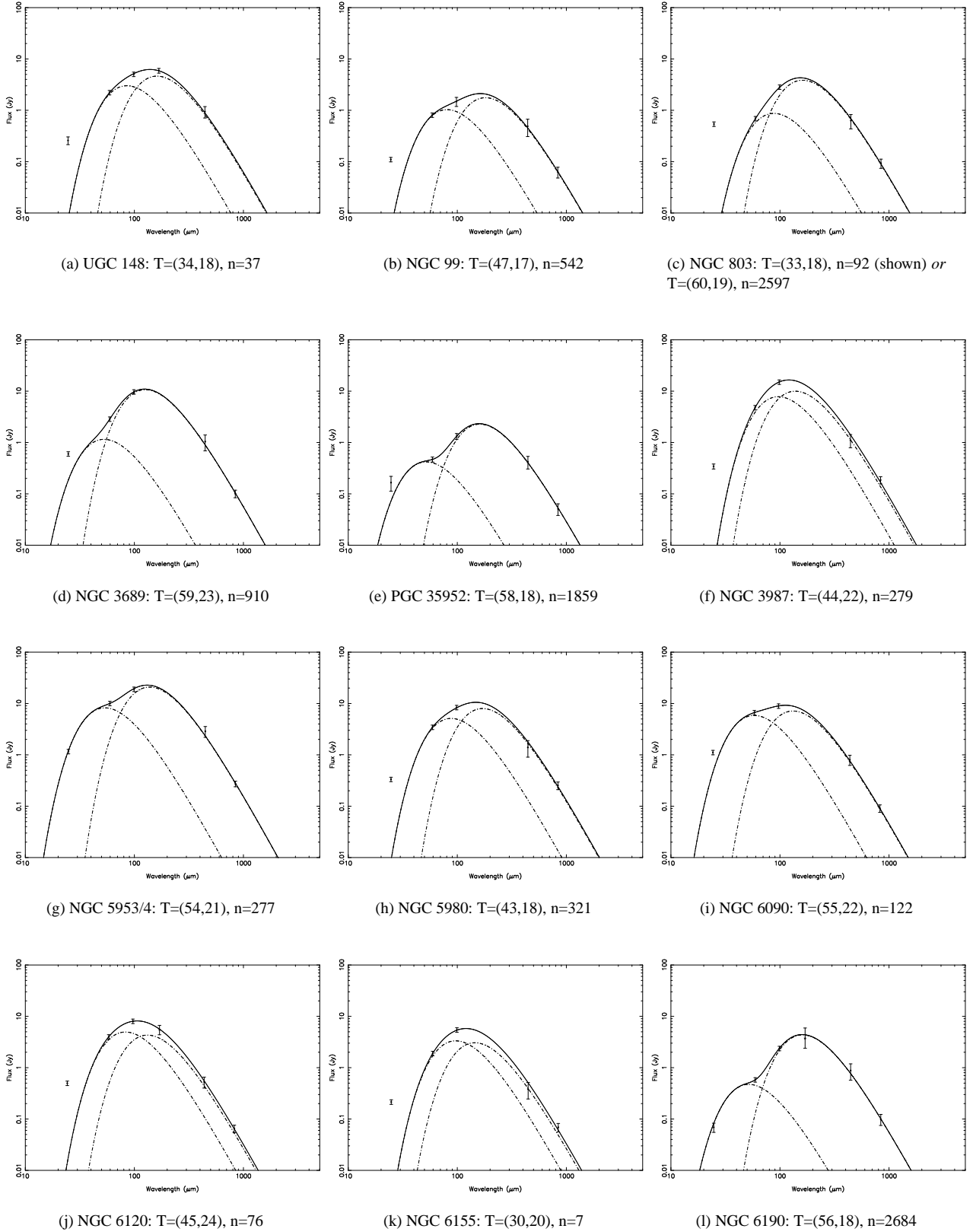

 \begin{center}
 \subfigure[UGC 148: T=(34,18), n=37]{
 \includegraphics[angle=270, width=5.5cm]{fig2_ugc148.ps}}
 \subfigure[NGC 99: T=(47,17), n=542]{
 \includegraphics[angle=270, width=5.5cm]{fig2_ngc99.ps}}
 \subfigure[NGC 803: T=(33,18), n=92 (shown) \textit{or} T=(60,19), n=2597]{
 \includegraphics[angle=270, width=5.5cm]{fig2_ngc803.ps}}
 \subfigure[NGC 3689: T=(59,23), n=910]{
 \includegraphics[angle=270, width=5.5cm]{fig2_ngc3689.ps}}
 \subfigure[PGC 35952: T=(58,18), n=1859]{
 \includegraphics[angle=270, width=5.5cm]{fig2_pgc35952.ps}}
 \subfigure[NGC 3987: T=(44,22), n=279]{
 \includegraphics[angle=270, width=5.5cm]{fig2_ngc3987.ps}}
 \subfigure[NGC 5953/4: T=(54,21), n=277]{
 \includegraphics[angle=270, width=5.5cm]{fig2_ngc5953-4.ps}}
 \subfigure[NGC 5980: T=(43,18), n=321]{
 \includegraphics[angle=270, width=5.5cm]{fig2_ngc5980.ps}}
 \subfigure[NGC 6090: T=(55,22), n=122]{
 \includegraphics[angle=270, width=5.5cm]{fig2_ngc6090.ps}}
 \subfigure[NGC 6120: T=(45,24), n=76]{
 \includegraphics[angle=270, width=5.5cm]{fig2_ngc6120.ps}}
 \subfigure[NGC 6155: T=(30,20), n=7]{
 \includegraphics[angle=270, width=5.5cm]{fig2_ngc6155.ps}}
 \subfigure[NGC 6190: T=(56,18), n=2684]{
 \includegraphics[angle=270, width=5.5cm]{fig2_ngc6190.ps}}
 \caption{\label{2compSEDfig}{Best-fitting two-component SEDs assuming $\beta=2$, fitted to the 60, 100, 450 and 850\mic fluxes (with the exception of (a), see Section~\ref{maps}). Solid lines represent the composite two-component SED and dot-dash lines indicate the warm and cold components. Any additional 170\mic (ISO) fluxes from the literature (Stickel et al. 2000, 2004) are also plotted, though (with the exception of (a)) not fitted. Note, captions show the fitted parameters as listed in Table~\ref{450tab}, so may be the averaged values (see Section~\ref{sed-fits}).}}
 \end{center}
\end{figure*}

\begin{figure*}
 \begin{center}
 \setcounter{subfigure}{12}
 \subfigure[IC 5090: T=(52,21), n=346]{
 \includegraphics[angle=270, width=5.5cm]{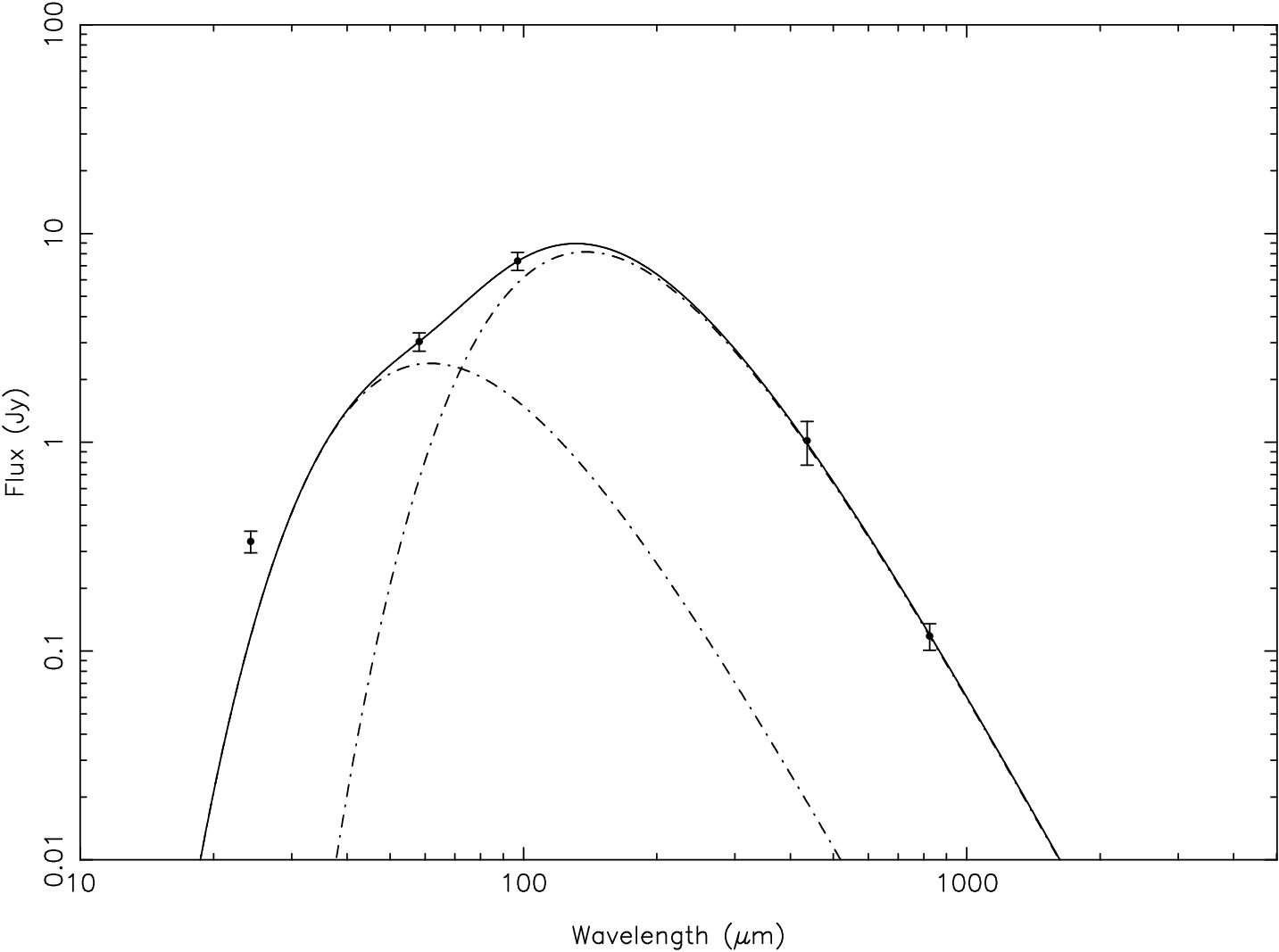}}
 \subfigure[IC 1368: T=(55,23), n=110]{
 \includegraphics[angle=270, width=5.5cm]{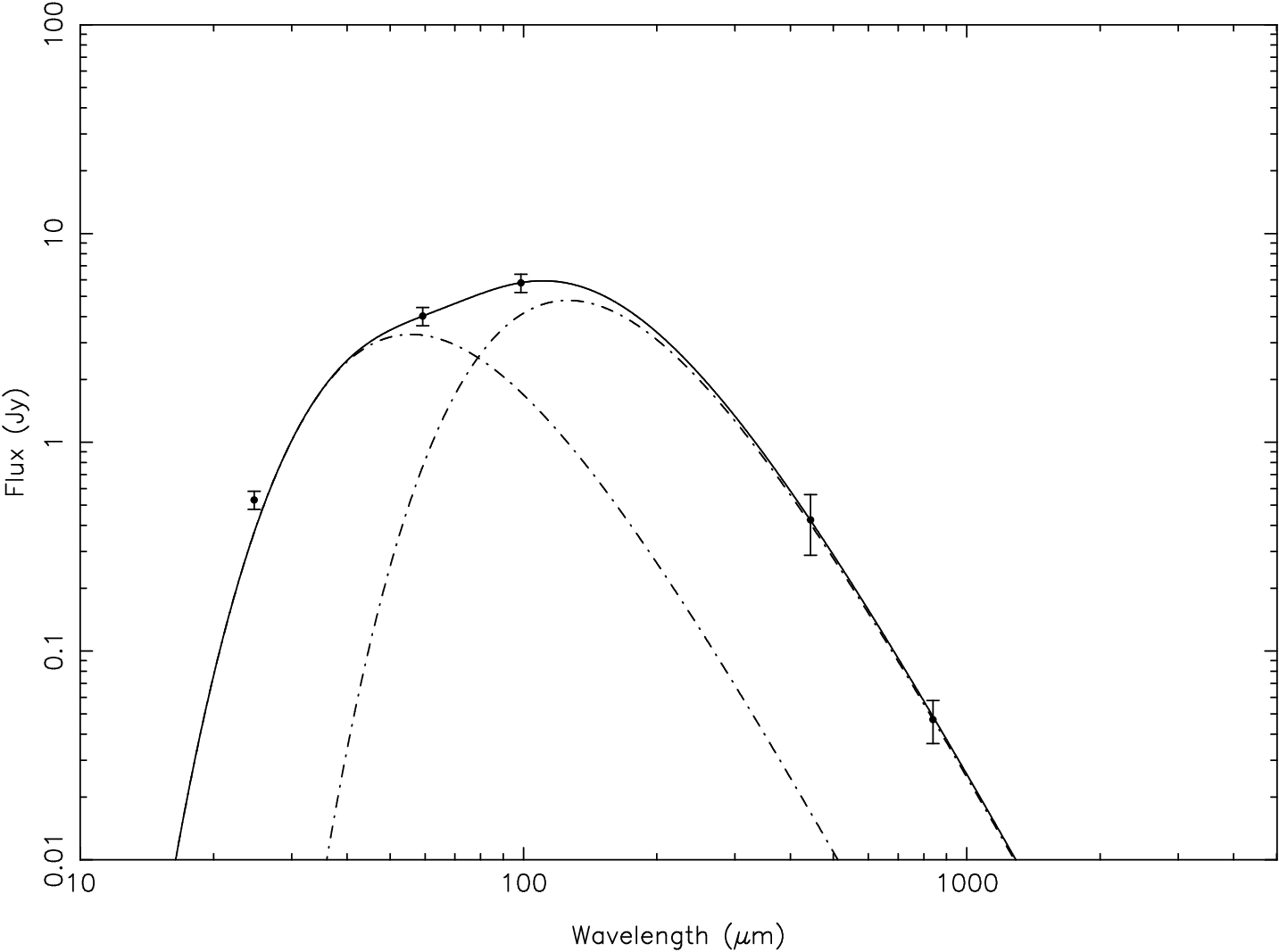}}
 \subfigure[NGC 7081: T=(32,20), n=6]{
 \includegraphics[angle=270, width=5.5cm]{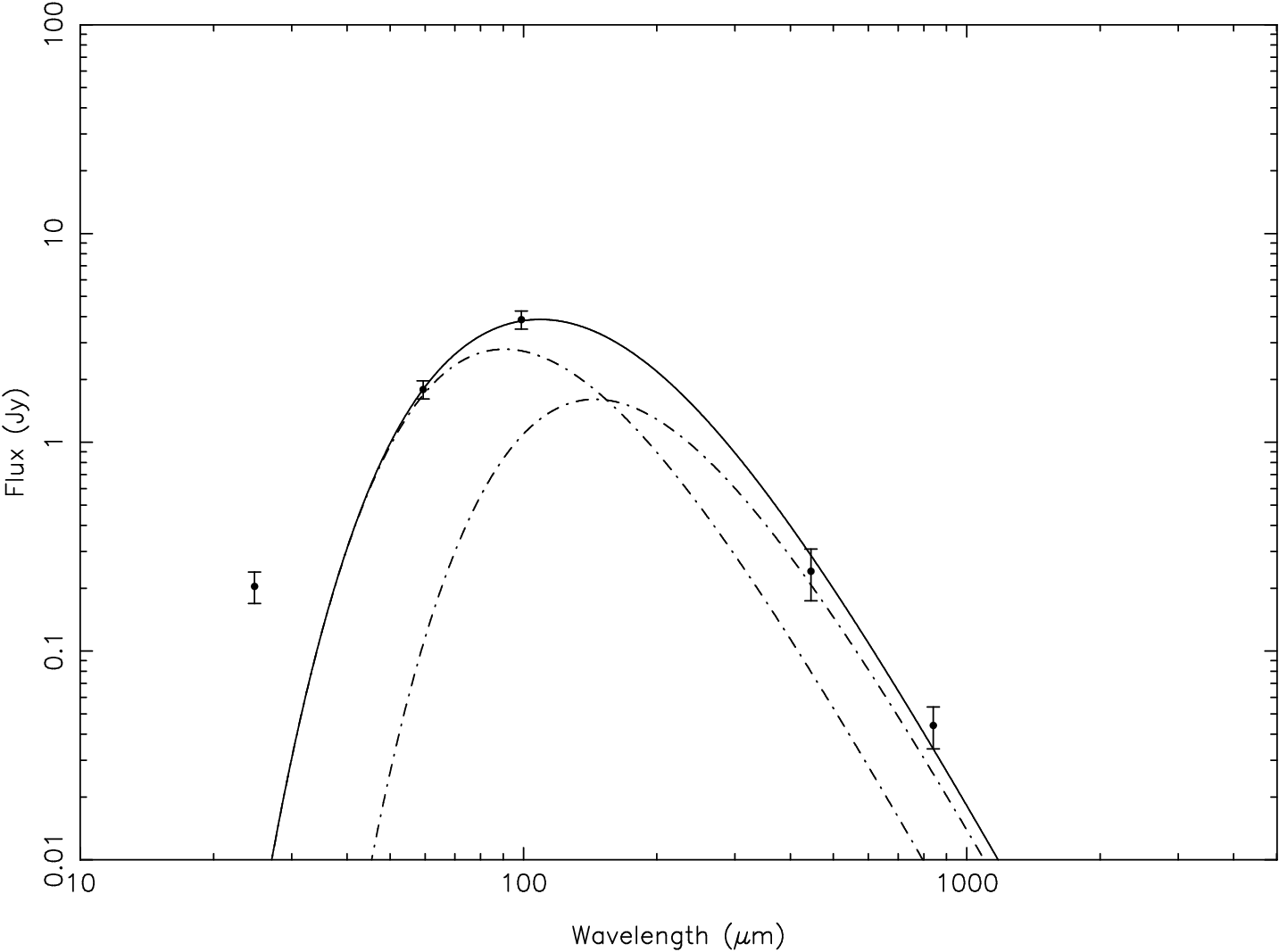}}
 \subfigure[NGC 7442: T=(54,20), n=665]{
 \includegraphics[angle=270, width=5.5cm]{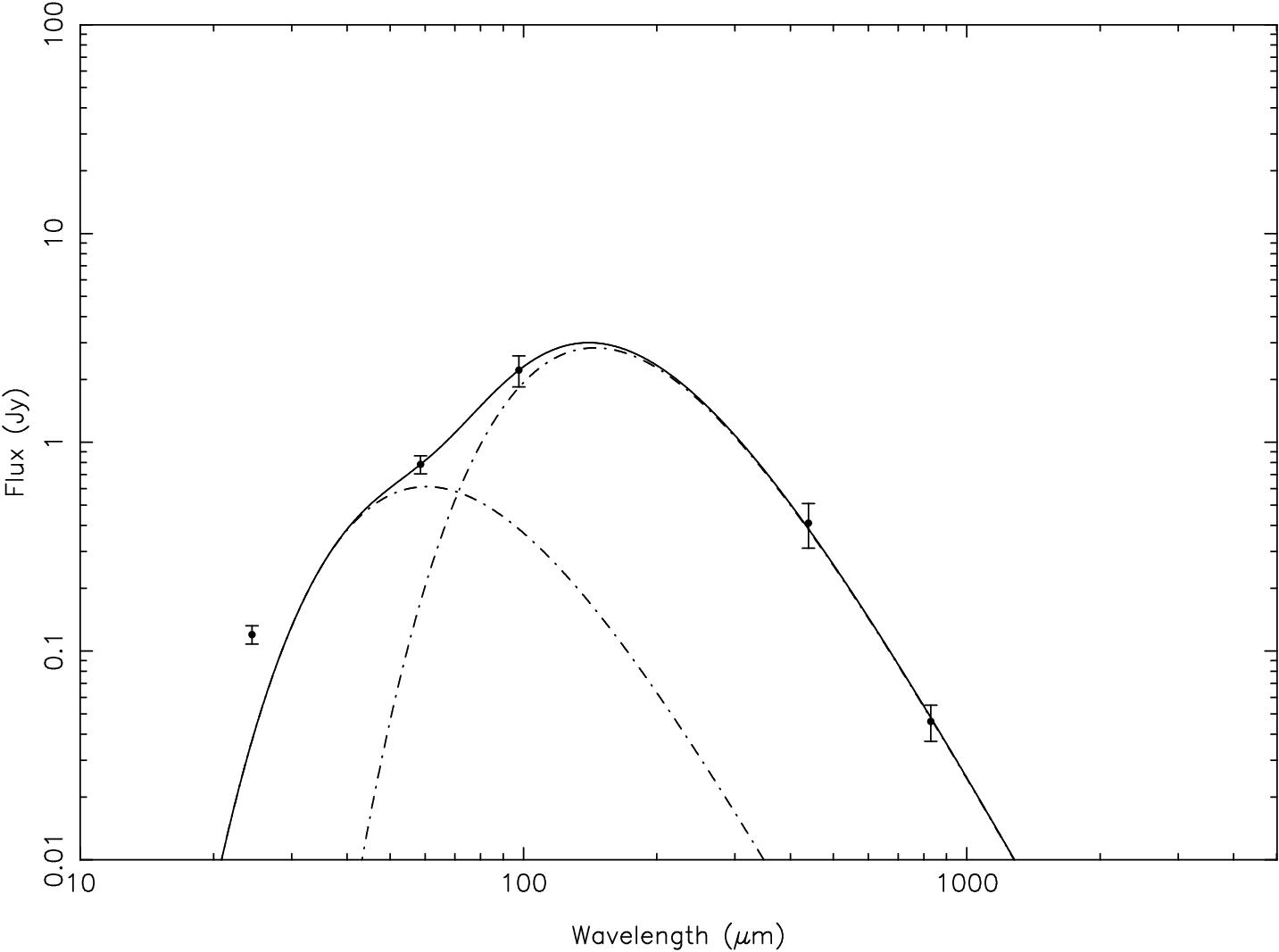}}\hfill
 \subfigure[UGC 12519: T=(28,17), n=12]{
 \includegraphics[angle=270, width=5.5cm]{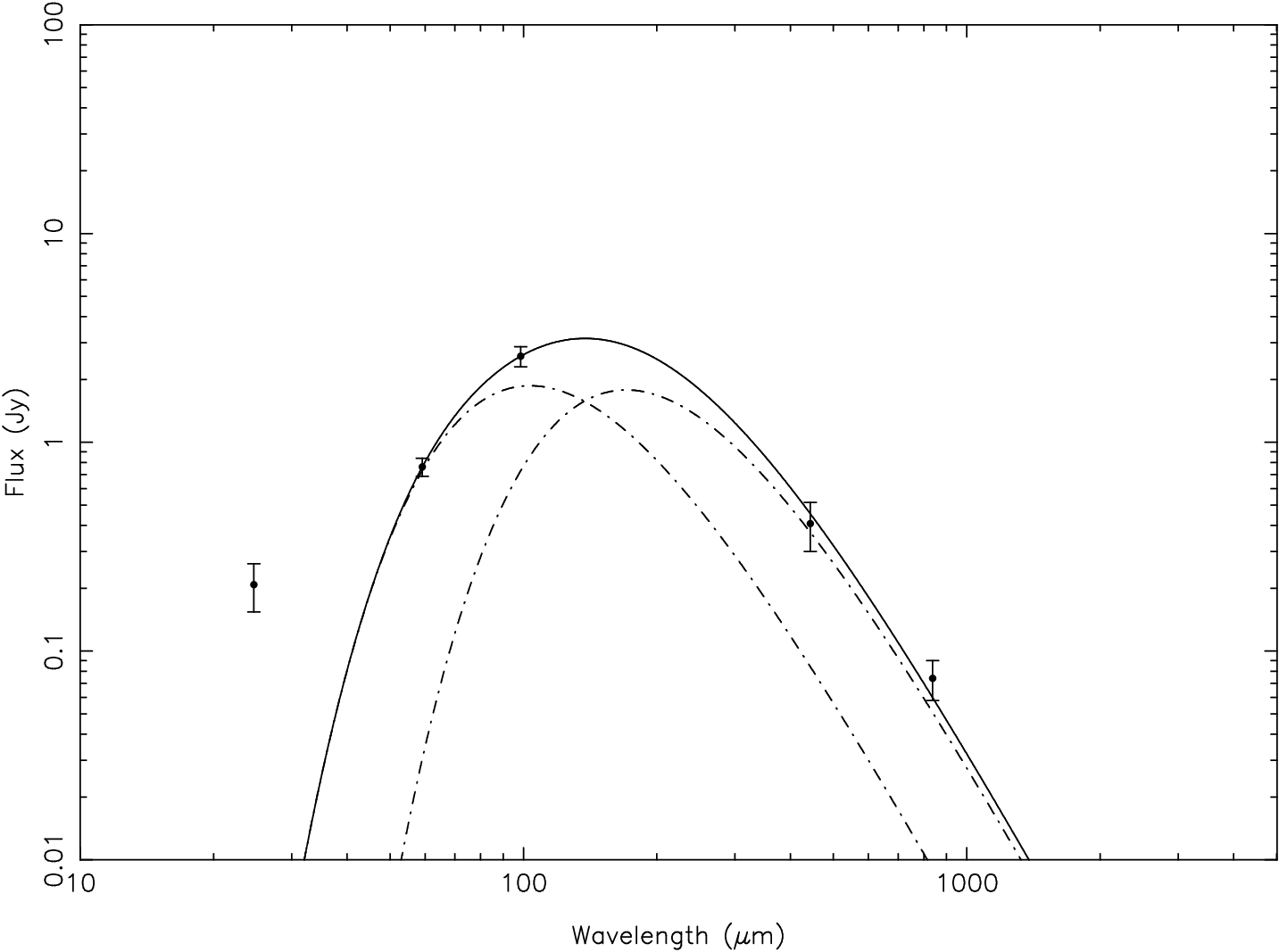}}\hfill
 \subfigure[NGC 7722: T=(54,20), n=1224]{
 \includegraphics[angle=270, width=5.5cm]{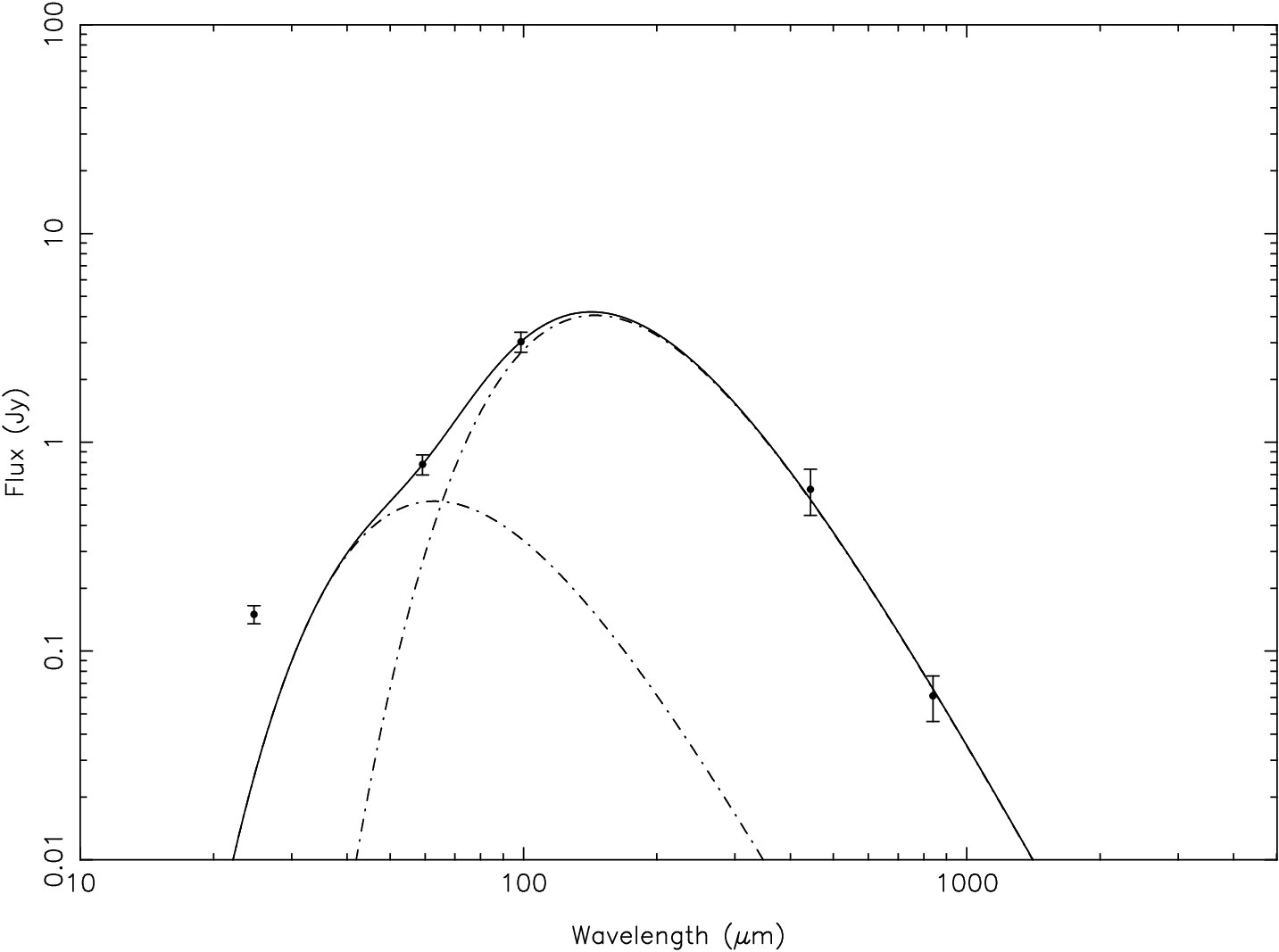}}
 \hfill
 \contcaption{}
 \end{center}
\end{figure*}

\textbf{NGC 5953/4} is also in the IRS SLUGS sample. While D00 used colour-corrected \textit{IRAS} fluxes as listed in the \textit{IRAS} BGS (from Soifer et al. (1989)) we present here, as for all the OS sample, fluxes from the \textit{IRAS} FSC.

\textbf{NGC 5980}. This observation suffered from a tilted sky, which potentially explains the extended 850\mic emission to the W of the galaxy. The aperture used to measure the 450\mic flux was smaller than at 850\mic in order to avoid a noisy bolometer, thus source flux at 450\mic may be underestimated.

\textbf{IC 1174}. This source is \textit{just} detected at 850\mic at the 3$\sigma$ level.

\textbf{UGC 10205}. The 850\mic map in Figure~\ref{egmaps} is a coadd of two observations. Since emission at the optical galaxy position is very clear in one observation (both at 850\mic \textit{and} 450\,\micron) and not in the other observation, and since we find no explanation for this, we simply coadd the two observations (Section~\ref{obs}). The submillimetre emission coincident with the main optical galaxy extent, and also the peak to the S, are not associated with any noisy bolometers. Peaks to the N and W of the galaxy lie in a region of the map which is slightly noisy. We note that this observation was carried out in less than ideal weather. 

\textbf{NGC 6090} is a closely interacting/merging pair, and is also in the IRS SLUGS sample.

\textbf{NGC 6103}. The submillimetre emission to the S of the galaxy is not associated with any noisy bolometers. This region contains a number of faint features seen in optical (DSS) and 2MASS images. We note that this object was observed in less than ideal weather.

\textbf{IC 1211}. We find in the literature no known radio sources associated with this elliptical galaxy (NVSS 1.4GHz 3$\sigma$ upper limit is \mbox{$<$1.2\,mJy}), and therefore cannot attribute the 850\mic flux detected here to contamination from synchrotron radiation.

\textbf{UGC 10325 (NED01)} is one galaxy of the pair \mbox{UGC 10325}. The SCUBA map is centred on this galaxy (NED01), but NED02 can be seen at the SE edge of the DSS image in Figure~\ref{egmaps}. Thus all fluxes given are for the individual galaxy \mbox{UGC 10325 NED01}.

\textbf{NGC 6127}. We find in the literature no known radio sources associated with this elliptical galaxy (NVSS 1.4GHz 3$\sigma$ upper limit is $<$1.2\,mJy), and therefore cannot attribute the 850\mic flux detected here to contamination from synchrotron radiation. The 4$\sigma$ submillimetre peak to the W of the galaxy, coincident with a knot in the optical, is not associated with any noisy bolometers.

\textbf{NGC 6120}. The submillimetre emission to the W of the galaxy, and at the S of the map, is associated with  noisy bolometers.

\textbf{NGC 6126}. The submillimetre source (which we measured as a point source) is offset to the S of the optical extent of the galaxy. We note that, at minimum contrast, a small satellite/companion object can be seen in this region in the DSS and 2MASS images. The 3$\sigma$ peak to the S of the map is not associated with any noisy bolometers and is coincident with a small object visible in the optical. This observation, however, was carried out in poor weather.

\textbf{NGC 6131}. The submillimetre emission to the very NW of the galaxy (beyond the main optical extent) may be associated with a noisy bolometer.

\textbf{NGC 6137}. We estimate that $\sim$\,20\% of the 850\mic flux from this elliptical galaxy could be due to synchrotron contamination from a radio source associated with the galaxy (Vlahakis et al., in prep.). Although only the submillimetre emission to the W of the galaxy coincides with a noisy bolometer we note that this observation had a tilted sky.

\textbf{NGC 6146}. We estimate that as much as 80\% of the 850\mic flux from this elliptical may be due to synchrotron contamination from a radio source associated with the galaxy (Vlahakis et al., in prep.).

\textbf{NGC 6155} The submillimetre map shows extended emission to the S and SE of the galaxy at 850\,\micron, coincident with a number of small galaxies/condensations in the optical. M\'arquez et al. (1999) find one of the spiral arms in this galaxy is directed towards the SE. None of the submillimetre peaks in this map are associated with noisy bolometers. 

A large aperture was used to measure all the flux associated with this object, and these results are listed in Table~\ref{fluxtab}. However at 450\mic any flux appears confined to the main optical extent (though the map at 450\mic is very noisy), and thus the flux measurement at 450\mic was made using a smaller aperture. Using these values of the 850\mic and 450\mic flux we found a two-component SED could not be fitted (Section~\ref{sed-fits}); the $S_{450}/S_{850}$ ratio is simply too low, most likely because we have measured extended emission at 850\,\micron. Thus we also measured the $S_{850}$ using a smaller aperture the same size as used at 450\,\micron, and find \mbox{S$_{850}=0.069\pm0.013$\,Jy}. For this smaller aperture we find that a two-component model can just be fitted to the data, and those parameters we list in Table~\ref{450tab}. 

\textbf{NGC 6166} is an elliptical and is located in a very busy field -- it is the dominant galaxy in the cluster Abell 2199. The presence of dust lanes is well documented in the literature. We note that our SCANPI measurements are in good agreement with those of Knapp et al. (1989). Using all available radio fluxes from the literature we estimate that as little as 4\% or as much as 100\% of the 850\mic flux from this elliptical may be due to synchrotron contamination from a radio source associated with the galaxy (depending whether a spectral index is assumed constant over the whole galaxy or whether it is assumed to have a flatter core) -- this is a preliminary analysis and discussion of this and the other five ellipticals detected in the OS sample will be the subject of a separate paper (Vlahakis et al., in prep.).

\textbf{NGC 6173}. We measure \mbox{S$_{100}$=0.20\,Jy} with SCANPI but the detection is unconvincing since the coadds do not agree. Therefore we give an upper limit at 100\mic in Table~\ref{fluxtab}.

\textbf{NGC 6190}. Some of the data for this object was very noisy and unusable. Consequently the remaining data may not be reliable. The submillimetre emission to the W of the galaxy lies in a region where there is a noisy bolometer in the 850\mic flux map. Thus the apertures used also unavoidably encompass some noisy bolometers, particularly at 450\,\micron, and results for this object should be used with caution. However, the rest of the 850\mic emission in the map is not associated with any noisy bolometers, so while the flux measurements may be unreliable this does not apply to the emission extent, which appears to follow the outer spiral arm structure.

\textbf{NGC 7081}. The submillimetre emission to the E and W of this galaxy are not associated with any noisy bolometers. The emission to the N and SE is coincident with regions of the map which are only slightly noisy, and since this observation was carried out in very good weather it is unlikely that any residual features of this noise remain in the S/N map (Section~\ref{snmaps}). Though from the optical DSS image only the central region of the galaxy (coincident with the main submillimetre peak) is clearly visible, there is evidence that this spiral has a very faint extended spiral arm structure. This is confirmed by optical images from SuperCOSMOS which clearly show very knotty and irregular faint spiral arms coincident with the peaks of submillimetre emission to the N and W of the galaxy.

\textbf{NGC 7442}. The 3$\sigma$ submillimetre peak to NW of main optical extent is coincident with faint optical knots and (unlike the 2$\sigma$ peaks elsewhere in the submillimetre map) is not associated with any noisy bolometers.

\textbf{NGC 7463}. This galaxy is part of a triple system with NGC 7464 (to the S of NGC 7463) and NGC 7465 (not shown in Figure~\ref{egmaps}). At 850\mic we clearly detect emission from both NGC 7463 and NGC 7464, which seem to be joined by a bridge of submillimetre emission. The flux listed in Table~\ref{fluxtab} is for NGC 7463 alone, measured in an aperture corresponding to its main optical extent. Unfortunately a very noisy bolometer to the SE prevents us measuring the flux from the eastern half of NGC 7464, but excluding this region we measure a flux for the pair of \mbox{0.051$\pm$0.012\,Jy}, though obviously a lower limit.

An $\IRAS$ source is associated with NGC 7465, which is resolved from the other members of the system at 60\mic (HIRES; Aumann, Fowler \& Melnyk 1990). Dust properties of this system (using SCUBA data observed as part of SLUGS) are studied in detail by Thomas et al. (2002).

\textbf{UGC 12519}. The 850\mic emission to the NE of this galaxy is coincident with a number of small objects seen in the optical and is not associated with any noisy bolometers. Although UGC 12519 is also detected at 450\mic the slightly smaller field of view of the short array means these NE objects lie just outside the 450\mic map.

\textbf{NGC 7722}. Along with a very high ratio of cold-to-warm dust this object also has a very prominent dust lane, extending over most of the NE `half' of the galaxy. The 850\mic emission clearly follows the dust lane evident in the optical. The 2$\sigma$ submillimetre peak to the NW of the galaxy is  not associated with any noisy bolometers.

\subsection{Spectral fits}
\label{sed-fits}
In this section we describe the dust models we fit to the spectral energy distributions (SEDs) of the OS sample galaxies and present the results of these fits. Comparison of the results of the OS and IRS samples is discussed in Section~\ref{properties}.

D00 found that for the IRS sample the 60\,\micron, 100\mic and 850\mic fluxes could be fitted by a single-temperature dust model. However, with the addition of the 450\mic data (DE01) they found that a single dust emission temperature no longer gives an adequate fit to the data, and that in fact two temperature components are needed, in line with the paradigm that there are two main dust components in galaxies (Section~\ref{cold-dust}). For the OS galaxies we have fitted a two-component dust model where there is a 450\mic flux available. Since only 19 of the galaxies have 450\mic flux densities we have also fitted an isothermal model to the data for all the galaxies in the OS sample which were detected at 850\,\micron.

We fitted two-component dust SEDs to the 60\,\micron, 100\mic (\textit{IRAS}), 450\mic \& 850\mic (SCUBA) fluxes, by minimising the sum of the $\chi^2$ residuals. This two-component model expresses the emission at a particular frequency as the sum of two modified Planck functions (`grey-bodies'), each having a different characteristic temperature, such that
\begin{equation} \label{eq:2comp}
S_{\nu}=N_{w} \times \nu^{\beta}B(\nu,T_{w})+ N_{c} \times \nu^{\beta}B(\nu,T_{c})
\end{equation}
for the optically thin regime. Here $N_{w}$ and $N_{c}$ represent the relative masses in the warm and cold components, $T_{w}$ and $T_{c}$ the temperatures, $B(\nu,T)$ the Planck function, and $\beta$ the dust emissivity index. DE01 used the high value for the ratio of $S_{450}/S_{850}$ and the tight correlation between $S_{60}/S_{450}$ and  $S_{60}/S_{850}$ for the IRS galaxies to argue that $\beta\approx2$. The OS galaxies follow a similar tight correlation (Section~\ref{prop:ir-opt}). For the OS sample we find the mean $S_{450}/S_{850}$=8.6 with $\sigma$=3.3, which is slightly higher than found for the IRS sample (where $S_{450}/S_{850}$=7.9 with $\sigma$=1.6) and with a slightly less tight distribution (though still consistent with being produced by the uncertainties in the fluxes). Both the OS and IRS values are somewhat higher than that found for the Stevens et al. (2005) sample of 14 local spiral galaxies, where the mean $S_{450}/S_{850}$=5.9 and $\sigma$=1.0. Since the OS galaxies have a similar high value for this ratio to the IRS galaxies we follow DE01 in assuming $\beta$=2.

We constrained $T_{w}$ by the \textit{IRAS} 25\mic flux (the fit was not allowed to exceed this value), though we did not actually fit this data point, and we allowed $T_{c}$ to take any value lower than $T_{w}$. This method is the same as that used in DE01 for the IRS sample, but while many of the IRS galaxies with 450\mic data also had fluxes at several other wavelengths in the literature we note that for the OS sample galaxies we have only four data points to fit. Since this is not enough data points to provide a well-constrained fit the values of $\chi^{2}_{min}$ may be unrealistically low. In Table~\ref{450tab} we list the parameters producing the best fits or, where more than one set of parameters produces an acceptable fit, we list an average of all those parameters. In practice we find that it is only $T_{w}$ (and hence $N_{c}$/$N_{w}$) for which there is sometimes a fairly large range of acceptable values, and this is likely due to our not fitting any data points below 60\,\micron. We show all our fitted two-component SEDs in Figure~\ref{2compSEDfig} (for the best-fitting, not averaged, parameters). Any additional 170\mic (ISO) fluxes available from the literature (Stickel et al. 2000, 2004) are also plotted in Figure~\ref{2compSEDfig}, though (with the exception of UGC 148 (see Section~\ref{maps})) \textit{not} fitted; where there are several 170\mic measurements available we plot the mean value.

\begin{figure}
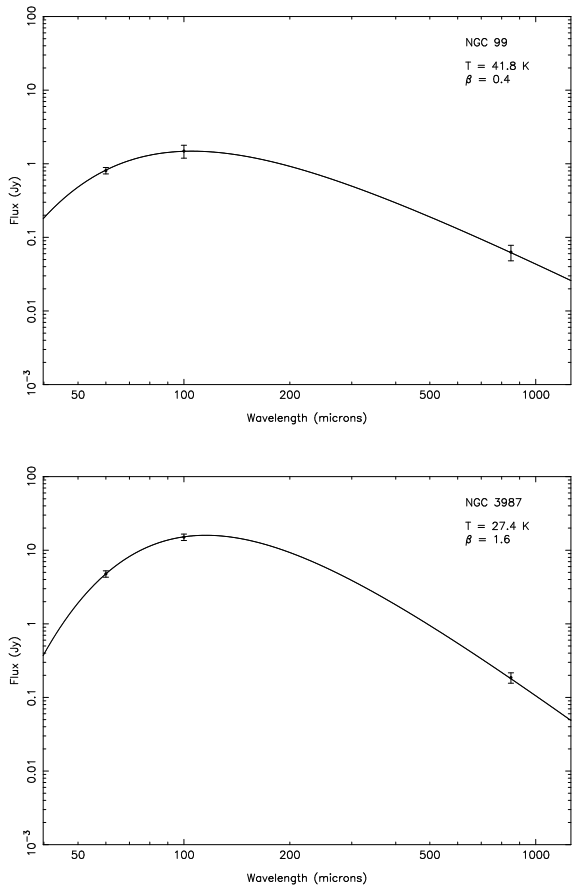

 \begin{center}
 \includegraphics[angle=270, width=7.5cm]{fig3_ngc99.ps}\\[4ex]
 \vfill
 \includegraphics[angle=270, width=7.5cm]{fig3_ngc3987.ps}
 \vfill
 \caption{\label{SEDfig}{Two representative isothermal SEDs.}}
 \end{center}
\end{figure}

We find a mean warm component temperature \mbox{$T_{w}=47.4\pm2.4$\,K} and a mean cold component temperature \mbox{$T_{c}=20.2\pm0.5$\,K}. The fitted warm component temperatures are in the range \mbox{$28<T_{w}<59$\,K} and cold component temperatures are in the range \mbox{$17<T_{c}<24$\,K}. Thus, the cold component temperature is close to that expected for dust heated by the general ISRF (Cox et al. 1986), one of the two components in the current paradigm (Section~\ref{cold-dust}).

We find a mean \mbox{$N_{c}/N_{w}=532\pm172$} (or higher if we include the higher value found for NGC 803 (see notes to Table~\ref{450tab} and Section~\ref{maps})). For the IRS sample DE01 found a large variation in the relative contribution of the cold component to the SEDs (described by the parameter $N_{c}$/$N_{w}$); for the OS sample we find an even larger variation. Objects NGC 6190 and PGC 35952 in Figure~\ref{2compSEDfig}, for example, clearly exhibit very `cold' SEDs with a strikingly prominent cold component (with \mbox{$\approx$\,2000} times as much cold dust as warm dust). Comparison of the two samples is discussed in detail in Section~\ref{properties}.

\begin{table*}
\centering
\begin{minipage}{12cm}
\caption{\label{lumtab}\small{Luminosities and masses.}}
\begin{tabular}{lrrrrrr}
\hline
 (1) & (2) & (3) & (4) & (5) & (6) & (7)\\
Name & log $L_{60}$ & log $L_{850}$ & log $L_{fir}$ & log $M_{d}$ & log $M_{HI}$ & log $L_{B}$ \\
& (W\,Hz$^{-1}$sr$^{-1}$) & (W\,Hz$^{-1}$sr$^{-1}$) & ($L_{\odot}$) & ($M_{\odot}$) & ($M_{\odot}$) & ($L_{\odot}$)\\
\hline
UGC 148 & 22.80 & 21.20         & 10.22 & 7.05     & 9.82 & 10.39 \\
NGC 99 & 22.57 & 21.46          & 9.94 & 7.17      & 10.29 & 10.37 \\
PGC 3563 & 22.24 & 21.13        & 9.76 & 6.99    & ... & 10.03 \\
NGC 786 & 22.56 & 21.34         & 9.99 & 7.14     & ... &  9.94 \\
NGC 803 & 21.69 & 20.82         & 9.37 & 6.76     & 9.78 & 10.13 \\
UGC 5129 & 21.86 & $<$20.96     & $<$9.46 & $<$7.09 & 9.34 & 10.01 \\
NGC 2954 &  $<$21.60 & $<$20.81 & $<$9.23 & ... & 8.09 & 10.25 \\
UGC 5342 & 22.46 & 21.03        & 9.84 & 6.81    & ... & 10.40 \\
PGC 29536 & $<$22.40 & $<$21.76 & $<$9.94 & ... & ... & 10.74 \\
NGC 3209 & $<$21.99 & $<$21.13 & $<$9.66 & ... & ... & 10.45 \\
NGC 3270 & 22.58 & 21.58       & 10.23 & 7.53    & 10.49 & 10.78 \\
NGC 3323 & 22.81 & 21.48       & 10.23 & 7.30    & 9.68 & 10.12 \\
NGC 3689 & $^{\displaystyle s}$22.54 & 21.09 & 10.09 & 7.04    & 9.14 & 10.29 \\
UGC 6496 &  ... &  ...         & ... &  ...  & ... & 10.10 \\
PGC 35952 & 22.08 & 21.11       & 9.61 & 6.96    & 9.73 & 10.07 \\
NGC 3799/3800$^{\scriptstyle p}$ & 22.93 & 21.38 &  10.39  & 7.27 & 9.34 & ... \\
NGC 3812 & $<$21.68 & $<$20.91 & $<$9.17 & ... & ... &  9.95 \\
NGC 3815 & 22.19 & 20.96       & 9.70 & 6.83    & 9.62 & 10.15 \\
NGC 3920 & 22.21 & 20.86       & 9.63 & 6.68    & ... &  9.87 \\
NGC 3987 & 23.20 & 21.79       & 10.74 & 7.72    & 9.75 & 10.63 \\
NGC 3997 & 22.63 & $<$20.93    & $<$9.96 & $<$7.06 & 9.83 & 10.30 \\
NGC 4005 & ... & $<$20.69 & ... & $<$6.82 & 9.22 & 10.35 \\
NGC 4015 & 21.88 & $<$21.18    & $<$9.49 & $<$7.31 & ... & ... \\
UGC 7115 & $<$22.17 & 21.59    & $<$9.80 & $\dag$7.71 & ... & 10.45 \\
UGC 7157 & $<$22.15 & $<$21.27   & $<$9.65 & ... & ... & 10.32 \\
IC 797 & 21.72 & 20.78          & 9.27 & 6.63       & 8.50 &  9.77 \\
IC 800 & 21.52 & 20.82          & 9.07 & 6.63       & 8.51 &  9.67 \\
NGC 4712 & 22.18 & 21.50       & 9.87 & 7.43    & 10.18 & 10.50 \\
PGC 47122 &  $<$21.95 & $<$21.46 & $<$9.71 & $<$7.58 & ... & 10.32 \\
MRK 1365 & 23.32 & 21.20        & 10.61 & 7.00     & 9.23 & 10.00 \\
UGC 8872 &  $<$22.04 & $<$21.02 & $<$9.44 & ... & ... & 10.29 \\
UGC 8883 & 22.36 & $<$21.31     & $<$9.86 & $<$7.44 & ... & 10.00 \\
UGC 8902 & 23.07 & 21.81        & 10.57 & 7.69    & ... & 10.80 \\
IC 979 & $^{\displaystyle s \ast}$22.27 & 21.75 & $^{\ast}$9.85 & $^{\ast}$7.56      & ... & 10.64 \\
UGC 9110 & U & $<$21.21         & ... & ... & 9.72 & 10.27 \\
NGC 5522 & 22.84 & 21.39        & 10.22 & 7.17    & 9.77 & 10.51 \\
NGC 5953/4$^{\scriptstyle p}$ & 22.80 & 21.22 & 10.17 & 7.03  & 9.32 & ... \\
NGC 5980 & 22.97 & 21.84        & 10.44 & 7.65    & ... & 10.53 \\
IC 1174 & $<$21.80 & 20.95      & $<$9.19 & $\dag$7.08    & ... & 10.18 \\
UGC 10200 & 21.95 & $<$20.10    & $<$9.21 & $<$6.22 & 9.54 &  9.05 \\
UGC 10205 & 22.44 & 21.61       & 10.10 & 7.53   & 9.57 & 10.55 \\
NGC 6090 & 23.93 & 22.06        & 11.20 & 7.78    & 8.82 & 10.73 \\
NGC 6103 & 22.97 & 21.88        & 10.45 & 7.71    & ... & 10.83 \\
NGC 6104 & 22.77 & $<$21.59     & $<$10.35 & $<$7.71 & ... & 10.62 \\
IC 1211 & $<$21.78 & 21.16      & $<$9.52 & $\dag$7.29 & ... & 10.37 \\
UGC 10325 & 22.92 & 21.34       & 10.35 & 7.20 & 9.92 & ... \\
NGC 6127 & $<$21.57 & 21.51     & $<$9.17 & $\dag$7.64 & ... & 10.66 \\
NGC 6120 & 23.74 & 21.95        & 11.11 & 7.80    & ... & 10.73 \\
NGC 6126 & $<$22.37 & 21.56     & $<$9.89 & $\dag$7.69   & ... & 10.61 \\
NGC 6131 & 22.49 & 21.36        & 10.07 & 7.27    & 9.83 & 10.37 \\
NGC 6137 & $<$22.41 & 21.62     & $<$9.94 & $\dag$7.75  & ... & 11.04 \\
NGC 6146 & $<$22.18 & 21.55     & $<$9.83 & $\dag$7.68  & ... & 11.01 \\
NGC 6154 & $<$21.95 & $<$21.37  & $<$9.44 & ...  & 9.86 & 10.38 \\
NGC 6155 & 22.25 & 21.04        & 9.78 & 6.93    & 8.95 &  9.82 \\
UGC 10407 & 23.28 & 21.48       & 10.63 & 7.32   & ... & 10.62 \\
NGC 6166 & $^{\displaystyle s}$22.14 & 22.00 & 10.03 & 7.96 & ... & 11.30 \\
NGC 6173 & $<$22.33 & $<$21.48  & $<$9.63 & ... & ... & 11.14 \\
NGC 6189 & 22.59 & 21.57        & 10.19 & 7.48    & 10.07 & 10.68 \\
NGC 6190 & 22.02 & 21.24        & 9.69 & 7.18    & 9.48 &  9.97 \\
\end{tabular} 
\end{minipage}
\end{table*}

\begin{table*}
\begin{minipage}{12cm}
\contcaption{}
\begin{tabular}{lrrrrrr}
\hline
 (1) & (2) & (3) & (4) & (5) & (6) & (7)\\
Name & log $L_{60}$ & log $L_{850}$ & log $L_{fir}$ & log $M_{d}$ & log $M_{HI}$ & log $L_{B}$ \\
& (W\,Hz$^{-1}$sr$^{-1}$) & (W\,Hz$^{-1}$sr$^{-1}$) & ($L_{\odot}$) & ($M_{\odot}$) & ($M_{\odot}$) & ($L_{\odot}$)\\
\hline
NGC 6185 & 22.47 & $<$21.72     & $<$10.06 & $<$7.85 & ... & 10.96 \\
UGC 10486 & $<$22.06 & $<$21.24 & $<$9.63 & ... & ... & 10.31 \\
NGC 6196 & $<$22.24 & $<$21.53  & $<$9.86 & ... & ... & 10.85 \\
UGC 10500 & $^{\displaystyle s \ast}$21.85 & $<$21.09 & $<$9.56 & $<$7.22 & ... & 10.20 \\
IC 5090 & 23.64 & 22.23         & 11.09 & 8.08     & ... & 10.46 \\
IC 1368 & 23.00 & 21.07         & 10.30 & 6.83     & ... & 10.14 \\
NGC 7047 & 22.35 & 21.45        & 9.98 & 7.38    & 9.08 & 10.50 \\
NGC 7081 & 22.49 & 20.88        & 9.89 & 6.72    & 9.51 &  9.75 \\
NGC 7280 & $<$20.80 & $<$20.34 & $<$8.51 & ... & 8.16 &  9.85 \\
NGC 7442 & 22.83 & 21.60       & 10.32 & 7.47    & 9.75 & 10.48 \\
NGC 7448 & 22.84 & 21.19 & 10.19 & 7.04 & 9.75 & 10.39\\
NGC 7461 & $<$21.70 & $<$20.81 & $<$9.33 & ... & ... &  9.90 \\
NGC 7463 & U & 20.60           & ... & $^{\scriptscriptstyle T}$6.73    & 9.33 & 10.12 \\
III ZW 093 & 23.26 & $<$21.99    & $<$11.06 & $<$8.12 & 9.95 & 11.60 \\
III ZW 095 & $<$21.92 & $<$21.24 & $<$9.93 & ... & ... &  9.90 \\
UGC 12519 & 22.37 & 21.36      & 9.95 & 7.26   & 9.53 & 10.27 \\
NGC 7653 & 22.59 & 21.52       & 10.17  & 7.43    & ... & 10.49 \\
NGC 7691 & 22.15 & $<$20.82    & $<$9.68 & $<$6.95 & 9.59 & 10.24 \\
NGC 7711 &  $<$21.60 & $<$20.86 & $<$9.20 & ... & ... & 10.57 \\
NGC 7722 & 22.31 & 21.20        & 9.94 & 7.15    & 9.51 & 10.48 \\
\hline
\end{tabular} \\
(1) Most commonly used name. \\
(2) 60\mic luminosity. \\
(3) 850\mic luminosity. \\
(4) FIR luminosity, calculated by integrating measured SED from 40--1000\,\micron. \\
(5) Dust mass, calculated using a single temperature, $T_{d}$, as listed in Table~\ref{fluxtab} ($T_{d}$ derived from fitted SED to the 60, 100 and 850\mic data points). Upper limits are calculated using $T_{d}$=20\,K.\\
(6) HI mass; refs.: Chamaraux, Balkowski \& Fontanelli (1987), Haynes $\&$ Giovanelli (1988, 1991), Huchtmeier $\&$ Richter (1989), Giovanelli $\&$ Haynes (1993), Lu et al. (1993), Freudling (1995), DuPrie $\&$ Schneider (1996), Huchtmeier (1997), Theureau et al. (1998),  Haynes et al. (1999). \\
(7) Blue luminosity, calculated from corrected blue magnitudes taken from the LEDA database. \\
\vspace{0.5pt}\\
$^{\scriptstyle p}$ A close or interacting pair which was resolved by SCUBA; parameters given refer to the combined system, as in Table~\ref{pairstab}.\\
$^{\ast}$ Values should be used with caution (see $^{\ast}$ notes to Table~\ref{fluxtab}).\\
$\dag$ Object was only detected at 850\mic (and not in either $\IRAS$ band), so these dust masses should be used with caution; these are all early types; dust masses calculated using $T_{d}$=20\,K. \\
$^{\scriptscriptstyle T}$ Dust mass calculated using $T_{d}$=20\,K, since no fitted value of $T_{d}$.\\
U Unresolved by \textit{IRAS}.\\
\vspace{0pt}\\
Notes on HI fluxes:-\\
NGC 7463 Giovanelli $\&$ Haynes (1993) note confused HI profile, many neighbours.\\
UGC 12519 HI flux from Giovanelli $\&$ Haynes (1993) gives HI mass of \mbox{log $M_{\odot}$=9.65}.\\ 
NGC 7691 Haynes et al. (1999) gives \mbox{log $M_{\odot}$=9.88}.\\
NGC 5953/4 HI flux from Freudling (1995); Huchtmeier $\&$ Richter (1989) give \mbox{log $M_{\odot}$} in the range \mbox{8.82 -- 9.20}.\\
NGC 3799/3800 flux for NGC\,3800 but sources may be confused (Lu et al. 1993).\\
NGC 803 Giovanelli $\&$ Haynes (1993) give \mbox{log $M_{\odot}$=9.68}, and note optical disk larger than beam.\\
NGC 6090 Huchtmeier $\&$ Richter (1989) give values up to \mbox{log $M_{\odot}$=10.24}.\\
NGC 6131 confused with neighbour.\\
NGC 6189 Haynes et al. (1999) gives \mbox{log $M_{\odot}$=10.26}.\\
NGC 7081 Huchtmeier $\&$ Richter (1989) give values up to \mbox{log $M_{\odot}$=9.69}.\\
NGC 4712 Huchtmeier $\&$ Richter (1989) give \mbox{log $M_{\odot}$=9.91}.\\
\end{minipage}
\end{table*}

\begin{figure*}
\begin{center}
\includegraphics[angle=0, width=14cm]{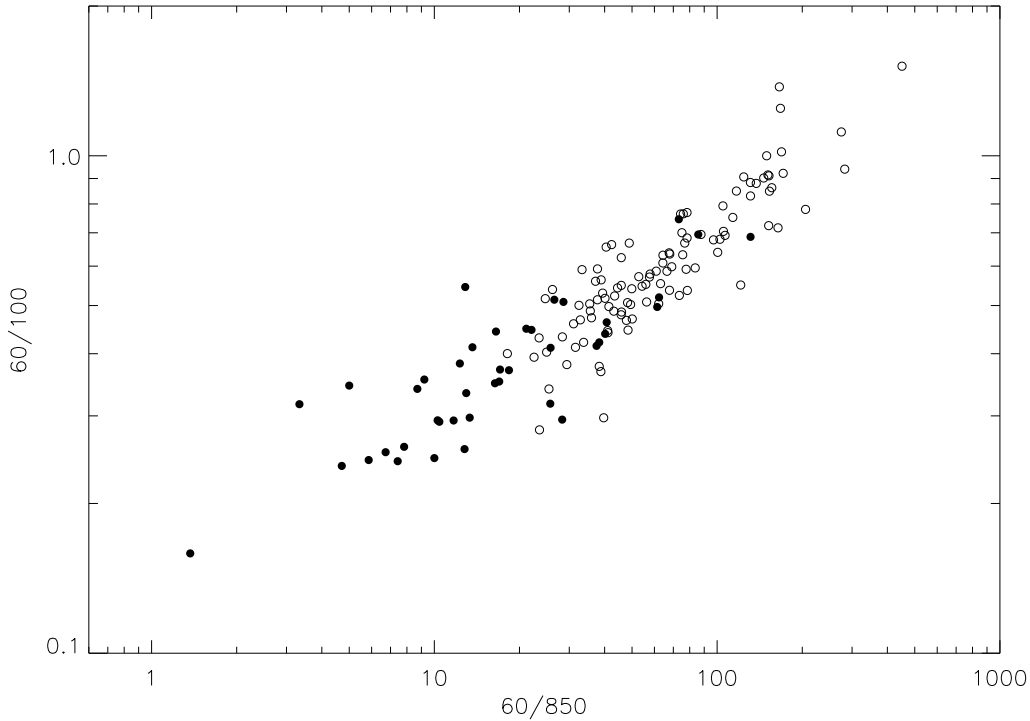}
\caption{\label{colplot}{Colour-colour plot: $S_{60}/S_{100}$ versus $S_{60}/S_{850}$ colours for the optically-selected (this work) and \textit{IRAS}-selected (D00) SLUGS (filled and open points respectively).}}
\end{center}
\end{figure*}

Since 450\mic fluxes are only available for only $\sim$ one third of the sample we have in addition fitted single-component SEDs for all sources in the OS sample. In these fits we have allowed $\beta$ to vary as well as $T_{d}$, as it is rarely possible to get an acceptable fit with $\beta$=2. The best fitting $T_{d}$ and $\beta$ are listed in Table~\ref{fluxtab}. We include fitted parameters only for those objects with detections in all 3 wavebands (60\,\micron, 100\mic and 850\,\micron). The sample mean and error in the mean for the best-fitting temperature is \mbox{$\bar{T}_{d}=31.6\pm0.6$\,K} and for the dust emissivity index $\bar{\beta}=1.11\pm0.05$. Figure~\ref{SEDfig} shows two representative isothermal SEDs. As an example of the potential dangers of fitting single-temperature SEDs, we note one of these objects (NGC 99) is also fitted with the two-component model (shown in Figure~\ref{2compSEDfig}). NGC 99 can clearly be well-fitted by both an isothermal dust model with very flat $\beta$ ($\beta$=0.4 in this case) \textit{and} a two-component model with much steeper $\beta$ ($\beta$=2); this is also the case for NGC 6190 and PGC 35952 described above. We note that the low values of $\beta$ found from the isothermal fits are not the \textit{true} values of $\beta$ but rather is evidence that galaxies across all Hubble types contain a significant proportion of dust that is colder than these fitted temperatures, and it is likely that these objects (as for NGC 99) require a two-component model to adequately describe their SED.

\subsection{Dust masses}
\label{sec:dmass}
Dust masses for the OS galaxies are calculated using the measured 850\mic fluxes and dust temperatures ($T_{d}$) from the isothermal fits (listed in Table~\ref{fluxtab}) using
\begin{equation} \label{eq:dmass}
M_{d}=\frac{S_{850}D^{2}}{\kappa_{d}(\nu)B(\nu,T_{d})}
\end{equation}
where $\kappa_{d}$ is the dust mass opacity coefficient at 850\,\micron, $B(\nu,T_{d})$ is the Planck function at 850\mic for the temperature $T_{d}$ and D is the distance.

As discussed in D00 we assume a value for $\kappa_{d}(\nu)$ of 0.077m$^{2}$kg$^{-1}$, which is consistent with the value derived by James et al. (2002) from the global properties of galaxies. Though the true value of $\kappa_{d}(\nu)$ is uncertain, as long as dust has similar properties in all galaxies then our relative dust masses will be correct. The uncertainties in the relative dust masses then depend only on errors in $S_{850}$ and $T_{d}$. 

Values for dust masses (calculated using $T_{d}$ from our isothermal fits) are given in Table~\ref{lumtab}. We find a mean dust mass \mbox{$\bar{M_{d}}=(2.34\pm0.36)\times{10^{7}}$ M$_{\odot}$} (where the $\pm$ error is the error on the mean), which is comparable to that found for the IRS sample (D00). This, together with the fact that for the OS sample we find significantly lower values of $\beta$, poses a number of issues. As shown by DE01, if more than one temperature component is present our use of a single-temperature model will have given us values of $\beta$ which are lower than actually true, biased our $T_{d}$ estimates to higher temperatures, and lead to underestimates of the dust masses. 

For those galaxies for which we have made two-component fits (Table~\ref{450tab}) we also calculate the two-component dust mass ($M_{d2}$), using
\begin{equation} \label{eq:dmass2}
M_{d2}=\frac{S_{850}D^2}{\kappa_{d}}\times\left[\frac{N_{c}}{B(\nu,T_{c})}+\frac{N_{w}}{B(\nu,T_{w})}\right]
\end{equation}
where parameters are the same as in Equation~\ref{eq:dmass} and $T_{c}$, $T_{w}$, $N_{c}$ and $N_{w}$ are the fitted two-component parameters as in Equation~\ref{eq:2comp} (and listed in Table~\ref{450tab}). The mean two-component dust mass is found to be \mbox{${\bar M_{d2}}=(4.89\pm1.20)\times{10^{7}}$ M$_{\odot}$}, and the two-component dust masses are typically a factor of 2 higher than found from fitting single-temperature SEDs, though in some cases (such as NGC 99) as much as a factor of 4 higher.          

Given the lack of CO measurements for the OS sample galaxies, one potential problem with the above estimates of dust mass is any contribution to the SCUBA 850\mic measurements by CO(3-2) line emission. Seaquist et al. (2004) find, for a representative subsample of the IRS SLUGS galaxies from D00, that contamination of 850\mic SCUBA fluxes by CO(3-2) reduces the average dust mass by \mbox{25--38\%}, though this does not affect the shape of the dust mass function derived using the IRS SLUGS sample in D00. However, the OS galaxies are relatively faint submillimetre sources compared with the IRS sample. From the fractional contribution of CO(3-2) line emission derived by Seaquist et al. (a linear fit to the plot of SCUBA-equivalent flux produced by the CO line versus SCUBA flux) we estimate that for the OS sample the CO line contribution to the 850\mic flux is small and is well within the uncertainties on the 850\mic fluxes we give in Table~\ref{fluxtab}.

\begin{figure*}
\begin{center}
\includegraphics[angle=0, width=14cm]{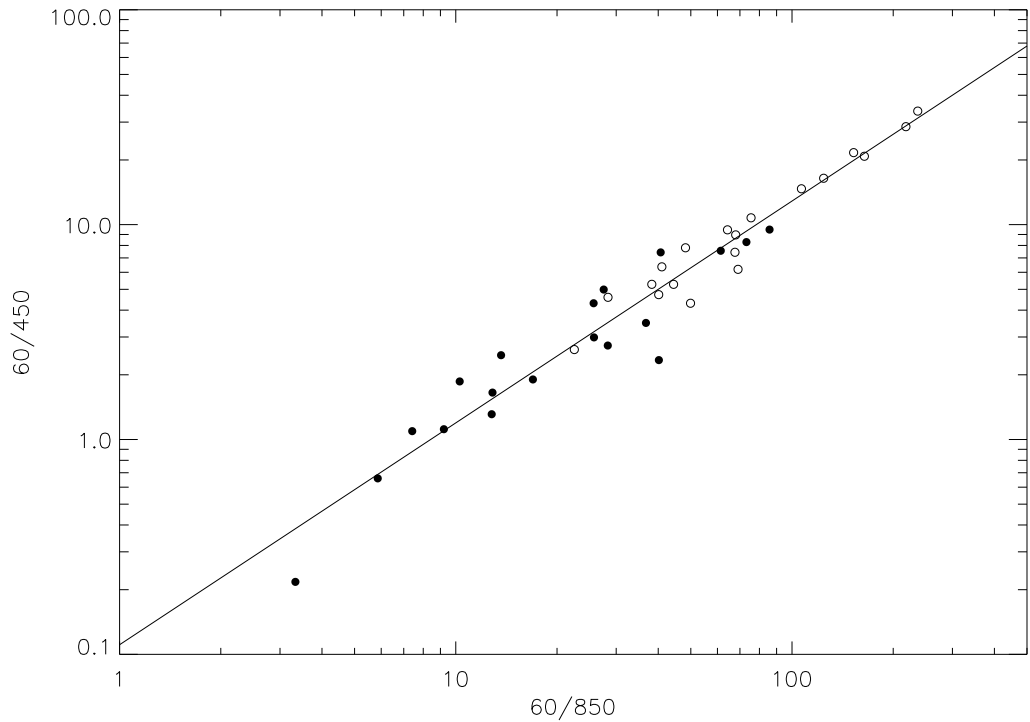}
\caption{\label{colplot-450}{Colour-colour plot: $S_{60}/S_{450}$ versus $S_{60}/S_{850}$ colours for the optically-selected (this work) and \textit{IRAS}-selected (D00) SLUGS (filled and open points respectively).}}
\end{center}
\end{figure*}

\subsection{Gas masses}
\label{sec:gasmass}

The neutral hydrogen masses listed in Table~\ref{lumtab} were calculated from HI fluxes taken from the literature\footnote{See notes to Table~\ref{lumtab}.} using
\begin{equation} 
M_{HI}=2.356\times10^5D^2S_{HI}
\end{equation}
where $D$ is in Mpc and $S_{HI}$ is in Jy km s$^{-1}$.

Only a small handful of objects in the OS sample had CO fluxes in the literature, and so in this work we will not present any molecular gas masses.

\subsection{Far-infrared luminosities}
\label{sec:fir}
The FIR luminosity ($L_{fir}$) is usually calculated using
\[FIR=1.26\times10^{-14}(2.58 S_{60}+S_{100})
\]
and
\begin{equation} \label{eq:fir}
L_{fir}=4\pi D^2\times FIR\times C
\end{equation}
as described in the Appendix of \textit{Catalogued Galaxies and Quasars Observed in the IRAS Survey} (Version 2, 1989), where $S_{60}$ and $S_{100}$ are the 60\mic and 100\mic \textit{IRAS} fluxes, D is the distance, and C is a colour-correction factor dependant on the ratio $S_{60}/S_{100}$ and the assumed emissivity index. The purpose of this correction factor is to account for emission outside the \textit{IRAS} bands, and is explained by Helou et al. (1988). 

However, since we have submillimetre fluxes we can use our derived $T_{d}$ and $\beta$ to integrate the total flux under the SED out to 1000\,\micron. This method gives more accurate values of $L_{fir}$ since it makes no general assumptions. We list in Table~\ref{lumtab} $L_{fir}$ calculated using this method and our fitted isothermal SEDs; $L_{fir}$ values calculated using our two-component SEDs are listed in Table~\ref{450tab}.

\subsection{Optical luminosities}
\label{sec:optlum}
The blue luminosities given in Table~\ref{lumtab} are converted (using M$_{B\odot}$=5.48) from blue apparent magnitudes taken from the Lyon-Meudon Extragalactic Database (LEDA; Paturel et al. 1989, 2003) which have already been corrected for galactic extinction, internal extinction and k-correction.

\section{The Submillimetre Properties of Galaxies}
\label{properties}

\begin{figure*}
 \begin{center}
 \subfigure[\label{beta:opt-irs}]{
 \includegraphics[angle=0, width=8.7cm]{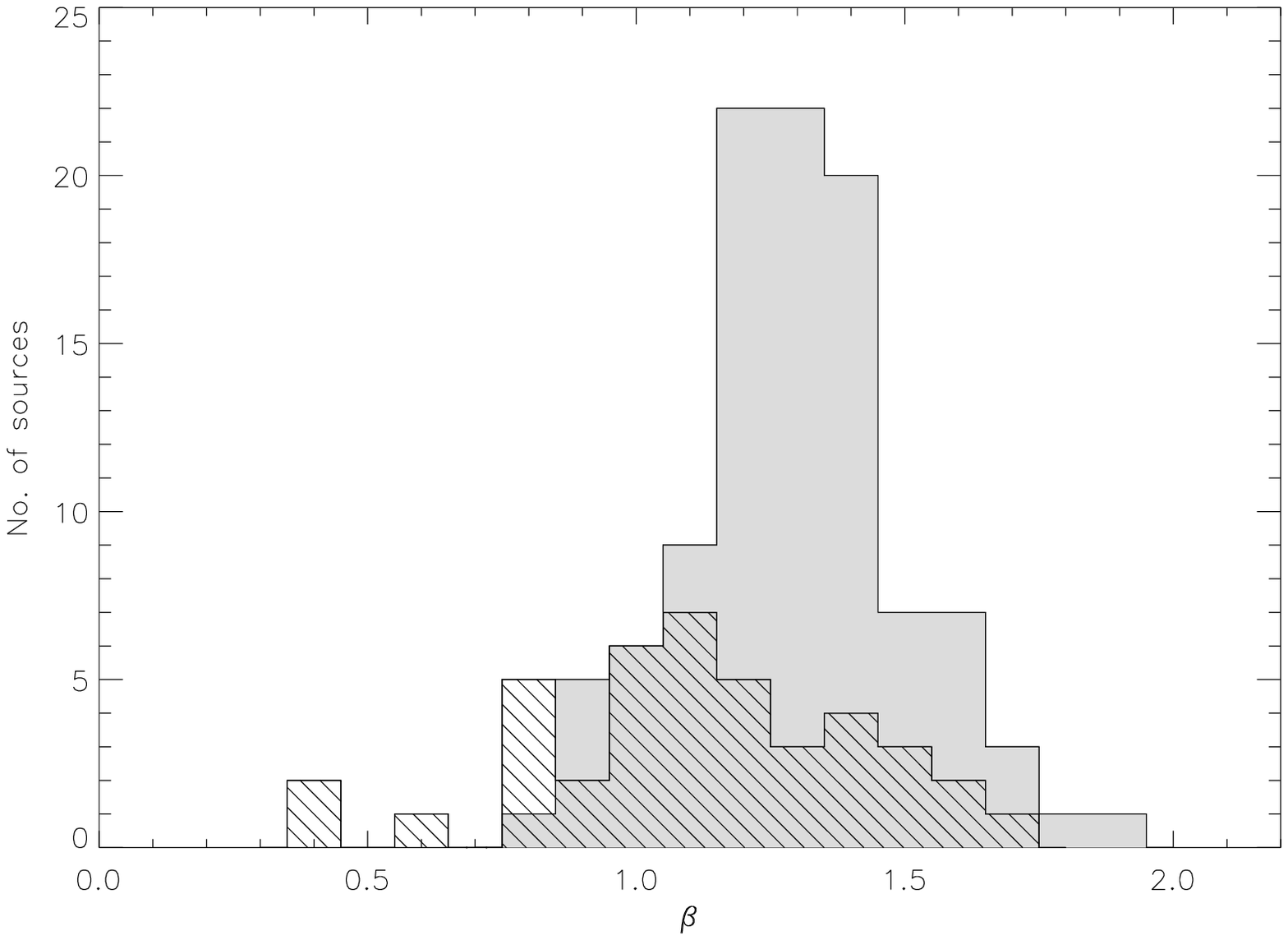}}
 \hfill
 \subfigure[\label{temp:opt-irs}]{
 \includegraphics[angle=0, width=8.7cm]{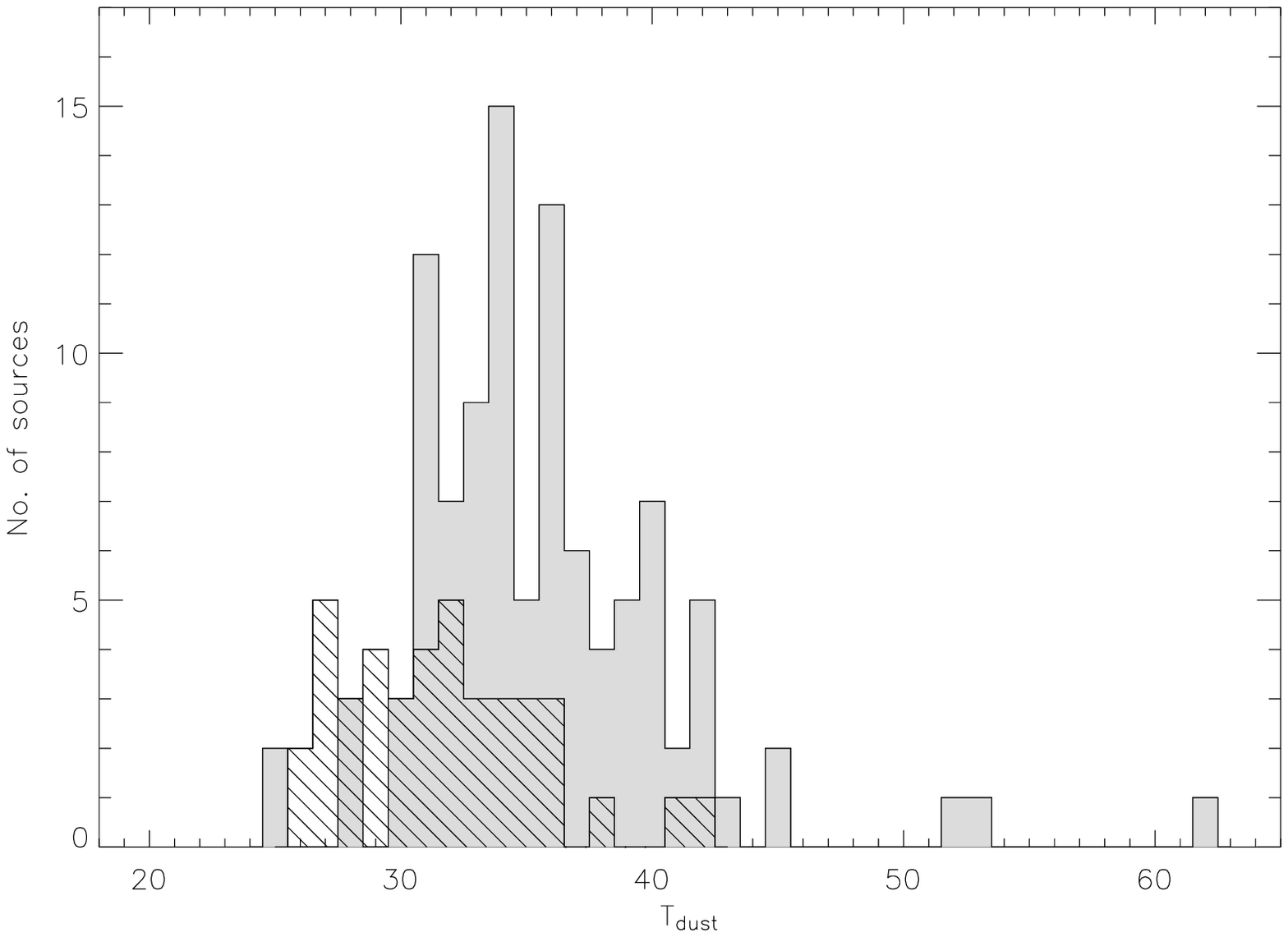}}
 \hfill
 \caption{\label{opt-iras-hist}{Distributions of (a) $\beta$ values and (b) $T_{d}$ values for the optically- and \textit{IRAS}-selected SLUGS (line-filled and shaded histograms respectively).}}
 \end{center}
\end{figure*}

\subsection{Optical selection versus IR selection}
\label{prop:ir-opt}
Figures~\ref{colplot} and~\ref{colplot-450} show the OS and IRS galaxies plotted on two-colour diagrams (filled and open symbols respectively). The IRS and OS galaxies clearly have different distributions, and in particular there are OS galaxies in parts of the diagram where there are no IRS galaxies. In Figure~\ref{colplot} $\sim$\,50\% of the OS galaxies are in a region of the colour-colour diagram completely unoccupied by IRS galaxies. This shows there are galaxies `missing' from IR samples, with important implications for the submillimetre LF (Section~\ref{lumfun}).

Figure~\ref{colplot-450} shows the $S_{60}/S_{450}$ versus $S_{60}/S_{850}$ colour-colour plot for the OS sample objects and IRS sample objects which have 450\mic fluxes. We confirm the very tight correlation found by DE01 (here the correlation coefficient \mbox{$r_{s}$\,=\,0.96}, \mbox{significance\,=\,9.20e-21})), and the scatter for the OS sample may be completely explained by the uncertainties on the fluxes. Importantly, this relationship holds for all the objects in the OS sample for which we have 450\mic fluxes, which include a wide range of galaxy types \mbox{(t-type=0 to 10)} and with $L_{fir}$ ranging over 2 orders of magnitude. The (least-squares) best-fitting line to the \textit{combined} \mbox{OS + IRS} samples shown in Figure~\ref{colplot-450} is given by
\[
\mathrm{log(S_{60}/S_{450})=(1.03\pm0.05)\,log(S_{60}/S_{850})-(0.955\pm0.070)}
\]
(or re-written $S_{60}/S_{450}=0.119(S_{60}/S_{850})^{1.03}$) and is very similar to that found by DE01, confirming the finding for the IRS sample that, within the uncertainties, the ratio $S_{450}/S_{850}$ is constant. DE01 conclude, from the results of simulations of the 450/850\mic flux ratio and from the fitted $\beta$ values for those galaxies whose SEDs require a cold component, that $\beta\sim2$ for all galaxies, and that therefore the cold dust component in all galaxies has a similar temperature \mbox{($T_{c}\sim$\,20--21\,K)}. The fact that we also find the $S_{450}/S_{850}$ ratio constant for the OS sample suggests that these conclusions are true for all Hubble types (only \mbox{t-types$<$0} are unrepresented in the OS sub-sample with 450\mic data).

The positions of the OS galaxies in the colour diagrams suggest there is more cold dust in the OS galaxies than in the IRS galaxies. We can investigate this further with the results of our spectral fits. Figure~\ref{beta:opt-irs} shows the comparison between the distribution of $\beta$ values (found from the isothermal fits) for the OS and IRS samples. We find OS sample galaxies with $\beta$ values lower than any found in the IRS sample. The two-sided Kolmogorov--Smirnov (K-S) test shows that the distributions of the two samples are significantly different (the probability that the two samples come from the same distribution function is only 1.8e-5). Though this clearly demonstrates that the properties of the dust in the OS and IRS samples are different, rather than interpreting this as a physical difference in the emissivity behaviour of the grains ($\beta$) we believe that it is a difference in the two samples' ratios of cold/warm dust.

Figure~\ref{temp:opt-irs} shows the comparison between the distribution of dust temperatures (from isothermal fits) for the OS and IRS samples. We note that the OS sample has consistently colder $T_{d}$ compared to the IRS sample. Once again using a K-S test we find that the OS and IRS sample dust temperatures do not have the same distribution, with the probability of the two samples coming from the same distribution function being only 1.41e-4. 

\begin{figure}
 \begin{center}
 \subfigure[\label{temp-warm}]{
 \includegraphics[angle=0, width=8.35cm]{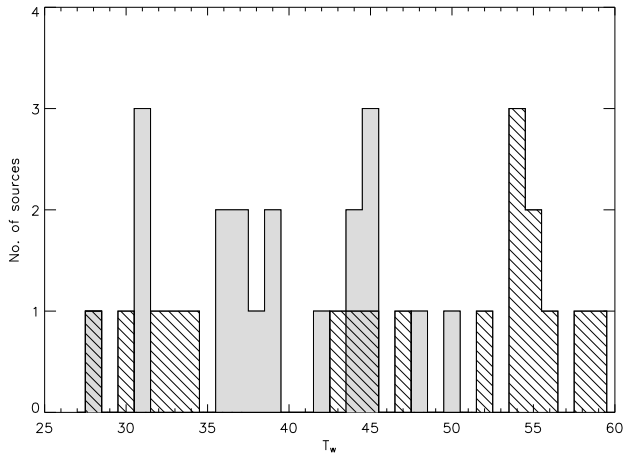}}\\
 \vfill
 \subfigure[\label{temp-cold}]{
 \includegraphics[angle=0, width=8.35cm]{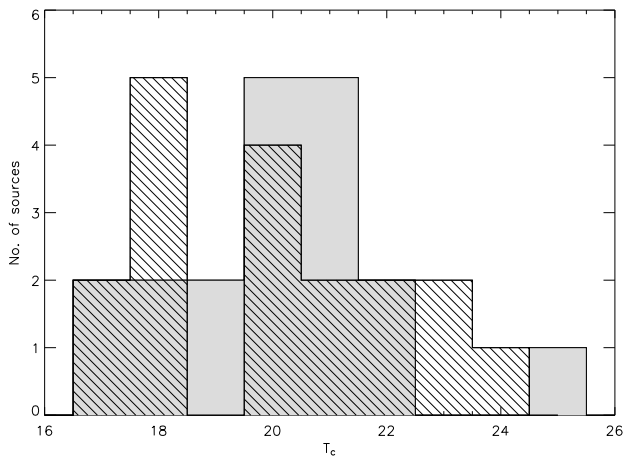}}
 \vfill
 \caption{\label{temp-warm-cold}{Distributions of warm component (a) and cold component (b) temperatures for the OS and IRS samples (line-filled and shaded histograms respectively).}}
 \end{center}
\end{figure}

For those objects in the OS and IRS samples for which two-component fits were possible, the distributions of the warm and cold component temperatures ($T_{w}$ and $T_{c}$) for the two samples are shown in Figures~\ref{temp-warm} and~\ref{temp-cold} respectively. The distributions of $T_{c}$ for the OS and IRS samples are statistically indistinguishable, while conversely the distributions of $T_{w}$ are not similar (probability of same distribution is 0.03). While the mean cold component temperature for the OS sample (\mbox{$\bar T_{c}=20.2\pm0.5$\,K}) is very similar to the value found for the IRS sample (mean \mbox{$\bar T_{c}=20.1\pm0.4$\,K}), the mean warm component temperature is rather higher (\mbox{$\bar T_{w}=47.4\pm2.4$\,K} for the OS sample as opposed to \mbox{$\bar T_{w}=39.3\pm1.4$\,K} for the IRS sample). 

\begin{figure}
 \begin{center}
 \includegraphics[angle=0, width=8.45cm]{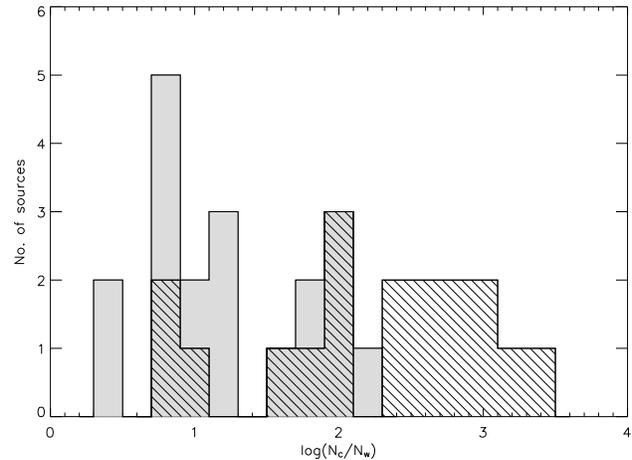}
 \caption{\label{norm-ratio}{Distribution of log($N_{c}/N_{w}$) for the OS and IRS samples (line-filled and shaded histograms respectively).}}
 \end{center}
\end{figure}

Figure~\ref{norm-ratio} shows the distribution of $N_{c}/N_{w}$ for the OS and IRS samples. The OS and IRS samples clearly have different distributions -- the (K-S test) probability of the two samples having the same distribution is 8.4e-4. For the OS sample the mean \mbox{$N_{c}/N_{w}=532\pm172$} (or higher, see Section~\ref{sed-fits}), for the IRS sample the mean \mbox{$N_{c}/N_{w}=38\pm11$}. For the OS sample there is a much larger range of $N_{c}/N_{w}$ than for the IRS sample. Interestingly, few of the OS objects even have a $N_{c}$/$N_{w}$ low enough to fall within the range found for the IRS sample, strongly suggesting a prevalence of cold dust in the OS sample compared to the IRS sample.

\begin{figure}
 \begin{center}
 \includegraphics[angle=0, width=8.45cm]{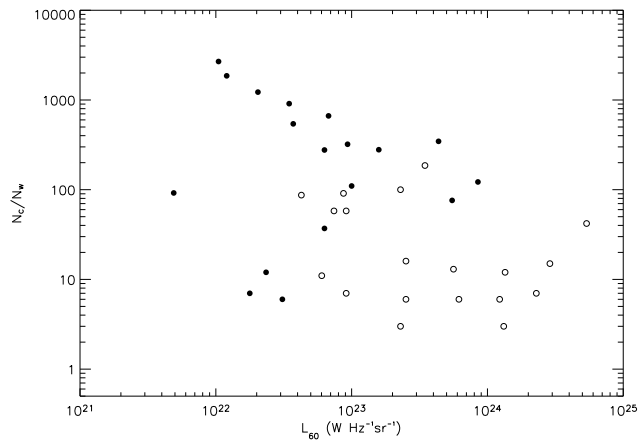}
 \caption{\label{norm-lum60}{$N_{c}/N_{w}$ versus 60\mic luminosity for the OS and IRS samples (filled and open points respectively).}}
 \end{center}
\end{figure}

The large difference between the distributions of $N_{c}/N_{w}$ for the OS and IRS galaxies implies that most OS sample galaxies contain much larger proportions of cold dust relative to warm dust than found for the IRS galaxies, additional evidence that \textit{IRAS} missed a population of cold-dust-dominated objects. The similarity of the temperature of the cold component for the OS and IRS sample and the difference in the distribution of $N_{c}/N_{w}$ supports the current paradigm for dust in galaxies. An alternative model for dust in galaxies would be one in which \IRAS\/ galaxies are ones in which the general ISRF is more intense, and therefore the majority of dust is hotter. The similarity of $T_{c}$ for the different samples argues against this and suggests that most dust in all galaxies is relatively cold and has a similar temperature. The temperature differences between galaxies arise from a second dust component, presumably the dust in regions of intense star formation. Our results for the OS sample indicate that the ratio of the mass of dust in this second component to the mass of dust in the first component can vary by roughly a factor of 1000. There are two other pieces of evidence in favour of the two-component model. First, the ISO 170\mic flux densities that exist for 3 of our two-component-fitted galaxies (Stickel et al. 2004; Section~\ref{sed-fits}) agree very well with our model SEDs (we did not use these data in making our fits, with one exception; see Section~\ref{sed-fits}). Second, the ratio of the mass of cold dust to the mass of warm dust correlates inversely with 60\mic luminosity (Figure~\ref{norm-lum60}; \mbox{$r_{s}$\,=\,$-$0.41}, \mbox{significance\,=\,1.24e-2}); in the two-component model one might expect the most luminous \IRAS\/ sources to be dominated by the warm component.

The difference in the distributions of $T_{w}$ does not, however, fit in with this general picture. In the two-component model one would expect $T_{w}$ and $T_{c}$ to be constants, with the only thing changing between galaxies being the proportion of cold and warm dust. The difference in the distributions of $T_{w}$ may indicate that this model is too simplistic. Two things may be relevant here. First, as can be seen in Figure~\ref{2compSEDfig} it is those OS galaxies with very prominent cold components which typically account for the highest warm component temperatures (for example PGC 35952 or NGC 6090). Second, the model SEDs with high values of $T_{w}$ also generally provide a good fit to the 25\mic flux density, whereas the model values with low values of $T_{w}$ tend to underestimate the 25\mic flux density. This last point suggests that to fully understand dust in galaxies one cannot ignore the measurements at wavelengths $<$\,60\,\micron; however, if we did include these measurements we would then definitely need more than two dust components. This is clearly demonstrated by Sievers et al. (1994) who, for NGC 3627, fit a three-component model. A two-component model is nonetheless adequate for our purposes, since we are interested in the cold component rather than a third hot component.

 \begin{figure}
 \begin{center}
 \includegraphics[angle=0, width=8.45cm]{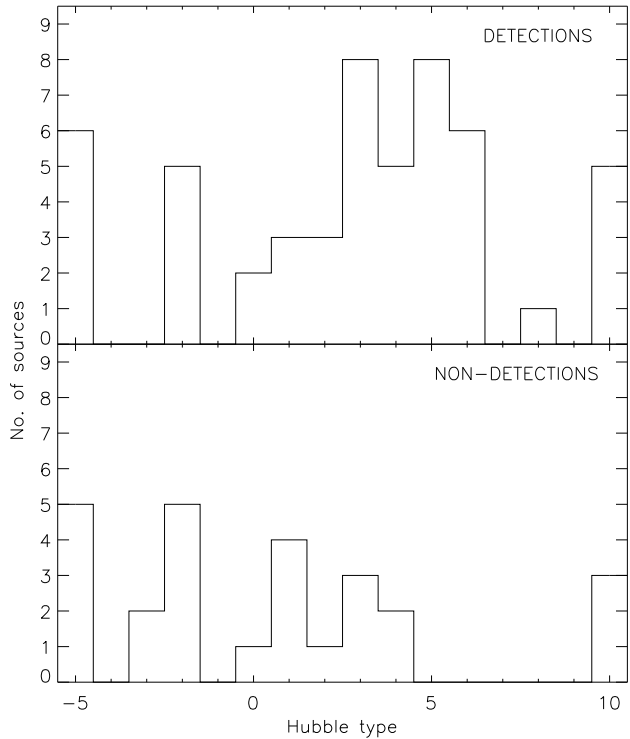}
 \caption{\label{type-hist}{Distribution of Hubble types for the OS sample 850\mic detections (upper panel) and non-detections (lower panel).}}
 \end{center}
\end{figure}

\begin{figure}
 \begin{center}
 \includegraphics[angle=0, width=8cm]{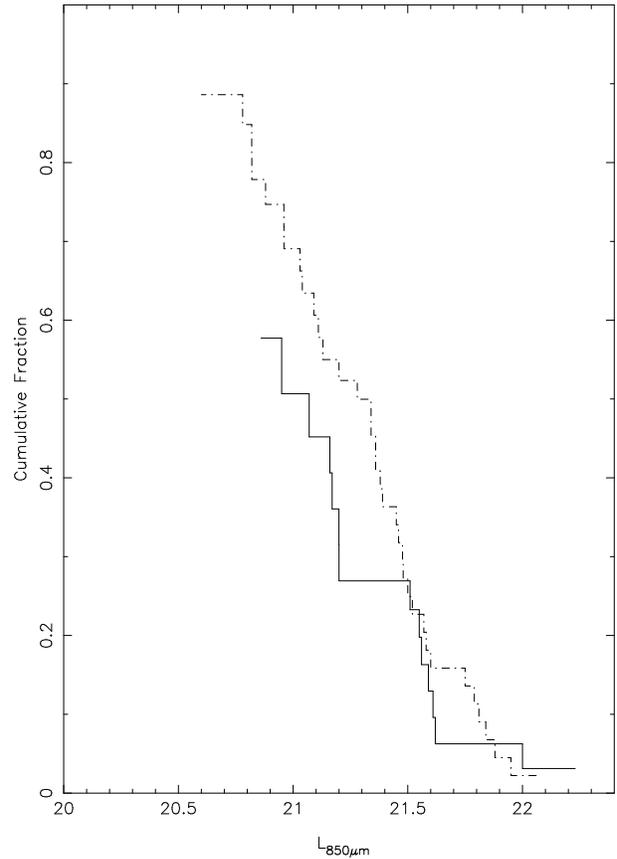}
 \caption{\label{steve-lum-plot}{Cumulative luminosity distributions for early-type galaxies
(solid line) and late-type galaxies (dot-dashed). The
maximum values for both samples are less than one
because of the upper limits that fall below the lowest
actual measurement.}}
 \end{center}
\end{figure}

\begin{figure}
 \begin{center}
 \includegraphics[angle=0, width=8.45cm]{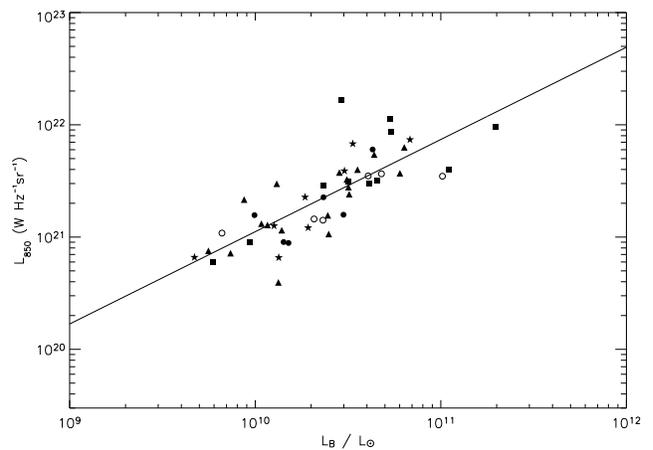}
 \caption{\label{850-opt}{850\mic luminosity versus optical luminosity $L_{B}$ for the OS sample, with different Hubble types indicated by different symbols: E-S0(t=-5 to 0), Early-type spirals(t=1 to 4), S?(t=5), Late-type spirals(t=6 to 10): circles, triangles, stars, and squares respectively. The 6 detected ellipticals are highlighted as open circles.}}
 \end{center}
\end{figure}

\subsection{Submillimetre properties along the Hubble sequence}
\label{prop:hubble}
In this section we investigate the submillimetre properties of galaxies as a function of Hubble type (\textit{t}). We first compare the distributions of Hubble type for the detections (D) and non-detections (ND) in our OS sample (Figure~\ref{type-hist}). We use the K-S test to find that the probability of their having the same distribution is $\simeq2\%$. Thus Figure~\ref{type-hist} suggests that early-type galaxies are less likely to be submillimetre sources than later types.

\begin{figure*}
 \begin{center}
 \subfigure[\label{bhist-split}]{
 \includegraphics[angle=0, width=8.7cm]{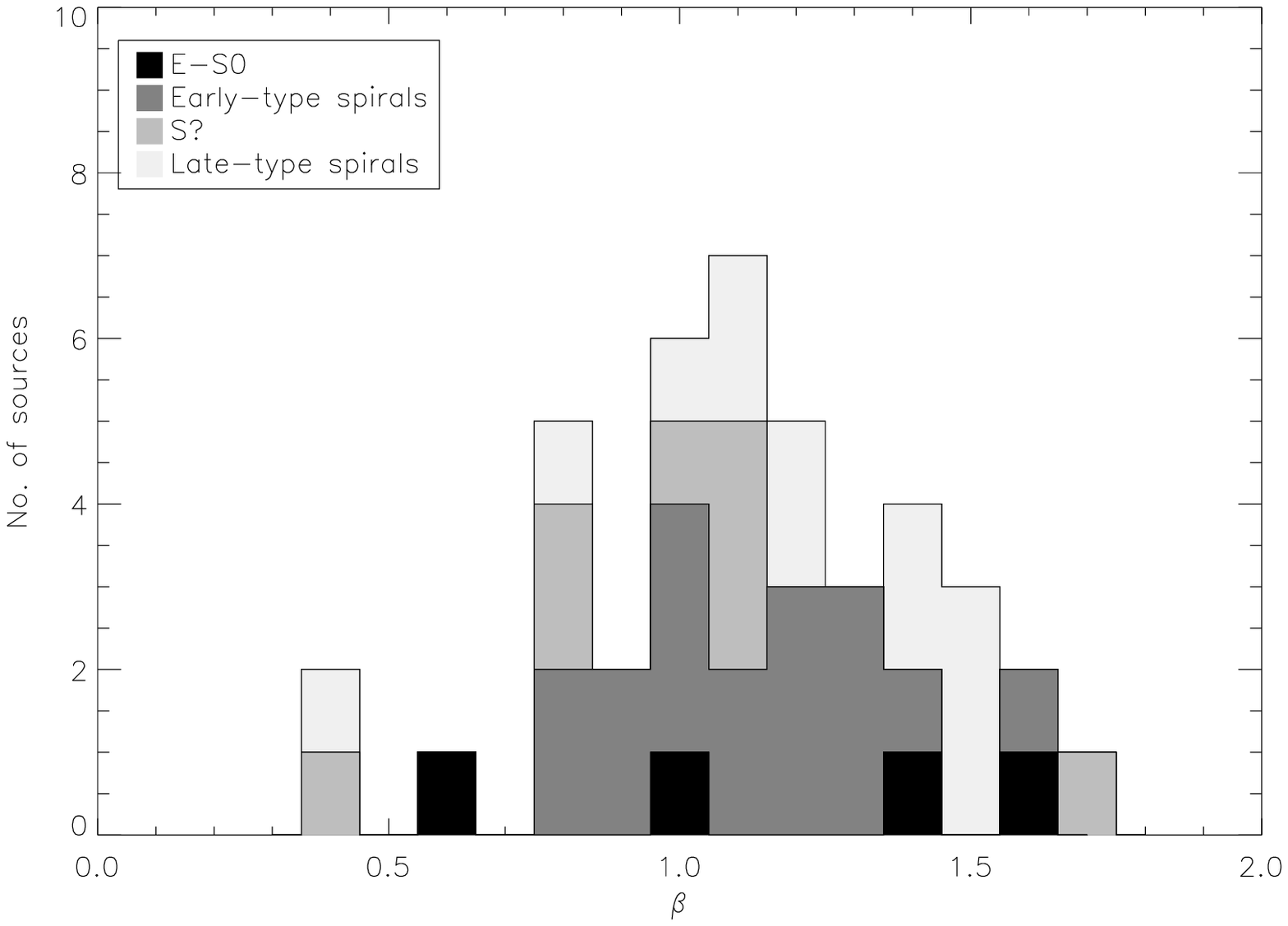}}
 \subfigure[\label{temp-type-m}]{
 \includegraphics[angle=0, width=8.7cm]{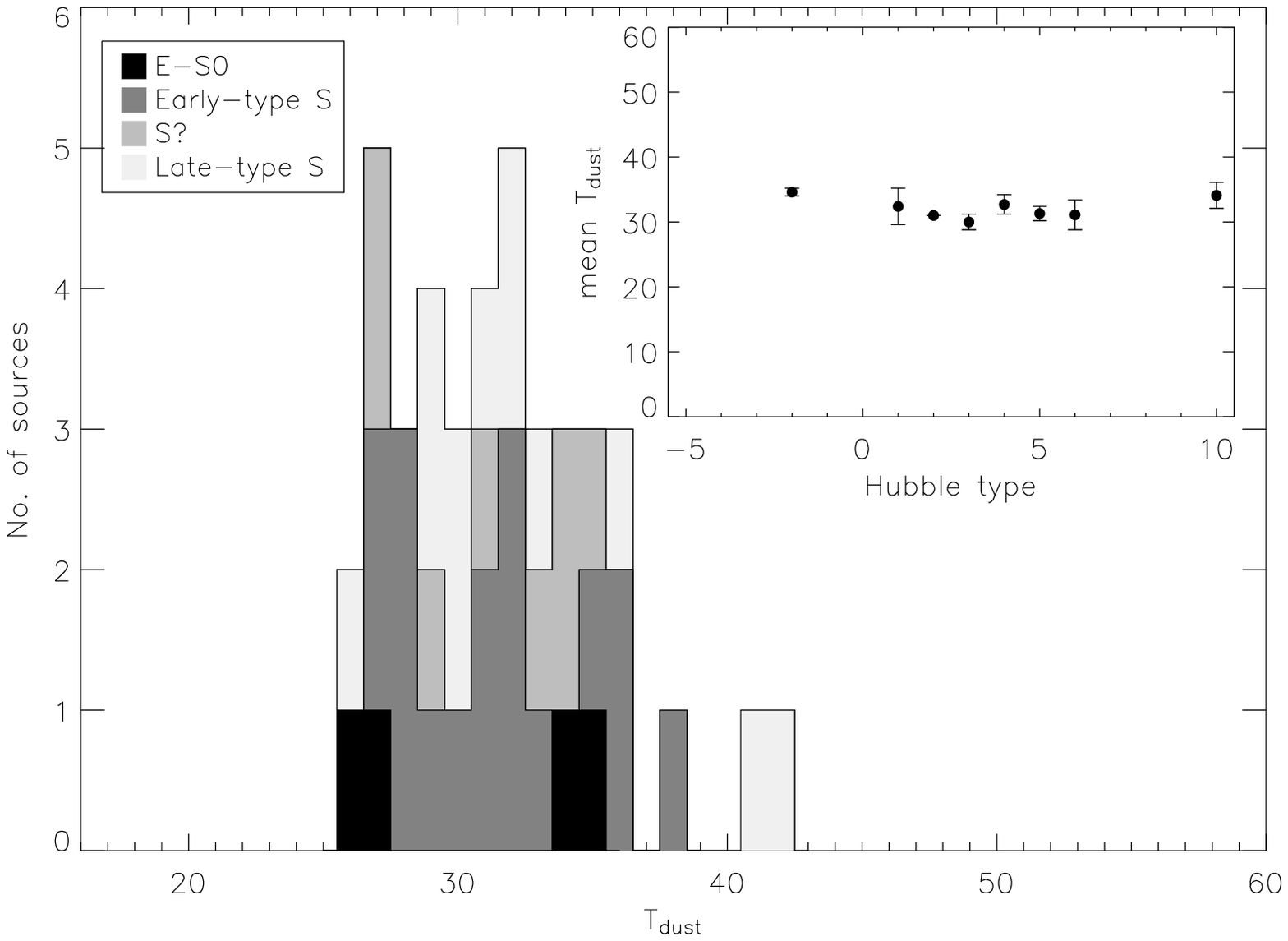}}
 \caption{\label{type-beta-temp}{Distribution of (a) $\beta$ values and (b) $T_{d}$ values for the OS SLUGS, with different Hubble types (as given in LEDA according to the RC2 code, and listed in Table~\ref{fluxtab}) indicated by different shaded regions: E-S0 (t=-5 to 0), Early-type spirals (t=1 to 4), S? (t=5), Late-type spirals (t=6 to 10). ((b): Inserted panel: mean $T_{d}$ for the OS SLUGS for different Hubble types, with error bars of error on the mean (bins with only 1 source are not plotted)).}}
 \end{center}
\end{figure*}  

To investigate this apparent morphological
difference further,
we estimated the submillimetre
luminosity distributions of early-
and late-type galaxies. A major complication is the
large number of upper limits. We used the
Kaplan-Meier estimator (Wall \& Jenkins 2003) to incorporate
information from both
the upper limits and the measurements. We defined early-type
galaxies as all those with $t\,\leq\,1$ and late-type
galaxies as those with $t\,>\,1$. We used this division, because
the greatest difference between the cumulative distributions
of Hubble type for detected and non-detected galaxies (Figure~\ref{type-hist}) was found at
t=1. Figure~\ref{steve-lum-plot} shows the cumulative luminosity distributions
estimated in this way for the early-type and late-type galaxies.
There appears to be a tendency for the late-type galaxies
to be more luminous submillimetre sources. However, the tendency
is not very strong. We also used the ASURV
statistical package for censored data (Feigelson \& Nelson 1985) to compare the results for the two samples,
using the Gehan test and
the log-rank test (see Wall \& Jenkins 2003). We found
a marginally significant (10\%) difference using the log-rank
test but no significant difference using the Gehan test.
Figure~\ref{850-opt} shows a plot of 850\mic luminosity versus optical
luminosity. For clarity we simply divide our sample into 4 broad groups based on the galaxies' t-type parameter given in LEDA (which uses the standard numerical codes for the de Vaucoulers morphological type, as defined in RC2): E-S0 \mbox{(t=-5 to 0)}, Early-type spirals \mbox{(t=1 to 4)}, S? (t=5) and Late-type spirals \mbox{(t=6 to 10)}. The different Hubble types show similar relationships.
On further inspection of the data, the more marked dependence on
Hubble type visible in Figure~\ref{type-hist} appears to be at least partly
caused by the early-type galaxies being observed in worse
conditions. In summary, there appears to be some difference
in submillimetre properties as one moves along the Hubble sequence,
but it is not very strong.

We can also use the results of our spectral fits to investigate whether there are any trends with Hubble type. As above, we simply divide our sample into 4 broad groups based on the galaxies' t-type. Figures~\ref{bhist-split} and \ref{temp-type-m} show the distributions of $\beta$ and $T_{d}$ (derived from our single-component fits) for the OS sample.  We note that the objects of each type appear fairly evenly distributed across the bins from \mbox{$\beta$\,=\,0 to 2} (Figure~\ref{bhist-split}), and in order to test this statistically we divide the sample into two broad groups: early types ($-5\leq \textrm{t-type} \leq4$) and late types ($5\leq \textrm{t-type} \leq10$), and perform a K-S test on the two groups. We find that the distributions of the early and late type groups are not significantly different. The distribution of isothermal dust temperatures appears similar for all Hubble types (Figure~\ref{temp-type-m}); we find no significant differences between the early and late types. We also investigated the distributions of the warm and cold component temperatures found from our two-component fits to look for any differences between early and late types; for example Popescu et al. (2002) find a tendency for the temperatures of the cold dust component to become colder for later types. We divided our 18 two-component fitted temperatures into Hubble types as in Popescu et al. (2002) and, due to our smaller number of sources, also into two broad groups of early ($0\leq \textrm{t-type} \leq4$) and late ($6\leq \textrm{t-type} \leq10$) types, and compared the overall distributions and the median $T_{c}$ for each type grouping. We found no differences between either the overall distributions or the median values of $T_{c}$ or $T_{w}$ for the early and late types, though we note the limitations of such a small sample.

\begin{figure}
 \begin{center}
 \includegraphics[angle=0, width=8.45cm]{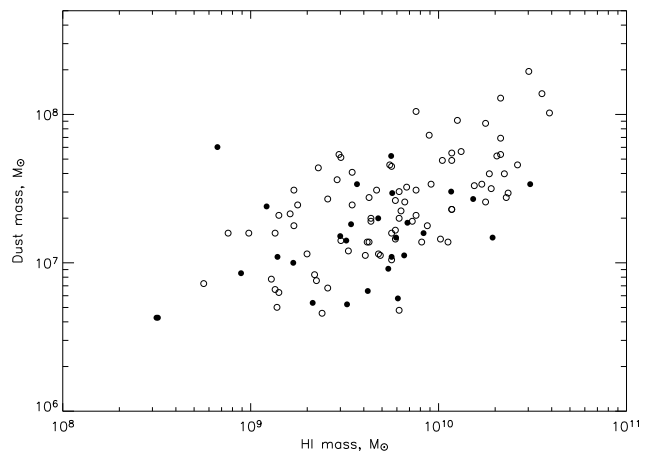}
 \caption{\label{d-HI-mass}{Dust mass versus HI mass for the OS and IRS samples (filled and open circles respectively).}}
 \end{center}
\end{figure} 

\subsection{Ellipticals}
\label{ellipticals}
It was once thought that ellipticals were entirely devoid of dust and gas, but optical absorption studies now show that dust is usually present (Goudfrooij et al. 1994; van Dokkum \& Franx 1995). Furthermore, dust masses for the \mbox{$\sim$\,15\%} of ellipticals detected by \textit{IRAS} (Bregman et al. 1998) have been found to be as much as a factor of 10--100 higher when estimated from their FIR emission compared to estimates from optical absorption (Goudfrooij \& de Jong 1995). 

At 850\mic we detect 6 ellipticals, from a total of 11 ellipticals in the OS sample, and find them to have dust masses in excess of \mbox{$10^{7}$ $M_{\odot}$}. However, a literature search revealed that for 4 of the 6 detections there are radio sources. We have used the radio data
to estimate the contribution of synchrotron emission at 850\,\micron. These estimates are often very uncertain because of the limited number of flux measurements available between 1.4GHz and 850\mic (353GHz). However, in some cases (Section~\ref{maps}) it is clear that some or all of the 850\mic emission may be synchrotron radiation. We are currently investigating ellipticals further with SCUBA observations of a larger sample. This will be the subject of a separate paper (Vlahakis et al., in prep.).

\subsection{The relationship between gas and dust}
\label{prop:gas-dust}

In D00 we found that both the mass of atomic gas and the mass of molecular gas are correlated with dust mass, but the correlation is tighter for the molecular gas. There are virtually no CO measurements for the OS sample, so here we have only estimated the mass of atomic gas. We compared the dust mass ($M_{d}$, calculated using dust temperatures from the isothermal fits) to the HI mass for the OS sample (Figure~\ref{d-HI-mass}) and find a very weak correlation. Though the correlation for the OS sample alone is very weak it is nonetheless consistent with the correlation found by D00 for the IRS sample; most of the OS points lie within the region covered by the IRS points, but though they cover the same range of HI masses we note that we do not have any HI masses for our OS sample objects with the higher dust masses. The weakness of the correlation for the OS sample is therefore likely due simply to the small number of OS sample 850\mic detections for which we have HI data (28 objects).

The mean neutral gas-to-dust ratio for the OS sample is $M_{HI}/M_{d}$=395$\pm71$, where the error given is the error on the mean. The (neutral $+$ molecular) gas-to-dust ratios for the IRS SLUGS sample and the Devereux \& Young (1990; herein DY90) sample of spiral galaxies are respectively $M_{{H_{2}}+HI}/M_{d}$=581$\pm43$ and  $M_{{H_{2}}+HI}/M_{d}$=1080$\pm70$, but since for the OS sample we have no CO measurements and therefore no measure of the mass of molecular hydrogen we can at this stage only compare the neutral gas-to-dust ratio for the OS sample. We therefore compare our OS value to mean neutral gas-to-dust ratios which we calculate, for the IRS sample and the DY90 sample respectively, to be $M_{HI}/M_{d}$=305$\pm24$ and $M_{HI}/M_{d}$=2089$\pm341$. There is a large difference between the values for both SLUGS samples and the value determined by DY90. This is almost certainly due to the fact that the DY90 dust masses were estimated from \IRAS\/ fluxes and therefore, for the reasons described in Section~\ref{intro}, will have `missed' the cold dust.

There is also a difference between the SLUGS values and the Galactic value of 160 for the (neutral $+$ molecular) gas-to-dust ratio (the value derived from Sodroski et al. (2004) by D00). The neutral gas-to-dust ratios for both the SLUGS samples are at least a factor of 2 larger than this Galactic value, and as shown by D00 when the molecular gas is included the value of the gas-to-dust ratio for the IRS sample is more than 3 times larger than the Galactic value. D00 attribute this discrepancy to a missed `cold dust' component \mbox{($T_{d}\le 20$\,K)} in the IRS sample. We have already noted in this paper that the single-temperature fits lead to dust masses approximately  a factor of 2 lower than the more realistic two-component fits (Section~\ref{sec:dmass}). Using the dust masses calculated using our two-component fits ($M_{d2}$; Table~\ref{450tab}), for the 13 galaxies for which there are HI masses we find the mean neutral gas-to-dust ratio for the OS sample is then $M_{HI}/M_{d2}$=192$\pm44$. This is in good agreement with the Galactic value, although if there is a significant amount of molecular gas this value would obviously be higher.

\section{Luminosity and Dust Mass Functions}
\label{lumfun}
The `accessible volume' method (Avni \& Bahcall 1980) will, in principle, produce unbiased estimates of the submillimetre luminosity function (LF) and dust mass function (DMF) provided that no population of galaxies is unrepresented by the sample used to derive the LF and DMF. In Paper I (D00) we produced a first estimate of the LF and DMF from the IRS sample. However, since our new observations of the OS sample have shown the existence of a population of galaxies with low values of the $S_{60}/S_{100}$ and $S_{60}/S_{850}$ flux ratios (Figure~\ref{colplot} and discussion in Section~\ref{sed-fits}) \textit{of which there is not a single representative in the IRS sample}, our earlier estimates of the LF and DMF are likely to be biased.

In this section we use our new (OS sample) results to produce new estimates of the submillimetre LF and DMF.

\subsection{Method}
\label{lumfun:method}
We derive the local submillimetre LF and DMF by two different methods: 1) directly from the OS SLUGS sample, and 2) by extrapolating the spectral energy distributions of the galaxies in the \textit{IRAS} PSCz catalogue out to 850\,\micron. The PSCz catalogue (Saunders et al. 2000) is a complete redshift survey of $\sim$15000 \IRAS\/ galaxies in the \IRAS\/ Point Source Catalogue. Serjeant \& Harrison (2005; herein SH05) used the PSCz galaxies and the IRS SLUGS submm:far-IR two-colour relation to extrapolate the SEDs of the PSCz galaxies out to 850\mic and produce an 850\mic LF. Importantly, this method allows us to probe a wider range of luminosities than probed directly by the SLUGS samples. 

We estimate the LF for both methods using

\begin{equation} \label{accvol}
\Phi(L)\Delta L=\sum_{i}\frac{1}{V_{i}}
\end{equation}
(Avni $\&$ Bahcall 1980). Here $\Phi(L)\Delta L$ is the number density of objects (Mpc$^{-3}$) in the luminosity range $L$ to $L+\Delta L$, the summation is over all the objects in the sample lying within this luminosity range, and $V_{i}$ is the accessible volume of the $i$th object in the sample. Throughout we use an $H_{0}$ of \mbox{75 km\,s$^{-1}$Mpc$^{-1}$} and a `concordance' universe with $\Omega_{M}$=0.3 and $\Omega_{\Lambda}$=0.7. We estimate the dust mass function (the space density of galaxies as a function of dust mass) in the same way as the LF, substituting dust mass for luminosity in Equation~\ref{accvol}. The details of these two methods, hereafter referred to as `directly measured' and `PSCz-extrapolated', are discussed in Sections~\ref{method:850LF} and~\ref{method:pscz} respectively.

\begin{table}
\caption{\label{optlf}\small{Directly measured OS SLUGS luminosity and dust mass functions}}
\begin{tabular}{cccc}
\hline
\multicolumn{4}{c}{850\mic luminosity function} \\
\smallskip \\
$log L_{850}$ & $\phi$(L) & $\sigma_{\phi}$ & \\
(W\,Hz$^{-1}$sr$^{-1}$) & (Mpc$^{-3}$dex$^{-1}$) & (Mpc$^{-3}$dex$^{-1}$) & \\
\smallskip \\
20.75 & 9.17e-3 & 3.47e-3 & \\
21.01 & 3.83e-3 & 1.15e-3 & \\
21.27 & 2.10e-3 & 6.32e-4 & \\
21.52 & 1.20e-3 & 3.10e-4 & \\
21.78 & 6.03e-4 & 2.46e-4 & \\
22.04 & 9.14e-5 & 5.28e-5 & \\
\medskip \\
$\alpha$ & $L_{\ast}$ & $\phi_{\ast}$ & $\chi^2_{\nu}$ \\
    & (W\,Hz$^{-1}$sr$^{-1}$) & (Mpc$^{-3}$dex$^{-1}$) \\
\smallskip \\
$-$1.71$^{+0.60}_{-0.57}$ & $4.96^{+6.1}_{-2.5}\times10^{21}$ & 1.67$^{+5.21}_{-1.18}\times10^{-3}$ & 0.31 \\
\medskip\\
\multicolumn{4}{c}{850\mic dust mass function} \\
\smallskip \\
$log M_{d}$ & $\phi$(M) & $\sigma_{\phi}$ & \\
($M_{\odot}$) & (Mpc$^{-3}$dex$^{-1}$) & (Mpc$^{-3}$dex$^{-1}$) & \\
\smallskip \\
6.75 & 9.08e-3 & 3.03e-3 & \\
6.99 & 3.99e-3 & 1.33e-3 &\\
7.23 & 3.09e-3 & 8.57e-4 & \\
7.48 & 9.25e-4 & 3.08e-4 &\\
7.72 & 8.14e-4 & 2.45e-4 &\\
7.96 & 5.69e-5 & 4.02e-5 &\\
\medskip \\
$\alpha$ & $M_{\ast}$ & $\phi_{\ast}$ & $\chi^2_{\nu}$ \\
    & ($M_{\odot}$) & (Mpc$^{-3}$dex$^{-1}$) \\
\smallskip \\
$-$1.67$^{+0.24}_{-0.25}$ & 3.09$^{+1.09}_{-0.64}\times10^{7}$ & 3.01$^{+1.62}_{-1.38}\times10^{-3}$ & 1.17 \\
\medskip \\
\hline
\medskip
\end{tabular}
\end{table}

\begin{table}
\caption{\label{PSCZlf}\small{PSCz-extrapolated luminosity function}}
\begin{tabular}{cccc}
\hline
\smallskip \\
log $L_{850}$ & $\phi$(L) & $\sigma_{\phi}^{down}$ & $\sigma_{\phi}^{up}$ \\
(W\,Hz$^{-1}$sr$^{-1}$) & (Mpc$^{-3}$dex$^{-1}$) & \multicolumn{2}{c}{(Mpc$^{-3}$dex$^{-1}$)} \\
\smallskip \\
18.52 & 3.42e-02 & 2.42e-02 & 2.74e-02 \\
18.75 & 6.30e-02 & 2.38e-02 & 2.83e-02 \\
18.99 & 3.90e-02 & 1.62e-02 & 9.45e-03 \\
19.23 & 3.20e-02 & 1.06e-02 & 6.17e-03 \\
19.47 & 2.35e-02 & 3.50e-03 & 7.86e-03 \\
19.70 & 3.08e-02 & 8.14e-03 & 3.42e-03 \\
19.94 & 1.85e-02 & 2.81e-03 & 5.72e-03 \\
20.18 & 1.26e-02 & 1.98e-03 & 1.34e-03 \\
20.42 & 1.16e-02 & 1.14e-03 & 6.74e-04 \\
20.65 & 1.02e-02 & 1.70e-03 & 4.41e-04 \\
20.89 & 6.67e-03 & 1.12e-03 & 8.25e-04 \\
21.13 & 4.30e-03 & 6.77e-04 & 3.01e-04 \\
21.36 & 2.73e-03 & 7.59e-04 & 1.63e-04 \\
21.60 & 1.34e-03 & 4.61e-04 & 1.00e-04 \\
21.84 & 4.43e-04 & 1.75e-04 & 1.36e-04 \\
22.08 & 1.17e-04 & 6.67e-05 & 2.36e-05 \\
22.31 & 1.85e-05 & 7.86e-06 & 1.45e-05 \\
22.55 & 4.27e-06 & 2.64e-06 & 3.81e-07 \\
22.79 & 2.91e-07 & 1.28e-07 & 9.39e-07 \\
23.03 & 9.86e-08 & 5.62e-08 & 3.49e-08 \\
\medskip \\
$\alpha$ & $L_{\ast}$ & $\phi_{\ast}$ & $\chi^2_{\nu}$ \\
    & (W\,Hz$^{-1}$sr$^{-1}$) & (Mpc$^{-3}$dex$^{-1}$) \\
\smallskip \\
$-$1.38$^{+0.02}_{-0.03}$ & 3.73$^{+0.29}_{-0.32}\times10^{21}$ & 4.17$^{+0.41}_{-0.45}\times10^{-3}$ & 1.0 \\
\smallskip \\
\hline
\medskip
\end{tabular}
\end{table}

\subsubsection{Directly measured 850\,$\mu$m luminosity function and dust mass function}
\label{method:850LF}
We calculated the directly measured LF and DMF from the 52 objects in the OS sample which were detected at 850\,\micron. For the DMF we use the dust masses listed in Table~\ref{lumtab}, which were calculated using the isothermal SED-fitted temperatures or, where no fit was made (11 objects), using a dust temperature of 20K. For the OS sample the accessible volume is the maximum volume in which the object would still be detected at 850\mic and still be included in the CfA sample. Since objects with \mbox{$cz<1900$\,km\,s$^{-1}$} were excluded from our sample this volume is not included in our calculation of $V_{i}$. When calculating the maximum redshift at which an object would still be detected at 850\mic we used the noise appropriate for the observation of that object. We corrected the LF by the factor 97/81 to account for the CfA galaxies we did not observe at all at 850\mic (Section~\ref{sample}).

The corrected directly measured 850\mic LF and DMF are shown as star symbols in Figures~\ref{lumfun-plot} and~\ref{dmfun} respectively, and are given in tabular form in Table~\ref{optlf}. The errors on the directly measured LF and DMF are standard Poisson errors. One effect that may lead to our estimates of the LF and DMF being slight underestimates is that we noticed that the OS galaxies not detected at 850\mic were generally observed under worse weather conditions than the sources that were detected.

\subsubsection{\textit{IRAS} PSCz-extrapolated 850\,$\mu$m luminosity function and dust mass function}
\label{method:pscz}

\begin{figure*}
 \begin{center}
  \includegraphics[angle=270, width=13cm]{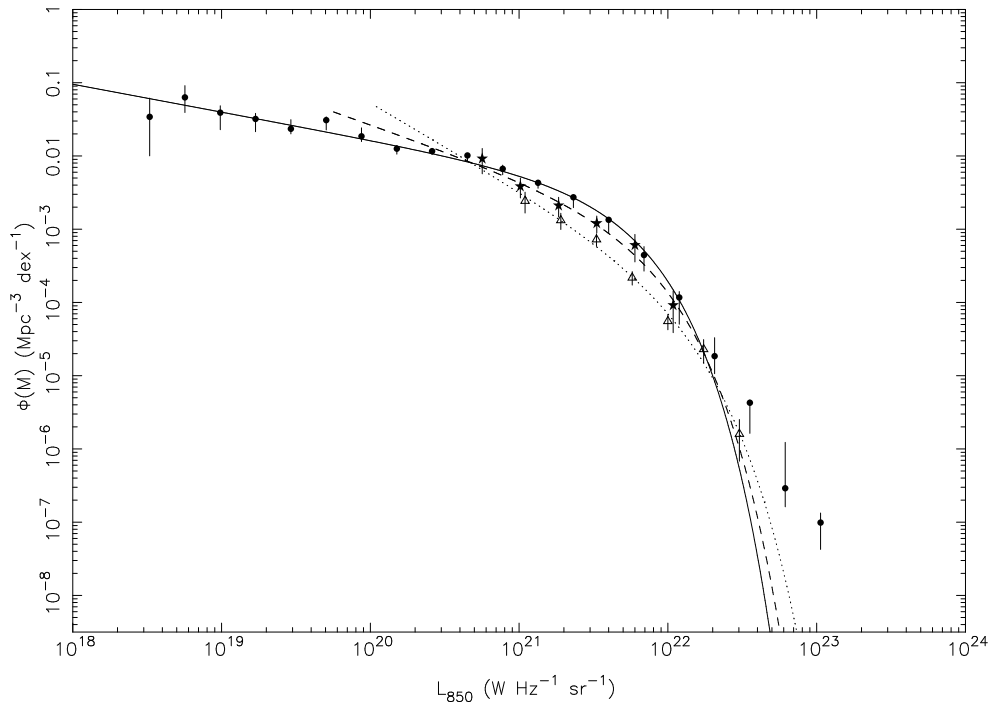}
  \caption{\label{lumfun-plot}{PSCz-extrapolated 850\mic luminosity function (filled circles) with best-fitting Schechter function (solid line). The parameters for the Schechter function are $\alpha=-1.38$, $L_{\ast}=3.7\times10^{21}$ W\,Hz$^{-1}$sr$^{-1}$. Also shown are the directly measured 850\mic luminosity function for the OS SLUGS sample (filled stars) with best-fitting Schechter function (dashed line) and the results for the IRS SLUGS sample from Dunne et al. (2000) (open triangles and dotted line).}}
 \end{center}                                                      
\end{figure*} 

\begin{figure*}
 \begin{center}
 \subfigure[]{\label{dmfun-a}
  \includegraphics[angle=270, width=13cm]{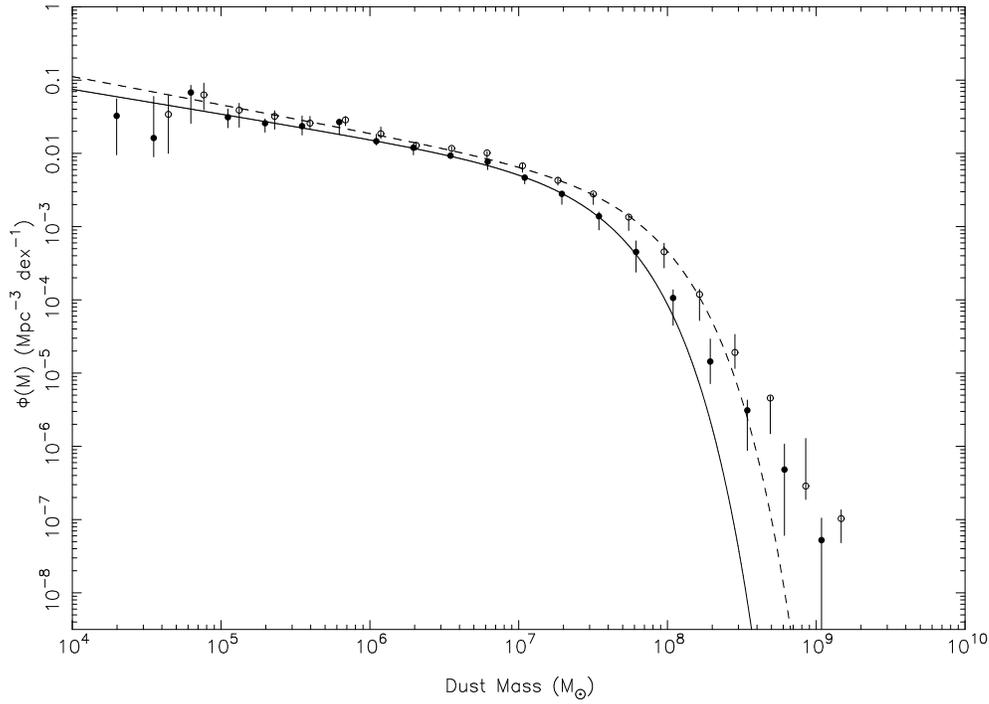}}\\
 \subfigure[]{ 
  \includegraphics[angle=270, width=13cm]{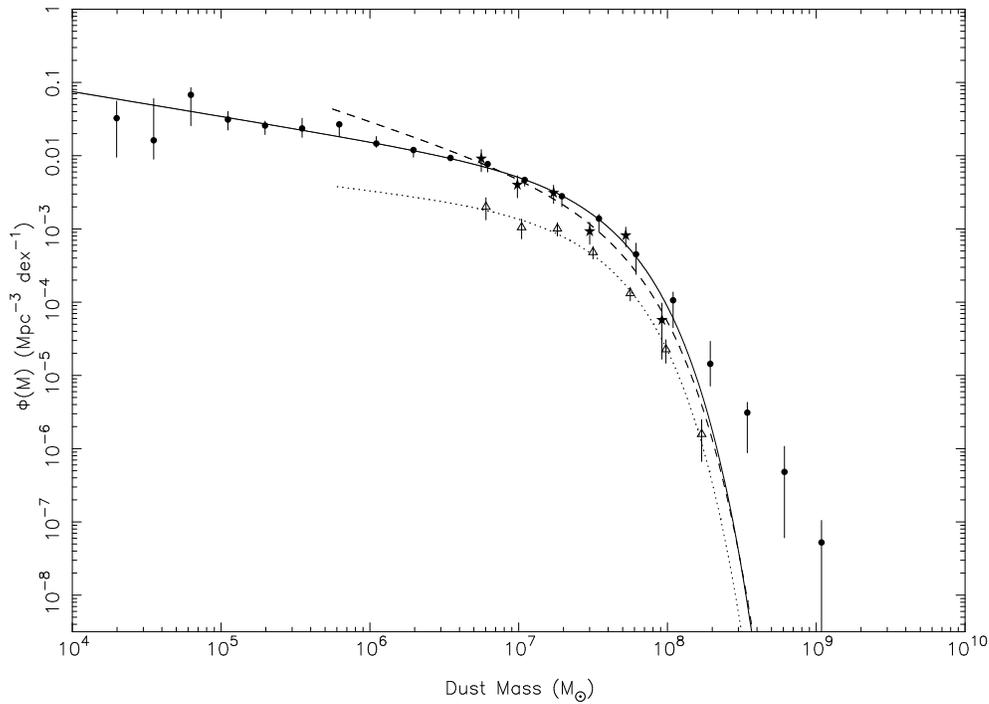}}
  \caption{\label{dmfun}{(a) PSCz-extrapolated dustmass function (filled circles) with best-fitting Schechter function (solid line). The dust masses were calculated using $T_{d}$ derived from the \textit{IRAS} 100/60 colour and $\beta$=2. The parameters for the Schechter function are $\alpha=-1.34$, $M_{\ast}=2.7\times10^7 M_{\odot}$. The dashed line and open circles are for a  `cold dustmass function' in which dust masses are calculated using $T_{d}$=20K and $\beta$=2. The best-fitting Schechter parameters are $\alpha=-1.39, M_{\ast}=5.3\times10^7 M_{\odot}$.
(b) Directly measured dustmass function for the OS SLUGS sample (filled stars) with best-fitting Schechter function (dashed line). The dust masses were calculated using $T_{d}$ from isothermal SED fitting. Also shown are the results for the IRS SLUGS sample from Dunne et al. (2000) (open triangles and dotted line). The filled circles and solid line show the PSCz-extrapolated dustmass function as in (a).}}
 \end{center}
\end{figure*}

In order to better constrain the LF at the lower luminosity end more data points are needed, probing a wider range of luminosities than probed directly by the SLUGS samples. We achieve  this using a method described by SH05, whereby the 850\mic LF is determined by extrapolating the spectral energy distributions of the $\sim$15000 \textit{IRAS} PSCz survey galaxies (Saunders et al. 2000) out to 850\,\micron. Since for the two SLUGS samples we find a strong correlation between the $S_{60}/S_{100}$ and $S_{60}/S_{850}$ colours (Figure~\ref{colplot}) we can use a linear fit to this colour-colour relation to make the extrapolation from 60\mic to 850\mic flux density. SH05 derived the submm:far-IR two-colour relationship from the IRS SLUGS sample. However, we have shown in this paper that the OS and IRS samples have quite different properties. In order to determine the sensitivity of the LF/DMF to the colour relationship we have derived colour relationships for the combined \mbox{OS + IRS} sample, the OS sample alone, and the IRS sample alone (Table~\ref{colplot-params}).

\begin{table}
\caption{\label{colplot-params}\small{Linear fit parameters for the SLUGS colour-colour plot (log($S_{60}/S_{100}$) vs log($S_{60}/S_{850}$)) shown in Figure~\ref{colplot}.}}
\begin{tabular}{ccc}
\hline
SLUGS data fitted  & \multicolumn{2}{c}{linear fit (y=mx+c)} \\
                   & m & c \\
\hline
OPT+\textit{IRAS} & $0.365\pm0.014$ & $-0.881\pm0.024$\\
OPT               & $0.296\pm0.031$ & $-0.797\pm0.039$\\
\textit{IRAS}     & $0.421\pm0.023$ & $-0.981\pm0.042$\\
\hline
\end{tabular}
\end{table}

In order to produce unbiased estimates of the LF and DMF we have excluded some PSCz galaxies. Firstly we exclude all those objects that do not have redshifts, those that have velocities \mbox{$<300$\,km\,s$^{-1}$} (to ensure that peculiar velocities are unimportant), and those that have redshifts $>0.2$ (thus excluding any ultra-luminous, high-redshift objects). We then exclude all objects with upper limits at 100\,\micron, since for these objects we cannot apply the SH05 method. Finally, we use the {\textit{IRAS} Point Source Catalogue flags, as listed in the PSCz catalogue, to exclude sources which are likely to be either solely or strongly contaminated by Galactic cirrus. It is important to exclude these sources because they are very cold sources and so potentially can have a large effect on the 850\mic LF. If two or more of the flags indicate Galactic cirrus (using flag value limits indicated in the \textit{IRAS} Explanatory Supplement) we exclude that object. As a check on the validity of this method we inspected by eye (using the IRSA ISSA Image Server) a sample of $\sim$40 objects randomly chosen from those excluded as Galactic cirrus, and a further sample of $\sim$40 objects randomly chosen from those that made it into our final sample. We found that 98\% of the sources with cirrus flags and 7\% of the sources without cirrus flags showed signs of significant cirrus, although for two thirds of the sources with cirrus flags there still appeared to be a genuine source present. In total, from the $\sim$14500 galaxies with redshifts in the \textit{IRAS} PSCz catalogue we exclude $\sim$4300 objects because of either 100\mic upper limits or Galactic cirrus. This leaves 10252 galaxies in our PSCz-selected sample.

\begin{table}
\caption{\label{PSCZdmf}\small{PSCz-extrapolated dustmass functions}}
\begin{tabular}{cccc}
\hline
\multicolumn{4}{c}{PSCz-extrapolated single temperature dustmass function}
\smallskip \\
log $M_{d}$ & $\phi$(M) & $\sigma_{\phi}^{down}$ & $\sigma_{\phi}^{up}$ \\
($M_{\odot}$) & (Mpc$^{-3}$dex$^{-1}$) & \multicolumn{2}{c}{(Mpc$^{-3}$dex$^{-1}$)} \\
\smallskip \\
4.30 & 3.26e-02 & 2.30e-02 & 2.30e-02 \\
4.55 & 1.62e-02 & 7.26e-03 & 4.37e-02 \\
4.80 & 6.78e-02 & 4.22e-02 & 1.70e-02 \\
5.05 & 3.11e-02 & 8.74e-03 & 9.10e-03 \\
5.30 & 2.58e-02 & 6.36e-03 & 3.77e-03 \\
5.54 & 2.36e-02 & 5.72e-03 & 8.87e-03 \\
5.79 & 2.68e-02 & 8.69e-03 & 2.48e-03 \\
6.04 & 1.46e-02 & 1.62e-03 & 3.60e-03 \\
6.29 & 1.19e-02 & 2.41e-03 & 6.62e-04 \\
6.54 & 9.29e-03 & 3.93e-04 & 8.97e-04 \\
6.79 & 7.68e-03 & 1.68e-03 & 2.53e-04 \\
7.04 & 4.67e-03 & 8.33e-04 & 3.80e-04 \\
7.29 & 2.80e-03 & 7.87e-04 & 1.03e-04 \\
7.54 & 1.38e-03 & 4.83e-04 & 1.94e-04 \\
7.79 & 4.51e-04 & 2.12e-04 & 1.87e-04 \\
8.04 & 1.06e-04 & 6.10e-05 & 3.12e-05 \\
8.29 & 1.44e-05 & 7.21e-06 & 1.48e-05 \\
8.54 & 3.10e-06 & 2.22e-06 & 1.18e-06 \\
8.79 & 4.82e-07 & 4.21e-07 & 5.94e-07 \\
9.04 & 5.24e-08 & 5.64e-08 & 5.24e-08 \\
\medskip \\
$\alpha$ & $M_{\ast}$ & $\phi_{\ast}$ & $\chi^2_{\nu}$ \\
   & ($M_{\odot}$) & (Mpc$^{-3}$dex$^{-1}$) \\
\smallskip \\
$-$1.34$^{+0.13}_{-0.08}$ & 2.74$^{+1.23}_{-1.13}\times10^{7}$ & 5.16$^{+3.90}_{-1.74}\times10^{-3}$ & 0.65 \\
\medskip \\
\multicolumn{4}{c}{PSCz-extrapolated 20K `cold' dustmass function}
\smallskip \\
log $M_{d}$ & $\phi$(M) & $\sigma_{\phi}^{down}$ & $\sigma_{\phi}^{up}$ \\
\smallskip \\
4.65 & 3.41e-02 & 2.41e-02 & 2.76e-02 \\
4.88 & 6.28e-02 & 2.37e-02 & 2.82e-02 \\
5.12 & 3.88e-02 & 1.61e-02 & 9.42e-03 \\
5.36 & 3.20e-02 & 1.06e-02 & 6.16e-03 \\
5.60 & 2.59e-02 & 4.59e-03 & 6.08e-03 \\
5.84 & 2.86e-02 & 4.67e-03 & 3.24e-03 \\
6.07 & 1.85e-02 & 2.60e-03 & 4.39e-03 \\
6.31 & 1.29e-02 & 1.82e-03 & 1.13e-03 \\
6.55 & 1.17e-02 & 1.11e-03 & 6.77e-04 \\
6.79 & 1.02e-02 & 1.58e-03 & 4.37e-04 \\
7.03 & 6.75e-03 & 1.21e-03 & 5.96e-04 \\
7.26 & 4.29e-03 & 6.15e-04 & 4.09e-04 \\
7.50 & 2.79e-03 & 7.83e-04 & 1.07e-04 \\
7.74 & 1.35e-03 & 4.65e-04 & 1.07e-04 \\
7.98 & 4.54e-04 & 1.81e-04 & 1.37e-04 \\
8.22 & 1.19e-04 & 6.64e-05 & 2.08e-05 \\
8.45 & 1.91e-05 & 7.57e-06 & 1.47e-05 \\
8.69 & 4.56e-06 & 3.07e-06 & 3.27e-07 \\
8.93 & 2.87e-07 & 9.82e-08 & 9.94e-07 \\
9.17 & 1.03e-07 & 5.48e-08 & 3.26e-08 \\
\medskip \\
$\alpha$ & $M_{\ast}$ & $\phi_{\ast}$ & $\chi^2_{\nu}$ \\
    & ($M_{\odot}$) & (Mpc$^{-3}$dex$^{-1}$) \\
\smallskip \\
$-$1.39$^{+0.03}_{-0.02}$ & 5.28$^{+0.45}_{-0.55}\times10^{7}$ & 4.04$^{+0.74}_{-0.50}\times10^{-3}$ & 1.28 \\
\medskip \\
\hline
\end{tabular}
\end{table}

For the PSCz-extrapolated sample the accessible volume is the maximum volume in which the object could still be seen and still be included in the \textit{IRAS} PSCz catalogue. Since objects with \mbox{$cz<300$\,km\,s$^{-1}$} were excluded from our sample this volume is not included in our calculation of $V_{i}$. For the PSCz-extrapolated DMF the dust masses were calculated using $T_{d}$ derived from the \textit{IRAS} 100/60 colour and $\beta$=2.

For completeness, the effect of excluding real 60\mic sources must be taken into account by applying a correction factor to the LF and DMF. This correction factor will be uncertain, since some excluded sources will be real and some not. Therefore we correct using our best estimate of real sources as follows. We corrected for two thirds of the sources we excluded as being contaminated by cirrus. The correct correction factor for the sources that were excluded because they have 100\mic upper limits is even more uncertain. These are probably all genuine sources, but they will generally have warmer colours than the sources that were not excluded. We arbitrarily corrected for 50\% of these. Including the correction for $\sim$100 sources without redshifts, the final correction factor for excluded sources is 1.27. This is obviously very uncertain, however at the most it could be 1.43 and at the least it could be 1.00. This produces maximum errors of $+$13\% and $-$21\% on the LF and DMF in addition to the errors described below. We made a correction for evolution out to z=0.2 using a density evolution $\propto (1+z)^{7}$ (Saunders et al. 1990). We confirmed that the strength assumed for the evolution made virtually no difference to our results.

The PSCz-extrapolated 850\mic LF and DMF are shown as filled circles in Figures~\ref{lumfun-plot} and~\ref{dmfun} respectively, and are given in tabular form in Tables~\ref{PSCZlf} and~\ref{PSCZdmf}. For comparison we also produce a `cold' PSCz-extrapolated DMF, produced as above but with dust masses calculated using $T_{d}$=20K and $\beta$=2; this is shown as open circles in Figure~\ref{dmfun-a} and listed in Table~\ref{PSCZdmf}.

While the errors on the directly measured LF and DMF are standard Poisson errors, the errors on the PSCz-extrapolated LF and DMF are derived from a combination of Poisson errors and the errors resulting from the fact that the 850\mic luminosities have been derived using the best-fitting linear relation to our SLUGS colour-colour plot (Figure~\ref{colplot}). In order to take into account how our `choice' of linear fit affects the LF we produce, we additionally generate two `extremes' of the PSCz-extrapolated LF and DMF using two alternative fits to the SLUGS colour-colour plot: 1) a fit to the OS data only, and 2) a fit to the IRS data only (linear fit parameters listed in Table~\ref{colplot-params}). We then use the maximum difference between these `extreme' LF values and our actual PSCz-extrapolated LF data points as the errors on our LF due to our `choice' of colour-colour linear relation. We then also take into account the number statistics, and thus add in quadrature the standard Poisson errors and the `choice of colour-colour fit' errors to obtain our total errors listed in Tables~\ref{PSCZlf} and~\ref{PSCZdmf}. In addition to these errors there are, at most, upper and lower errors of +13\% and $-$21\% from our choice of correction factors.

\subsection{Results and discussion}
\label{lumfun:results}
The directly measured OS LF and PSCz-extrapolated LF agree remarkably well over the range of luminosities covered by the SLUGS samples, yet we find that in comparison the IRS sample of D00 (plotted as triangles in Figures~\ref{lumfun-plot} and~\ref{dmfun}) consistently underestimates the submillimetre LF by a factor of 2 and the DMF by a factor of 4. The fact that we see this underestimate compared to our OS sample, which by definition should be free from any dust temperature selection effects, is  strong evidence that a population of `cold' dusty galaxies was indeed `missed' by \textit{IRAS} and that therefore the IRS sample was missing $\sim$half the galaxies. The bigger difference between the DMFs is probably due to the fact that, unlike the IRS sample, for the OS sample we do not have fitted isothermal SEDs for all galaxies and therefore have calculated dust masses using an assumed $T_{d}$=20K for $\sim$20\% of the sample (Section~\ref{method:850LF}).

We fit both the directly measured and PSCz-extrapolated 850\mic LFs and DMFs with Schechter functions of the form
\[
\Phi(L)dL=\phi(L)\left(\frac{L}{L_{\ast}}\right)^\alpha e^{-(L/L_{\ast})} dL/L_{\ast}
\]
(Press \& Schechter 1974; Schechter 1975).
The best-fitting parameters for the PSCz-extrapolated 850\mic LF and DMF are listed in Tables~\ref{PSCZlf} and~\ref{PSCZdmf} respectively, along with the reduced chi-squared values ($\chi^{2}_{\nu}$) for the fits; likewise best-fitting parameters for the directly measured 850\mic LF and DMF are shown in Table~\ref{optlf}.

We find that both the directly measured and PSCz-extrapolated LFs and DMFs are well-fitted by Schechter functions. For the PSCz-extrapolated LF and DMF the best-fitting Schechter function \mbox{($\alpha$\,=\,$-$1.38)} fits the data points extremely well across most of the luminosity range -- however, we note that the PSCz-extrapolated functions are much less well fitted at the high luminosity end. Investigation of the 3 or 4 high end luminosity bins has found several anomalies for the objects in these bins, the most striking of which is the fact that in each bin there is typically a small number of objects with accessible volumes 2 or 3 orders of magnitude lower than the rest of the objects in that bin, and thus it is these few objects in each of these bins which are the main contributors to the high space density. 

There are many possible explanations for the excess at the high luminosity end. One possible explanation could be that the objects in these bins are multiple systems. At larger distances \IRAS\/ galaxies are mostly very luminous starbursts and are frequently in interacting pairs. The density of galaxy pairs at these distances might by substantially higher than the local galaxy density which may produce an excess in the LF at high luminosities. Several authors find this excess at high luminosities or high masses. For example Lawrence et al. (1999) find a similar excess in their 60\mic LF, as do Garcia-Appadoo, Disney \& West (in preparation) for their HI Mass Function, who find that the higher HI masses are typically multiple systems. One can also think of ways our use of a global colour-colour relation might have produced a spurious excess if, for example, the galaxies at the highest luminosities have systematically different colours. This would not, however, explain the excess seen in the 60\mic LF.

In our earlier work the 850\mic LF derived from the IRS sample (D00) was found to have a slope steeper than $-$2 at the low luminosity end, suggesting that the submillimetre sky should be infinitely bright (a submillimetre `Olbers' Paradox'). Using the OS sample we find the slope of the PSCz-extrapolated 850\mic LF is $-$1.38, showing that the LF does flatten out at luminosities lower than those probed by the IRS sample, thus solving the submillimetre `Olbers' Paradox'.

\section{Conclusions}
\label{conc}
Following our previous SCUBA survey of an \IRAS-selected sample of galaxies we have carried out the first systematic survey of the local submillimetre Universe free from dust temperature selection effects -- a submillimetre survey of a sample of 81 galaxies selected from the CfA optical redshift survey. We obtained the following results:

(i) We detected 52 out of 81 galaxies at 850\mic and 19 galaxies at 450\,\micron. Many of these galaxies have 850\mic emission which appears extended with respect to the DSS optical emission, and which appears to correspond to very faint optical features.

(ii) We fitted two-component dust spectral energy distributions to the 60, 100, 450 and 850\mic flux densities for 18 of the galaxies which were detected at 850\mic \textit{and} at 450\,\micron. We find that the \IRAS\/ and submillimetre fluxes are well-fitted by a two-component dust model with dust emissivity index $\beta$=2. The tight and fairly constant ratio of $S_{450}/S_{850}$ for both the OS galaxies and the IRS galaxies is evidence that $\beta\approx 2$. The temperatures of the warm component range from 28 to 59\,K; the cold component temperatures range from 17 to 24\,K.

(iii) We find the ratio of the mass of cold dust to the mass of warm dust is much higher for our optically-selected galaxies than for our previous work on \IRAS-selected galaxies (DE01), and can reach values of $\sim$1000. By comparing the results for the \IRAS- and optically-selected samples we show that there is a population of galaxies containing a large proportion of cold dust that is unrepresented in the \IRAS\/ sample.

(iv) We also fitted single-temperature dust spectral energy distributions (to the 60, 100 and 850\mic flux densities) for the 41 galaxies in the OS sample with detections in all 3 wavebands. The mean best-fitting temperature for the sample is $\bar{T}_{d}=31.6\pm0.6$K and the mean dust emissivity index is $\bar{\beta}=1.12\pm0.05$. These values are significantly lower than for the IRS sample. The very low value of $\beta$ is additional evidence that galaxies, across all Hubble types, contain a significant amount of cold dust.

(v) Using our isothermal fits we find a mean dust mass \mbox{$\bar{M_{d}}=(2.34\pm0.36)\times{10^{7}}$ M$_{\odot}$}, which is comparable to that found for the IRS sample. However, using our two-component fits we find a mean dust mass a factor of two higher.

(vi) We find little change in the properties of dust in galaxies along the Hubble sequence, except a marginally significant trend for early-type galaxies to be less luminous submillimetre sources than late-types.

(vii) We detect 6 out of 11 ellipticals in the sample and find them to have dust masses in excess of \mbox{$10^{7}$ $M_{\odot}$}. It is possible, however, that for some of these galaxies the submillimetre emission may be synchrotron emission rather than dust emission.  

(viii) We have derived local submillimetre luminosity and dust mass functions, both directly from the optically-selected SLUGS sample and by extrapolation from the \textit{IRAS} PSCz survey, and find excellent agreement between the two. By extrapolating the spectral energy distributions of the \textit{IRAS} PSCz survey galaxies out to 850\mic we have probed a wider range of luminosities than probed directly by the SLUGS samples. We find the LFs to be well-fitted by Schechter functions except at the highest luminosities. We have shown that, whereas the slope of the \textit{IRAS}-selected LF at low luminosities was steeper than $-$2 (a submillimetre `Olbers' Paradox'), the PSCz-extrapolated LF, as expected, flattens out at the low luminosity end and has a slope of \mbox{$-$1.38}. 

(ix) We find that as a consequence of the omission of a population of `cold' dusty galaxies from the \IRAS\/ sample the LF presented in our earlier work (D00) is too low by a factor of 2, and the DMF by a factor of 4.

In order to further investigate the properties of dust in galaxies follow-up optical imaging (to obtain deeper images than available from the DSS) for the whole OS sample detected at 850\mic is needed, in order to make a full comparison of the optical versus submillimetre emission. This is important since for many of the OS sample galaxies the 850\mic emission appears extended with respect to the DSS optical emission. Work on obtaining this data is in progress.

\section*{Acknowledgements}
We thank Diego Garcia-Appadoo for providing information about his HI Mass Function, and Jonathan Davies for his useful comments. We also thank Steve Serjeant for useful discussions. Many of the observations for this survey were carried out as part of the JCMT service programme, so we are grateful to Dave Clements, Rob Ivison and the many other observers and members of the JCMT staff who have contributed to this project in this way. This research has made use of the NASA/IPAC Extragalactic Database (NED) and the NASA/IPAC Infrared Science Archive which are operated by the Jet Propulsion Laboratory, California Institute of Technology, under contract with the National Aeronautics and Space Administration. We have also made use of the LEDA and DSS databases. Research by LD and SE is supported by the Particle Physics and Astronomy Research Council.


\begin{thebibliography}{}
\bibitem[]{} Alton P.B. et al., 1998a, A\&A, 335, 807
\bibitem[]{} Alton P.B., Bianchi S., Rand R.J., Xilouris E., Davies J.I., Trewhella M., 1998b, ApJ, 507, L125
\bibitem[]{} Alton P.B., Lequeux J., Bianchi S., Churches D., Davies J., Combes F., 2001, A\&A, 366, 451
\bibitem[]{} Archibald E.N. et al., 2002, MNRAS, 336, 1
\bibitem[]{} Aumann H.H., Fowler J.W., Melnyk M., 1990, ApJ, 99, 1674
\bibitem[]{} Avni Y., Bahcall J.N., 1980, ApJ, 235, 694
\bibitem[]{} Barger A.J., Cowie L.L., Sanders D.B., Fulton E., Taniguchi Y., Sato Y., Kawara K., Okuda H., 1998, Nat, 394, 248
\bibitem[]{} Barger A.J., Cowie L.L., Sanders D.B., 1999, ApJ, 518, L5
\bibitem[]{} Beichman C.A., Neugebauer G., Habing H.J., Clegg P.E., Chester T.J., eds, 1988, \textit{Infrared Astronomical Satellite (IRAS)} Explanatory Supplement, NASA RP-1190
\bibitem[]{} Bianchi S., Alton P.B., Davies J.I., Trewhella M., 1998, MNRAS, 298, L49
\bibitem[]{} Bianchi S., Davies J.I., Alton P.B., Gerin M., Casoli F., 2000, A\&A, 353, L13
\bibitem[]{} Blain A.W., Kneib J.-P., Ivison R.J., Smail I., 1999a, ApJ, 512, L87
\bibitem[]{} Blain A.W., Smail I., Ivison R.J., Kneib J.-P., 1999b, MNRAS, 302, 632
\bibitem[]{} Bothun G.D., Lonsdale C.J., Rice W., 1989, ApJ, 341, 129
\bibitem[]{} Braine J., Gu\'elin M., Dumke M., Brouillet N., Herpin F., Wielebinski R., 1997, A\&A, 326, 963
\bibitem[]{} Bregman J.N., Snider B.A., Grega L., Cox C.V., 1998, ApJ, 499, 670
\bibitem[]{} Catalogued Galaxies and Quasars Observed in the \IRAS\/ Survey, 1989, Version 2. Prepared by Fullmer L., Lonsdale C.J., JPL, Pasadena
\bibitem[]{} Chamaraux P., Balkowski C., Fontanelli P., 1987, A\&AS, 69, 263
\bibitem[]{} Chini R., Kr\"ugel E., Kreysa E., Gemuend H.-P., 1989, A\&A, 216, L5
\bibitem[]{} Chini R. \& Kr\"ugel E., 1993, A\&A, 279, 385
\bibitem[]{} Condon J.J., Cotton W.D., Broderick J.J., 2002, AJ, 124, 675
\bibitem[]{} Contursi A., Boselli A., Gavazzi G., Bertagna E., Tuffs R., Lequeux J., 2001, A\&A, 365, 11
\bibitem[]{} Cox C.V., Bregman J.N., Schombert J.M., 1995, ApJS, 99, 405
\bibitem[]{} Cox P., Kr\"ugel E., Mezger P.G., 1986, A\&A, 155, 380
\bibitem[]{} Davies J.I., Alton P., Trewhella M., Evans R., Bianchi S., 1999, MNRAS, 304, 495
\bibitem[]{} Devereux N.A., Young J.S., 1990, ApJ, 359, 42
\bibitem[]{} Dumke M., Braine J., Krause M., Zylka R., Wielebinski R., Gu\'elin M., 1997, A\&A, 325 124
\bibitem[]{} Dunne L., 2000, PhD thesis, University of Wales
\bibitem[]{} Dunne L., Eales S., Edmunds M., Ivison R., Alexander P., Clements D.L., 2000, MNRAS, 315, 115 (Paper I)
\bibitem[]{} Dunne L., Eales S.A., 2001, MNRAS, 327, 697 (Paper II)
\bibitem[]{} DuPrie K., Schneider S.E., 1996, AJ, 112 937
\bibitem[]{} Eales S.A., Lilly S.J., Gear W.K., Dunne L., Bond R.J., Hammer F., Le F\`evre O., Crampton D., 1999, ApJ, 515, 518
\bibitem[]{} Economou F., Jenness T., Currie M., Adamson A., Allan A., Cavanagh B., 2004, Starlink User Note 230, Starlink Project, CLRC
\bibitem[]{} Feigelson, E.D., Nelson, P.I., 1985, ApJ, 293, 192
\bibitem[]{} Frayer D.T., Ivison R.J., Smail I., Yun M.S., Armus L., 1999, AJ, 118, 139
\bibitem[]{} Freudling W., 1995, A\&AS, 112, 429
\bibitem[]{} Giovanelli R., Haynes M.P., 1993, AJ, 105, 1271
\bibitem[]{} Goudfrooij P., Hansen L., J\o rgensen H.E., N\o rgaard-Nielsen H.U., 1994, A\&AS, 105, 341
\bibitem[]{} Goudfrooij P., de Jong T., 1995, A\&A, 298, 784
\bibitem[]{} Gu\'elin M., Zylka R., Mezger P.G., Haslam C.G.T., Kreysa E., Lemke R., Sievers A.W., 1993, A\&A, 279, L37
\bibitem[]{} Gu\'elin M., Zylka R., Mezger P.G., Haslam C.G.T., Kreysa E., 1995, A\&A, 298 L29
\bibitem[]{} Haas M., Lemke D., Stickel M., Hippelein H., Kunkel M., Herbstmeier U., Mattila K., 1998, A\&A, 338, L33
\bibitem[]{} Haas M., Klaas U., Coulson I., Thommes E., Xu C., 2000, A\&A, 356, L83
\bibitem[]{} Haynes M.P., Giovanelli R., 1988, AJ, 95, 607
\bibitem[]{} Haynes M.P., Giovanelli R., 1991, ApJS, 77, 331 
\bibitem[]{} Haynes M.P., Giovanelli R., Chamaraux P., da Costa L.N., Freudling W., Salzer J.J., Wegner G., 1999, AJ, 117, 2039
\bibitem[]{} Helou G., Khan I.R., Malek L., Boehmer L., 1988, ApJS, 68, 151
\bibitem[]{} Hildebrand R.H., 1983, QJRAS, 24, 267
\bibitem[]{} Hippelein H., Haas M., Tuffs R.J., Lemke D., Stickel M., Klaas U., V\"olk H.J., 2003, A\&A, 407, 137
\bibitem[]{} Holland W.S. et al., 1999, MNRAS, 303, 659
\bibitem[]{} Huchra J., Davis M., Latham D., Tonry J., 1983, ApJS, 52, 89
\bibitem[]{} Huchtmeier W.K., 1997, A\&A, 319, 401
\bibitem[]{} Huchtmeier W.K., Richter O.-G., 1989, A General Catalog of HI Observations of Galaxies: The Reference Catalog, XIX. Springer-Verlag, Berlin, Heidelberg, New York
\bibitem[]{} Hughes D.H. et al., 1998, Nat, 394, 241
\bibitem[]{} James A., Dunne L., Eales S., Edmunds M.G., 2002, MNRAS, 335, 753
\bibitem[]{} Jenness T., Lightfoot J.F., 1998, in Albrecht R., Hook R.N., Bushouse H.A., eds, ASP Conf. Ser. Vol. 145, Astronomical Data Analysis Software \& Systems VII. Astron. Soc. Pac., San Francisco, p. 216
\bibitem[]{} Jenness T., Lightfoot J.F., 2000, {\small {SURF}} -- SCUBA User Reduction Facility. Starlink User Note 216, Starlink Project, CLRC
\bibitem[]{} Jenness T., Stevens J.A., Archibald E.N., Economou F., Jessop N.E., Robson E.I., 2002, MNRAS, 336, 14
\bibitem[]{} Knapp G.R., Guhathakurta P., Kim D.-W., Jura M., 1989, ApJS, 70, 329 
\bibitem[]{} Lawrence A. et al., 1999, MNRAS, 308, 897
\bibitem[]{} Lilly S.J., Eales S.A., Gear W.K., Hammer F., Le F\`evre O., Crampton D., Bond R.J., Dunne L., 1999, ApJ, 518, 641
\bibitem[]{} Lu N.Y., Hoffman G.L., Groff T., Roos T., Lamphier C., 1993, ApJS, 88, 383
\bibitem[]{} Lonsdale Persson C.J., Helou G., 1987, ApJ, 314, 513
\bibitem[]{} M\'arquez I. et al., 1999, A\&AS, 140,1
\bibitem[]{} Masi S. et al., 1995, ApJ, 452, 253
\bibitem[]{} Meijerink R., Tilanus R.P.J., Dullemond C.P., Israel F.P., van der Werf P.P., 2005, A\&A, 430, 427
\bibitem[]{} Mortier A.M.J. et al., 2005, MNRAS, 363, 509
\bibitem[]{} Moshir M. et al., 1990, Infrared Astronomical Satellite Catalogs, The Faint Source Catalog, version 2.0
\bibitem[]{} Neininger N., Gu\'elin M., Garc\'\i a-Burillo S., Zylka R., Wielebinski R., 1996, A\&A, 310, 725
\bibitem[]{} Papadopoulos P.P., Seaquist E.R., 1999, ApJ, 514, L95
\bibitem[]{} Paturel G., Fouque P., Bottinelli L., Gouguenheim L., 1989, A\&AS, 80, 299
\bibitem[]{} Paturel G., Petit C., Prugniel P., Theureau G., Rousseau J., Brouty M., Dubois P., Cambr\'esy L., 2003, A\&A, 412 45
\bibitem[]{} Popescu C.C., Tuffs R.J., V\"olk H.J., Pierini D., Madore B.F., 2002, ApJ, 567, 221
\bibitem[]{} Press W.H., Schechter P., 1974, ApJ, 187, 425
\bibitem[]{} Reach W.T. et al., 1995, ApJ, 451, 188
\bibitem[]{} Rowan-Robinson M., Crawford J., 1989, MNRAS, 238, 523
\bibitem[]{} Saunders W., Rowan-Robinson M., Lawrence A., Efstathiou G., Kaiser N., Ellis R.S., Frenk C.S., 1990, MNRAS, 242, 318
\bibitem[]{} Saunders W. et al., 2000, MNRAS, 317, 55
\bibitem[]{} Schechter P., 1975, PhD thesis, California Institute of Technology
\bibitem[]{} Seaquist E., Yao L., Dunne L., Cameron H., 2004, MNRAS, 349, 1428
\bibitem[]{} Serjeant S., Harrison D., 2005, MNRAS, 356, 192
\bibitem[]{} Sievers A.W., Reuter H.-P., Haslam C.G.T., Kreysa E., Lemke R., 1994, A\&A, 281, 681
\bibitem[]{} Smail I., Ivison R.J., Blain A.W., 1997, ApJ, 490, L5
\bibitem[]{} Sodroski T.J., et al., 1994, ApJ, 428, 638
\bibitem[]{} Sodroski T.J, Odegard N., Arendt R.G., Dwek E., Weiland J.L., Hauser M.G., Kelsall T., 1997, ApJ, 480, 173
\bibitem[]{} Soifer B.T., Boehmer L., Neugebauer G., Sanders D.B., 1989, AJ, 98, 766
\bibitem[]{} Spinoglio L., Andreani P., Malkan M.A., 2002, ApJ, 572, 105
\bibitem[]{} Stevens J.A., Amure M., Gear W.K., 2005, MNRAS, 357, 361
\bibitem[]{} Stickel M., et al., 2000, A\&A, 359, 865
\bibitem[]{} Stickel M., Lemke D., Klaas U., Krause O., Egner S., 2004, A\&A, 422, 39
\bibitem[]{} Theureau G., Bottinelli L., Coudreau-Durand N., Gouguenheim L., Hallet N., Loulergue M., Paturel G., Teerikorpi P., 1998, A\&AS, 130, 333
\bibitem[]{} Thomas H.C., Dunne L., Clemens M.S., Alexander P., Eales S., Green D.A., James A, 2002, MNRAS, 331 853
\bibitem[]{} Thomas H.C., Dunne L., Green D.A.,  Clemens M.S., Alexander P., Eales S., 2004, MNRAS, 348, 1197
\bibitem[]{} van Dokkum P.G., Franx M., 1995, AJ, 110, 2027
\bibitem[]{} Wall J.V., Jenkins C.R., 2003, Practical Statistics for Astronomers, CUP
\bibitem[]{} Xilouris E.M., Byun Y.I., Kylafis N.D., Paleologou E.V., Papamastorakis J., 1999, A\&A, 344, 868
\end{thebibliography}
\end{document}